%% file: RRP-master-arXiv-v2.tex
\pdfoutput=1

\documentclass[rmp,aps,twocolumn]{revtex4}
\newcommand {\ET}{BEDT-TTF }
\newcommand {\ETn}{BEDT-TTF}
\newcommand {\kET}{$\kappa$-BEDT-TTF }
\newcommand {\ETtwo}{(BEDT-TTF)$_2$ }
\newcommand {\kETX}{$\kappa$-(BEDT-TTF)$_2X$ }
\newcommand {\kETXn}{$\kappa$-(BEDT-TTF)$_2X$}

\newcommand {\kX}{$\kappa$-(BEDT-TTF)$_2X$ }
\newcommand {\kXn}{$\kappa$-(BEDT-TTF)$_2X$}

\newcommand {\Br}{$\kappa$-(BEDT-TTF)$_2$\-Cu\-[N\-(CN)$_2$]\-Br }
\newcommand {\Brn}{$\kappa$-(BEDT-TTF)$_2$\-Cu\-[N\-(CN)$_2$]\-Br}
\newcommand {\Cl}{$\kappa$-(BEDT-TTF)$_2$\-Cu\-[N\-(CN)$_2$]\-Cl }
\newcommand {\NCS}{$\kappa$-(BEDT-TTF)$_2$\-Cu\-(NCS)$_2$ }

\newcommand {\CN}{$\kappa$-(BEDT-TTF)$_2$\-Cu$_2$\-(CN)$_3$ }
\newcommand {\CNn}{$\kappa$-(BEDT-TTF)$_2\-$Cu$_2$\-(CN)$_3$}
\newcommand {\Cln}{$\kappa$-(BEDT-TTF)$_2$\-Cu\-[N\-(CN)$_2$]\-Cl}
\newcommand {\BrCl}{\ClBr}
\newcommand {\BrCln}{\ClBrn}
\newcommand {\ClBr}{$\kappa$-(BEDT-TTF)$_2$\-Cu\-[N\-(CN)$_2$]\-Cl$_{1-x}$Br$_x$ }
\newcommand {\ClBrn}{$\kappa$-(BEDT-TTF)$_2$\-Cu\-[N\-(CN)$_2$]\-Cl$_{1-x}$Br$_x$}
\newcommand {\NCSn}{$\kappa$-(BEDT-TTF)$_2$\-Cu\-(NCS)$_2$}

\newcommand {\dmit}{Pd(dmit)$_2$ }
\newcommand {\dmitn}{Pd(dmit)$_2$}
\newcommand {\dmittwo}{[Pd(dmit)$_2$]$_2$ }
\newcommand {\dmittwon}{[Pd(dmit)$_2$]$_2$}
\newcommand {\Sbtwo}{Et$_2$\-Me$_{2}$\-Sb\-[Pd(dmit)$_2$]$_2$ }
\newcommand {\EtMePn}{Et$_n$\-Me$_{4-n}\-Pn\-$[Pd(dmit)$_2$]$_2$ }

\newcommand {\etal}{{\it et al}. }
\newcommand {\etaln}{{\it et al}.}

\usepackage{bibunits,epsfig,latexsym,amssymb,amsmath,amsbsy,graphics,graphicx,mathrsfs,enumerate}
\usepackage{rotating}
\usepackage{epstopdf}

\begin{document}

\def \cation{$\kappa$-(ET)$_2$}~
\def \kpx{$\kappa$-(ET)$_2X$}
\def \kbr{$\kappa$-(ET)$_2$Cu[N(CN)$_2$]Br}
\def \deut8br{$\kappa$(d8)-(ET)$_2$Cu[N(CN)$_2$]Br}
\def \h8br{h8-(ET)$_2$Cu[N(CN)$_2$]Br}
\def \kcl{$\kappa$-(ET)$_2$Cu[N(CN)$_2$]Cl}
\def \kncs{$\kappa$-(ET)$_2$Cu(NCS)$_2$}
\def \kcn3{$\kappa$-(ET)$_2$Cu$_2$(CN)$_3$}
\def \me4p{$\beta'-$(Me$_4$P)[Pd(dmit)$_2$]$_2$}
\def \et2me2p{$\beta'-$(Et$_2$Me$_2$P)[Pd(dmit)$_2$]$_2$}
\def \>{\textgreater}
\def \<{\textless}
\def \q{\vec{q}}
\def \Q{\vec{Q}}
\def \kpcl{$\kappa$-Cl}
\def \kpbr{$\kappa$-Br}
\def \kpncs{$\kappa$-NCS}
\def \kpcn3{$\kappa$-(CN)$_3$}
\def \d8pbr{$\kappa$(d8)-Br}
\def \m{\mathrm{m}}
\def \max{\mathrm{max}}
\def \cross{\mathrm{cross}}
\def \M{\mathrm{M}}
\def \c{\mathrm{c}}
\def \lw{\mathrm{LW}}
\def \af{\mathrm{AF}}
\def \fm{\mathrm{FM}}
\def \sf{\mathrm{SF}}
\def \res{{\rho \sim T^2}}
\def \us{{\Delta v/v}}
\def \nmr{\mathrm{NMR}}
\def \ks{{K_s}}
\def \exp{\mathrm{exp}}
\def \chiqw{$\chi({\bf q},\omega)$}
\def \fm{\xi_\mathrm{FM}}
\def \afm{\xi_\mathrm{AF}}

\title{Quantum frustration in organic Mott insulators:
from spin liquids to unconventional superconductors}

\author{B. J. Powell}
\email{bjpowell@gmail.com}
\address{Department of Physics, University of Queensland, Brisbane,
4072, Australia}
\author{Ross H. McKenzie}
\email{r.mckenzie@uq.edu.au}
\address{Department of Physics, University of Queensland, Brisbane,
4072, Australia}

\date{\today}

\begin{abstract}
We review the interplay of frustration and strong electronic correlations in quasi-two-dimensional organic charge transfer salts, such as (\ETn)$_2X$ and Et$_n$Me$_{4-n}Pn$\dmittwon. These two forces drive a range of exotic phases including spin liquids, valence bond crystals, pseudogapped metals,
 and unconventional superconductivity. 
Of particular interest is that in several materials 
with increasing pressure
there is a first-order transition from a spin liquid Mott insulating
state to a superconducting state.
Experiments on these materials raise a number of profound questions about the quantum behaviour of frustrated systems, particularly the intimate
connection between spin liquids and superconductivity. Insights into these questions  have come from a wide range of theoretical techniques including first principles electronic structure, quantum many-body theory and quantum field theory.  In this review we introduce some of
the basic ideas of the field by discussing a simple frustrated Heisenberg model with four spins. 
We then describe the key experimental results, emphasizing
that for two materials, \CN and Et\-Me$_3$\-Sb\-[Pd(dmit)$_2$]$_2$,
  there is strong evidence for a spin liquid ground state,
and for another, Et\-Me$_3$\-P\-[Pd(dmit)$_2$]$_2$, there is evidence of a valence bond crystal 
ground state.
 We review theoretical attempts to explain these phenomena, arguing that
they can be captured by a Hubbard model on the anisotropic triangular
lattice at half filling, and that Resonating Valence Bond 
(RVB) wavefunctions
capture most of the essential physics. 
We review evidence that this Hubbard model can have a spin
liquid ground state for a range of parameters that are realistic for the relevant materials.
In particular, spatial anisotropy and ring exchange are
key to destabilising magnetic order.
We conclude by summarising the progress made thus far and identifying some of the key questions still to be answered. 
\end{abstract}

\maketitle

\tableofcontents

\input{intro}

\input{ET}

\input{dmit}

\input{nmr}

\input{lattice}

\input{gauge}

\input{quasiparticles}

\input{quarter}

\input{conclusions}

\section*{Acknowledgements}

We thank J. G. Analytis, A. Ardavan, A. Bardin, S. J. Blundell, P. Burn, C.-H. Chung, R. Coldea, J. O. Fjaerestad, A. C. Jacko, C. Janani, S.-C. Lo,
J. B. Marston, J. Merino, P.  Pairor, M. R. Pedersen, E. Scriven,
R. R. P. Singh, M. F. Smith, A. P. Stephenson,
and E. Yusuf for fruitful collaborations related to this review.
We thank L. Balents, L. Bartosch, M. de Souza, T. Grover,
R. Kato, H. H. Lai, P. A. Lee, S. Mazumdar,
O. Motrunich, B. Normand, T. Senthil,
and A. Vishwanath for helpful discussions. We thank C. Janani for drawing our attention to a number of typographical errors in an earlier draft of this manuscript.

BJP was the recipient of an Australian Research Council (ARC) Queen Elizabeth II Fellowship (project no. DP0878523). RHM was the recipient of an ARC Australian Professorial Fellowship (project no. DP0877875).



\bibliographystyle{jphysicsB}
\bibliography{rossrefs,expts}

\end{document}

%% file: intro.tex
\section{Introduction}

In the early 1970's, Anderson 
and Fazekas \cite{AndersonMRS73,FazekasPM74} 
proposed that the ground state of the antiferromagnetic
 Heisenberg spin-1/2 model on the triangular lattice did not break
spin rotational symmetry, i.e., had no net magnetic moment. 
A state of matter characterised by well
 defined local moments and the absence of long range order has become known as a spin liquid \cite{NormandCP09}. Such states, are known in one-dimensional (1d) systems, but 1d systems have some very special properties that are not germane to higher dimensions. Until very recently there has been a drought of experimental evidence for spins liquids in higher dimensions \cite{LeeScience08}.

In 1987 Anderson  \cite{AndersonScience87}, stimulated by the discovery of 
high-$T_c$ superconductivity in layered copper oxides, made a radical proposal that has given rise to lively debate ever since. We summarise
Anderson's proposal as:
\begin{quote}
The fluctuating spin singlet pairs produced by the exchange interaction in the Mott insulating state become charged superconducting pairs when the insulating state is destroyed by doping, frustration or reduced correlations.
\end{quote}
These fluctuations are enhanced by spin frustration and low dimensionality.
Furthermore, partly inspired by resonating valence bond (RVB) ideas
from chemical bonding \cite{AndersonPT08,Shiak08}, Anderson proposed a variational wave function for the
Mott insulator:  a BCS superconducting state from which all doubly
occupied sites are projected out.

In the decades since, there has been an enormous outgrowth of ideas about spin liquids and frustrated quantum systems, which we will review. We will also  consider the extent to which
several families of organic charge transfer salts
can be used as tuneable systems to test such ideas about the
interplay of superconductivity, Mott insulation, quantum
fluctuations, and  spin frustration.

A goal of this review is not to be exhaustive but rather to 
be pedagogical, critical, and constructive.
We will attempt to follow the goals for such reviews proposed long ago \cite{HerringPT68}.

\subsection{Motivation: frustration, spin liquids, and spinons}

\subsubsection{Key questions }              

A major goal for this review  will be to address the following questions:

\begin{enumerate}

\item Is there a clear relationship between superconductivity in 
organic charge transfer salts
 and in other strongly correlated electron systems?

\item Are there materials for which  the ground state of the Mott insulating phase 
is a spin liquid? 

\item  What is the relationship between spin liquids and superconductivity?
In particular, does the same fermionic pairing occur in both?

\item What are the quantum numbers (charge, spin, statistics) of the quasiparticles in each phase?

\item Are there deconfined spinons in the insulating phase of any of these materials? 

\item Can spin-charge separation occur in the metallic phase?

\item  In the metallic phase close to the Mott insulating phase 
is there an anisotropic pseudogap, as in the cuprates?

\item What is  the simplest low-energy
effective quantum many-body Hamiltonian on a lattice that can
describe all possible ground states of these materials?

\item  Can a RVB variational wave function give an
appropriate  theoretical description of the competition between the Mott insulating and the superconducting phase?

\item Is there any significant difference between destroying the Mott insulator by
hole doping and by increasing the bandwidth?

\item  For systems close to the isotropic triangular lattice, does the superconducting state have broken time-reversal symmetry?

\item How can we quantify the extent of frustration? Are there  differences between classical and quantum frustration? If so what are the differences?

\item What is the relative importance of frustration and 
static disorder due to impurities?

\item Is the ``chemical pressure" hypothesis valid?

\item Is there quantum critical behaviour associated with  quantum phase transitions    in these materials?

\item  Do these materials illustrate specific ``organising principles'' that are
useful for understanding other frustrated materials?

\end{enumerate}

At the end of the review we consider some possible answers to these questions.

\subsubsection{A hierarchy of theories: from quantum chemistry to field theory}

\begin{figure}
\begin{centering}
\includegraphics[width=8cm]{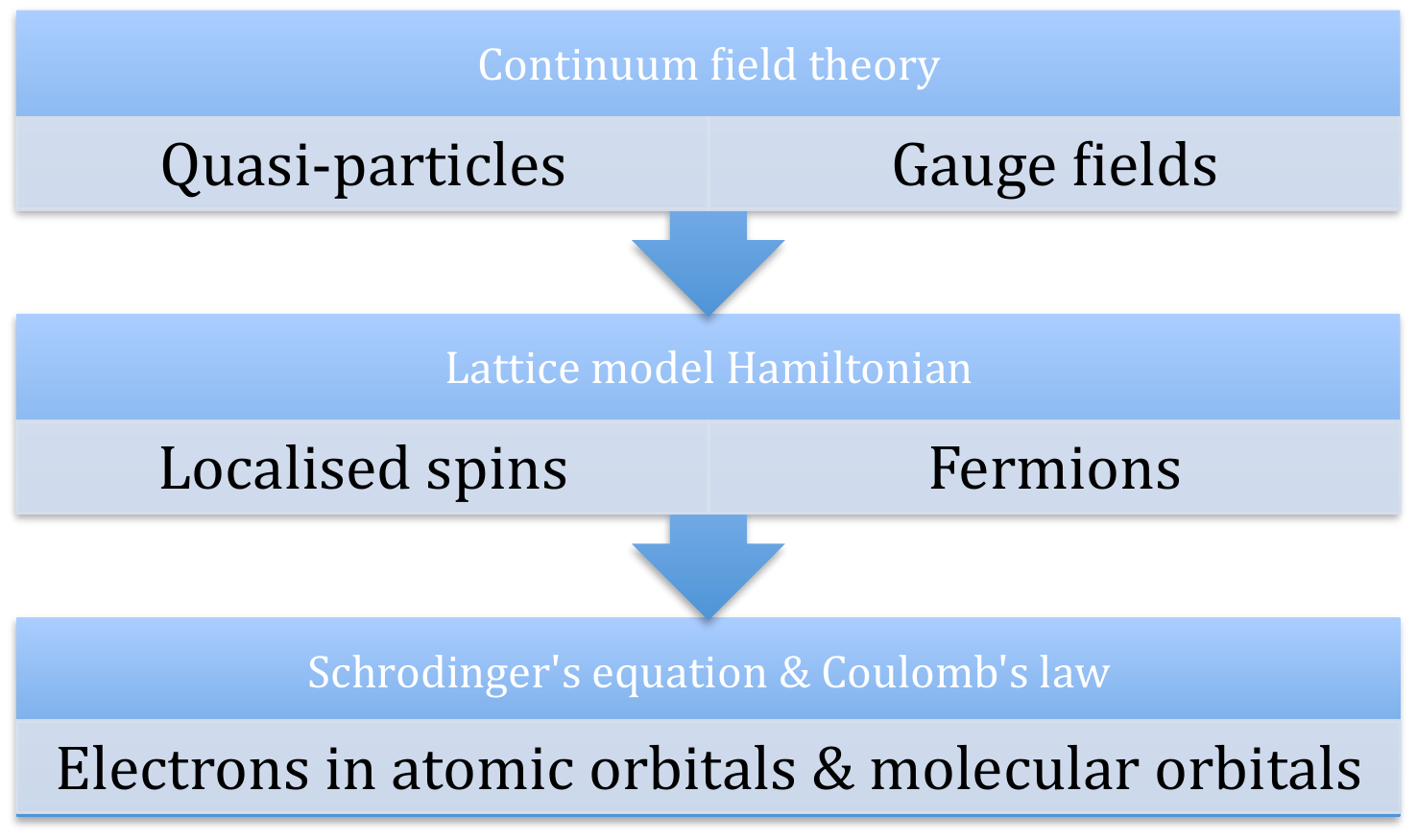}
\caption{The hierarchy of objects and descriptions associated with theories
of organic charge transfer salts.
The arrows point in the direction of decreasing length scales, 
increasing energy scales, and increasing numbers of degrees
of freedom.
At the level of quantum chemistry (Schr\"odinger's equation
and Coulomb's law) one can describe the electronic
states of single (or pairs of) molecules in terms of molecular orbitals (which can be approximately
viewed as superpositions  of atomic orbitals). Just a few of these molecular orbitals interact significantly with those
of  neigbouring molecules in the solid. Low-lying electronic states
of the solid can be described in terms of itinerant fermions  on a lattice
and an effective Hamiltonian such as a Hubbard model (see Section
\ref{sec:hubbard}). In the Mott insulating phase the electrons are localised on single lattice sites and can described by a Heisenberg spin model (see Section
\ref{sec:heisenberg}). The low-lying excitations of these lattice Hamiltonians
and long-wavelength properties of the
system may have a natural description in terms of quasi-particles which can be described by a continuum field theory such as a non-linear sigma model.
At this level unexpected objects may emerge such as gauge fields and quasi-particles with fractional statistics (see Section \ref{sec:gauge}).
}
\label{fig:fields}
\end{centering}
\end{figure}

The quantum many-body physics of condensed matter provides many striking  examples of emergent phenomena at different energy and length scales \cite{AndersonScience72,LaughlinPNAS00,Coleman,Wenbook,McKenzieNP07}.
Figure \ref{fig:fields} illustrates how this is played out
in the molecular crystals, which form the focus of this article, showing the stratification
of different theoretical treatments and the associated
objects.
It needs to be emphasized that when it comes to theoretical
descriptions going up the hierarchy is extremely
difficult, particularly determining the
quantum numbers of quasi-particles and the effective interactions
between them, starting from a lattice Hamiltonian.

\subsubsection{Organic charge transfer salts are an important class
of materials}              

Organic charge transfer salts have a number of features that make them a playground for the study of
 quantum many-body physics. They have of  several properties that are distinctly different from other
strongly correlated electron materials, such
as transition metal oxides and intermetallics. These properties include:

\begin{itemize}

    \item They are available in ultra-pure single crystals, which allow observation of quantum magnetic oscillations such as the de Haas van Alphen effect.

    \item The superconducting transition temperature and upper critical field are low enough that one can destroy the superconductivity and probe the metallic state in steady magnetic fields less than 20 Tesla.  As a result, one can observe rich physics in experimentally accessible magnetic fields and pressure ranges

\item Chemical substitution provides a means to tune the ground state.

\item Chemical doping (and the associated disorder)  is not necessary
 to induce transitions between different phases.

\item These materials are compressible enough that pressures of the order of 
kbars  can induce transitions between different ground states.

\end{itemize}

Consequently, over the past decade it has been possible to observe several unique effects due to strongly correlated electrons, sometimes phenomena that have not been seen in inorganic materials. These significant 
observations include:
\begin{itemize}

\item Magnetic field induced superconductivity.

\item A first-order transition between a Mott insulator and superconductor induced with deuterium substitution, anion substitution, pressure, or magnetic field.

\item A valence bond solid in a frustrated antiferromagnet.

\item A spin liquid in a frustrated antiferromagnet.

\item Novel critical exponents near the critical point of Mott metal-insulator transition.

\item Collapse of the Drude peak in the optical conductivity
(a signature of the destruction of quasi-particles) above temperatures of order of tens of Kelvin in the metallic phase.

\item Bulk measurement of the Fermi surface using angle-dependent magnetoresistance.

\item Low superfluid density in a weakly correlated metal.

\item Multi-ferroic states.

\item Superconductivity near a charge ordering transition.
\end{itemize}

\begin{figure}
\begin{centering}
\includegraphics[width=9cm]{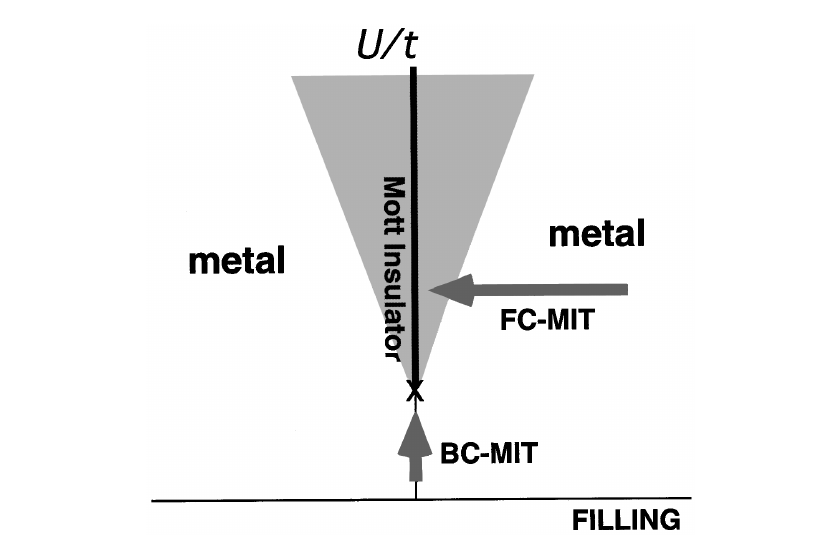}
\caption{
Schematic phase diagram associated with the Mott-Hubbard metal-insulator transition.
[Figure after Reference \cite{ImadaRMP98}].
 The Mott insulating phase occurs at half filling and when the
 on-site Coulomb repulsion $U$ is much larger than the hopping energy $t$
and the associated band width.
A transition to a metallic phase occurs either by doping away from half filling [FC-MIT= filling controlled metal-insulator transition] or by
decreasing the ratio $U/t$ [BC-MIT = bandwidth controlled metal-insulator transition]. In the cuprates a FC-MIT occurs whereas in the organic charge transfer salts considered in this review one might argue that BC-MIT occurs.
On the other hand, perhaps one should consider a third co-ordinate, the frustration, in addition to the filling and band width. This would lead to the notion ofa frustration controlled transition (FrC-MIT).
In the Hubbard model on the anisotropic lattice at half-filling for fixed
$U/t$ increasing the hopping $t'/t$
 can drive an insulator to metal transition
(compare Figure \ref{fig:rvbphasediag}).
[Copyright (1998) by the American Physical Society.]
}
\label{fig:motttrans}
\end{centering}
\end{figure}

Figure \ref{fig:motttrans}
illustrates schematically two possible different routes to destroying the
Mott insulating phase, either by varying the band filling or by
varying the bandwidth.
Another possible route is by varying the amount of frustration of
the spin interactions.
 An important consequence of Anderson RVB's theory of 
the filling controlled metal-insulator transition (FC-MIT) is that the
 ``preexisting magnetic singlet pairs of the insulating state become charged superconducting pairs when the insulator is doped sufficiently strongly"  \cite{AndersonScience87}. It is therefore important to understand whether this extends to the bandwidth controlled metal-insulator transition (BC-MIT) were one has equal numbers of ``holons" and ``doublons". More generally, an important question, that has not yet received adequate attention, is what are the similarities and differences between the FC-MIT and the BC-MIT?

\subsubsection{What are spin liquids?}
\label{sec:spinliquid}

This question has recently been reviewed in detail
 \cite{NormandCP09,BalentsNat10,Sachdev09}.
 There are several alternative definitions.
The definition that we think is the most illuminating, because it brings
out their truly exotic nature, is
the following. \begin{quote}{\it
A spin liquid
has a ground state in which there is no long-range magnetic order and no breaking of spatial symmetries (rotation or translation) and which is not adiabatically connected to the band (Bloch) insulator.}\end{quote}
One can write down many such quantum states. 
Indeed, Wen classified hundreds of them for the square lattice \cite{WenPRB02}.
But the key question is whether such a state can be the 
ground state of a physically realistic Hamiltonian.
A concrete example is the ground state
of the one-dimensional antiferromagnetic Heisenberg model with nearest-neighbour
interactions.
However, despite an exhaustive search
since Anderson's 1987 Science paper, \cite{AndersonScience87} it seems extremely difficult to find a
physically realistic Hamiltonian in two dimensions which
has such a ground state.

As far as we are aware there is still no
  definitive    counter-example to the following conjecture:
\begin{quote}
Consider a family of spin-1/2 Heisenberg models on a two-dimensional 
lattice with short range antiferromagnetic exchange 
interactions (pairwise, ring exchange and higher order terms are allowed).
 The Hamiltonian is invariant under $SU(2) \times L$, where $L$ is a 
space group and there is a non-integer total spin in
the repeat unit of the lattice Hamiltonian.
Let $\gamma$ be a parameter which can be used to
distinguish  different Hamiltonians in the family (e.g., it could be
the relative magnitude of different interaction terms in
the Hamiltonian). Then a non-degenerate  ground state is
only possible for discrete values of $\gamma$ (e.g., at 
a quantum critical point).
In other words, the ground state  
spontaneously breaks at least one of 
the two symmetries $SU(2)$ and $L$ over all continuous ranges 
of $\gamma$.\end{quote}

The requirement of non-integer spin in the repeat unit ensures   that 
the generalisation of the Lieb-Schultz-Mattis theorem to dimensions greater
than one \cite{HastingsPRB04,AletPA06} does not apply.
The theorem states that for spin-1/2 systems with one spin per unit cell on a two-dimensional lattice, if the ground state is non-degenerate
and there is no  symmetry breaking,
one cannot have a non-zero energy gap to the lowest excited state.
Note that, the triangular, kagome, and pyrochlore lattices
contain one, three, and four spins per unit cell respectively \cite{NormandCP09}. Hence, Hasting's theorem cannot be used to rule out a spin
liquid for the pyrochlore lattice.

One of the best candidate counter examples to the above conjecture
is the Heisenberg model on the triangular lattice with ring exchange \cite{LimingPRB00} which will be discussed in more detail
in Section \ref{sec:heisenberg}.


Sachdev \cite{SachdevPhysics09} pointed out that such Heisenberg models have possible ground states 
in four classes: Neel order, spiral order, a valence bond crystal, or 
a spin liquid. Examples of the first two occur on the 
square and the triangular lattices respectively. For both cases
spin rotational symmetry and lattice symmetry are broken. For a
valence bond crystal, only the spatial symmetry is broken.
It may be that valence bond crystal ground state occurs
on the anisotropic triangular lattice (cf. Section \ref{sec:heisenberg}).

Normand \cite{NormandCP09}
considered three different classes of spin liquids,
 each being defined by their excitation spectrum. If we denote the energy gap 
between the singlet ground state
and the lowest-lying triplet state by $\Delta_T$
 and the gap to the first excited singlet state by $\Delta_S$.
The three possible cases are:
\begin{enumerate}
\item $\Delta_S \neq 0$ and $\Delta_T \neq 0$. \label{typeI}

\item $\Delta_S = 0$ and $\Delta_T \neq 0$.\label{typeII}

\item $\Delta_S = \Delta_T = 0$.
\end{enumerate}
Normand refers to the first two as Type I and Type II respectively.
The third case is referred to as an Algebraic spin liquid.
The case $\Delta_T = 0$ and $\Delta_S \neq 0$ is not an option
because, by Goldstone's theorem, it would be associated with 
broken spin-rotational symmetry.

An important question is how to distinguish these different states experimentally. It can be shown that for a singlet
ground state at zero temperature singlet excited states do not contribute to the dynamical spin susceptibility.
If the susceptibility is written in the spectral representation,
\footnote{Here $\beta=1/k_B T$ is the inverse temperature and $S^\pm({\bf q})$ are the spin raising/lowering operators.}
\begin{equation}
\chi_{-+}({\bf q},\omega) = 
\sum_n \exp(-\beta(E_n - E_0)) 
\frac{ |\langle n| S^+(-{\bf q})|0\rangle|^2 }{E_n - E_0 - \omega}, 
\end{equation}
 it is clear that the matrix elements of the spin operators
between the singlet ground state and any singlet excited state must be zero.
 This means that at low temperatures, only triplet
excitations contribute to the uniform magnetic susceptibility,
the NMR relaxation rate, Knight shift, and
inelastic neutron scattering cross section.
In contrast, both singlet and triplet
excitations contribute to the specific heat capacity and the thermal conductivity at low temperatures.
Hence, comparing the temperature dependence of thermal and magnetic
properties should allow one to distinguish Type I spin liquids from Type II spin liquids.
Furthermore,
the singlet spectrum will not shift in a magnetic field but the triplets will split and the corresponding spectral weight be redistributed.

One important reason for wanting to understand these details of the spin liquid states is that the  spin excitation spectrum may well be important for understanding unconventional superconductivity. This has led to a lot of attention
being paid to a magnetic resonance seen by inelastic
neutron scattering the cuprates. It is still an
open question as to whether this triplet excitation
is correlated with
 superconductivity \cite{HwangNature04,CukNature04,ChubukovPRB06,HaoPRB09}.
 Strong coupling RVB-type theories focus on {\it singlet} excitations whereas weak-coupling antiferromagnetic
spin fluctuation theories focus on {\it triplet} excitations. This important 
difference is emphasized and discussed in a review on the cuprates \cite{Norman}.

\subsubsection{What are spinons?}

A key question is what are the quantum numbers and statistics of the
lowest lying excitations.  
In a Neel ordered antiferromagnet these            excitations are
``magnons" or ``spin waves" which have total spin one and obey Bose-Einstein
statistics \cite{Auerbach}. Magnons can be
viewed as a spin flip propagating through the background of Neel ordered
spins. They can also be viewed as the Goldstone modes associated with the spontaneously broken symmetry of the ground state.

In contrast, in a one-dimensional antiferromagnetic spin chain
(which has a spin liquid ground state)  the lowest
lying excitations are gapless
spinons which have total spin-1/2 and obey ``semion'' statistics,
which are intermediate between fermion and boson statistics
(i.e. there is a phase factor of $\pi/2$ associated
with particle exchange) \cite{HaldanePRL91}.
The spinons are ``deconfined" in the sense that if a pair of them are created
(for example, in an inelastic neutron scattering experiment)
with different momentum
then they will eventually move infinitely far apart.
Definitive experimental signatures of this deconfinement are seen in the dynamical structure factor $S(\omega,\vec{q})$ which shows a continuum of low-lying excitations rather than the sharp features associated with spin waves. 
This is clearly seen in the compound KCuF$_3$, which is composed of linear chains of spin-1/2 copper ions \cite{TennantPRB95}.
The most definitive evidence for such excitations in a real two-dimensional 
material comes from Cs$_2$CuCl$_4 $ \cite{ColdeaPRB03,KohnoNatPhys07} above 
the Neel ordering temperature.  Below the Neel temperature these excitations become confined into conventional magnons \cite{FjaerestadPRB07,StarykhPRB10}.
It is an open theoretical question as to whether there is any two-dimensional
Heisenberg model with such excitations at zero temperature, other than
at a quantum critical point \cite{SinghP10}.

What type of spinon statistics might be possible in two dimensions?
Wen used quantum orders and projective symmetry groups, 
to construct hundreds of symmetric spin liquids, 
having either SU(2), U(1), or Z$_2$ gauge structures at low energies \cite{WenPRB02}.
He divided the  spin liquids into four classes, based
on the statistics of the quasi-particles and whether they
were gapless: 
\begin{description}
\item[Rigid spin liquid:] spinons (and all other excitations) are fully gapped and may have either bosonic, fermionic, or fractional statistics. 

\item[Fermi spin liquid:] spinons are gapless and are described by a Fermi liquid theory (the spinon-spinon interactions vanish as
the Fermi energy is approached).

\item[Bose spin liquid:] low-lying gapless excitations are described by a free-boson theory.

\item[Algebraic spin liquid:] spinons are gapless, but they are not described by free  fermionic and free bosonic quasiparticles.
\end{description}


\subsubsection{Antiferromagnetic fluctuations}

It has been proposed that an instability to
a d-wave superconducting  state  can
occur in a metallic phase which is close to
an antiferromagnetic instability \cite{ScalapinoPRB86}.
This has been described theoretically by an
Eliashberg-type theory in which the effective pairing interaction
is proportional to the dynamical spin susceptibility, 
$\chi(\omega,\vec{q})$ \cite{MoriyaRPP03}.
If this quantity has a significant peak near
some wavevector then that will significantly enhance
the superconducting $T_c$ in a specific pairing channel.
 NMR relaxation rates are also determined by
$\chi(\omega,\vec{q})$ and so NMR can provide useful information about the magnetic fluctuations.
For example,
a signature of large antiferromagnetic fluctuations is the dimensionless Korringa
ratio that is much larger than one.

From a local picture one would like to know the strength
of the antiferromagnetic exchange $J$ between localised spins
in the Mott insulating and the bad metallic phase.
In RVB theory $J$ sets the scale for the superconducting
transition temperature. It is important to realise that this is very different from picture of a ``glue" in the Eliashberg-type theories 
where superconductivity arises due to the
formation of Cooper pairs between Fermi liquid
quasi-particles \cite{AndersonS07,MaierPRB08} (also see section \ref{sect:weak}).

\subsubsection{Quantum critical points}

\begin{figure}\begin{centering}\includegraphics[width=8cm]{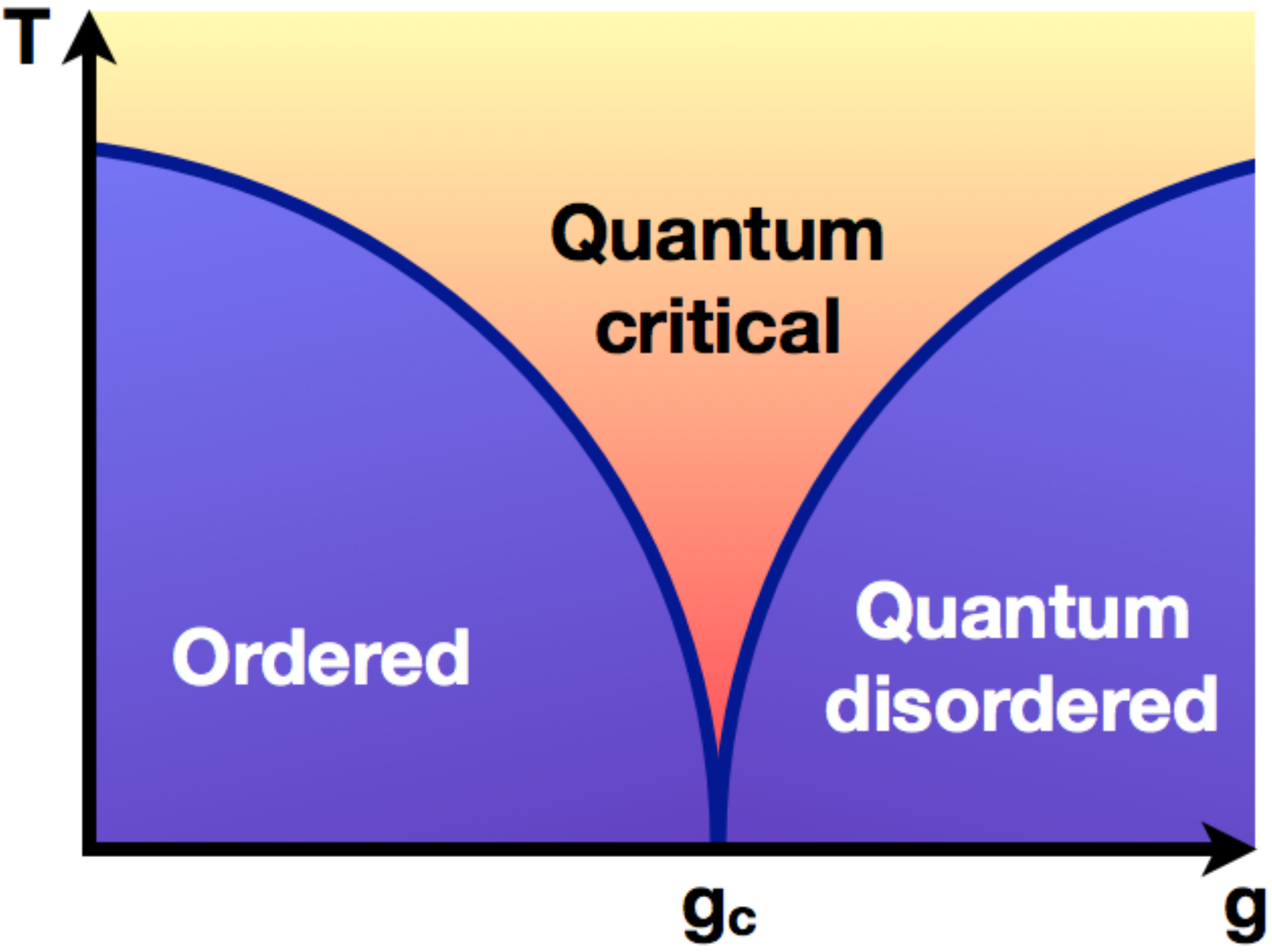}
\caption{
Schematic phase diagram associated with a quantum critical point.
The vertical axis is temperature and the horizontal axis represents
a coupling constant, $g$. Quantum fluctuations increase with increasing
$g$ and for a critical value $g_c$ there is a quantum phase transition from
an ordered phase (with broken symmetry) to 
a disordered phase, usually associated with an energy gap, $\Delta \sim
(g-g_c)^{z \nu}$ where $z$ is the dynamical critical exponent and
$\nu$ is the critical exponent associated with the correlation length
$\xi \sim |g-g_c|^{-\nu}$.
In the quantum critical region the only energy scale is the temperature
and the correlation length $\xi \sim 1/T^{1/z}$. In this region
there are
also no quasi-particles (i.e., any singularities in spectral
functions are not isolated poles but rather branch cuts).
}
\label{fig:qcp}
\end{centering}
\end{figure}

Figure \ref{fig:qcp} shows a schematic phase diagram associated
with a quantum critical point \cite{Sachdevbook,ColemanN05}.
We will discuss the relevance of such diagrams  to the
organic charge transfer salts below.
We will see that some of the theoretical models (such as the Heisenberg
model on an anisotropic triangular lattice) do undergo a quantum phase transition from a magnetically ordered to a quantum disordered phase with an energy 
gap to the lowest lying triplet excitation.

A particularly important question is whether  any signatures
of quantum critical behavior have been seen in organic charge transfer
sites. Most transitions at zero temperature are first-order.
Perhaps, the clearest evidence of quantum critical fluctuations come from the NMR spin relaxation rate
in $\kappa$-(BEDT-TTF)$_2$Cu$_2$(CN)$_3$, which will be discussed in Section \ref{sec:nlsigma}.

\subsection{Key consequences of frustration}

We briefly list some key consequences of frustration. Many of
these are discussed in more detail later in the review.

\begin{itemize}
\item Frustration enhances the number of low energy excitations. This increases the entropy at low temperatures \cite{RamirezARMS94}.
The temperature dependence of the magnetic susceptibility is flatter and the peak occurs at a lower temperature (Section \ref{sec:measures}).

\item Quantum fluctuations in the ground state are enhanced due to the
larger density of states at low energies.
These fluctuations can destroy magneticially ordered
phases 
(Section \ref{sec:heisenberg}).

\item Singlet excitations are stabilised and singlet pairing correlations are enhanced.  Resonating valence bond states have a larger overlap with the true ground state of the system  (Section \ref{sec:heisenberg}).

\item Intersite correlations are reduced which enhances the accuracy of single site approximations such as Curie-Weiss theory
and dynamical mean-field theory (Section \ref{ET-metal}).

\item In Heisenberg models frustrated spin interactions
 produce incommensurate correlations. These  can also
change the symmetry of the superconducting pairing \cite{PowellPRL07},
lead to new triplet excitations (phasons) \cite{ChandraJPCM90},
and the emergence of new gauge fields which are deconfining (Section \ref{sec:gauge}).

\item Frustration of kinetic energy (such as in non-bipartite lattices or
by next-nearest-neighbor hopping)
reduces nesting of the Fermi surface and stabilises the metallic state
(Section \ref{sec:hubbard}).
\end{itemize}

\subsubsection{Reduction of the correlation length}
\label{sec:corrlength}

The temperature dependence of the correlation length $\xi(T)$
and the static structure factor $S(\vec Q)$, associated
with the classical ordering 
wavevector  $\vec Q$ has been calculated  for both        the
triangular lattice and square lattice
Heisenberg models using high-temperature
 series expansions \cite{ElstnerPRL93,ElstnerJAP94}.
For the triangular lattice the correlation length has
values of about 0.5 and 2 lattice constants, at temperatures
 $T=J$ and $T=0.2J$, respectively. In contrast, the model on the square lattice has correlation lengths of
about 1 and 200 lattice constants, at 
 $T=J$ and $T=0.2J$, respectively.
At $T=0.2J$
the static structure factor 
has values of about 1 and 3000 for the triangular and square
lattices, respectively.
Hence, frustration leads to a significant reduction of
the spin correlation length.
These distinct differences in temperature 
dependence can be understood in terms of frustration
producing a `roton' like minimum in the triplet excitation spectra
of the triangular lattice model \cite{ZhengPRB06}.

We discuss later how the temperature dependence
of the uniform magnetic susceptibility 
of several frustrated charge transfer salts
 can be fit to that of the Heisenberg model
on the triangular lattice with
 $J=250$ K \cite{TamuraJPCM02,ShimizuPRL03,ZhengPRB05}.
This implies that $\xi \simeq 2 a$ at 50 K.
This is consistent with estimates of the spin-spin correlation length in organic charge transfer salts from low temperture NMR
relaxation rates \cite{YusufPRB07}.

\subsubsection{Competing phases}

One characteristic feature of strongly correlated electron systems that, we believe,  
should be discussed
more is how sensitive they are to small perturbations. This is particularly true in frustrated systems. A related issue is that there are often several competing phases which are very close in energy. This can make variational wave functions unreliable. Getting a good variational energy may not be a good indication that the wave function captures the key physics.
Below we give two   concrete examples to illustrate this point.

Firstly, consider the spin 1/2 Heisenberg model on the isotropic triangular on a lattice of 36 sites, and with exchange interaction $J$. Exact diagonalisation \cite{SindzingrePRB94} gives a ground state energy per site of $-0.5604 J$ and a net magnetic moment (with 120 degree order as in the classical model) of 0.4, compared to the classical value of 1/2.
In contrast, a variational short-range RVB wavefunction has zero magnetic moment and a ground state energy of $-0.5579 J$.
Yet, it is qualitatively incorrect because it predicts no magnetic order (and
thus no spontaneous symmetry breaking) in the thermodynamic limit. Note, however, that the energy difference is only $J/400$.
[For details and references see Table III in \cite{ZhengPRB06}].

The second example concerns the spin 1/2 Heisenberg model on the anisotropic triangular lattice, viewed as chains with exchange $J'$ and frustrated interchain coupling $J$. For $J'  \sim  3 J$ this describes the compound Cs$_2$CuCl$_4$.
The triplet excitation spectrum of the model has been calculated
both  with a small Dzyaloshinski-Moriya interaction $D$, and without ($D=0$). It is striking that even when $D  \sim J'/20$ it induces energy changes in the spectrum of energies as large as $J'/3$, including new energy gaps \cite{FjaerestadPRB07}.
For $J' \gg J$ the ground state turns out to be ``exquisitely sensitive''
to other residual interactions as well \cite{StarykhPRB10}.

\subsubsection{Alternative measures of frustration}
\label{sec:measures}

Balents recently considered how to quantify        
the amount of frustration in an antiferromagnetic material (or model)
and its tendency to have a spin liquid ground state  \cite{BalentsNat10}. 
He used a measure \cite{RamirezARMS94} 
$f=T_{CW}/T_N$, the ratio of the Curie-Weiss temperature
$T_{CW}$ to the Neel temperature, $T_N$
at which three-dimensional magnetic ordering occurs.

One limitation of this measure is that it does not separate out the effects of fluctuations (both quantum and thermal), dimensionality, and frustration.
 For strictly one or two dimensional systems, $T_N$ is zero. For quasi-two-dimensional systems the interlayer coupling determines $T_N.$ Thus, $f$ would be larger for a set of weakly coupled unfrustrated chains than for a layered triangular lattice in which the layers are moderately coupled together.

Section II of \cite{ZhengPRB05} contains a detailed discussion 
of two different measures of frustration for model Hamiltonians:
(1) the number of degenerate ground states, and
(2) the ratio of the ground state energy to the
base energy [the sum of all bond energies if they are independently fully satisfied.]
This measure was introduced previously for classical models \cite{LacorreJPC87}.

\begin{figure}\begin{centering}\includegraphics[width=8cm]{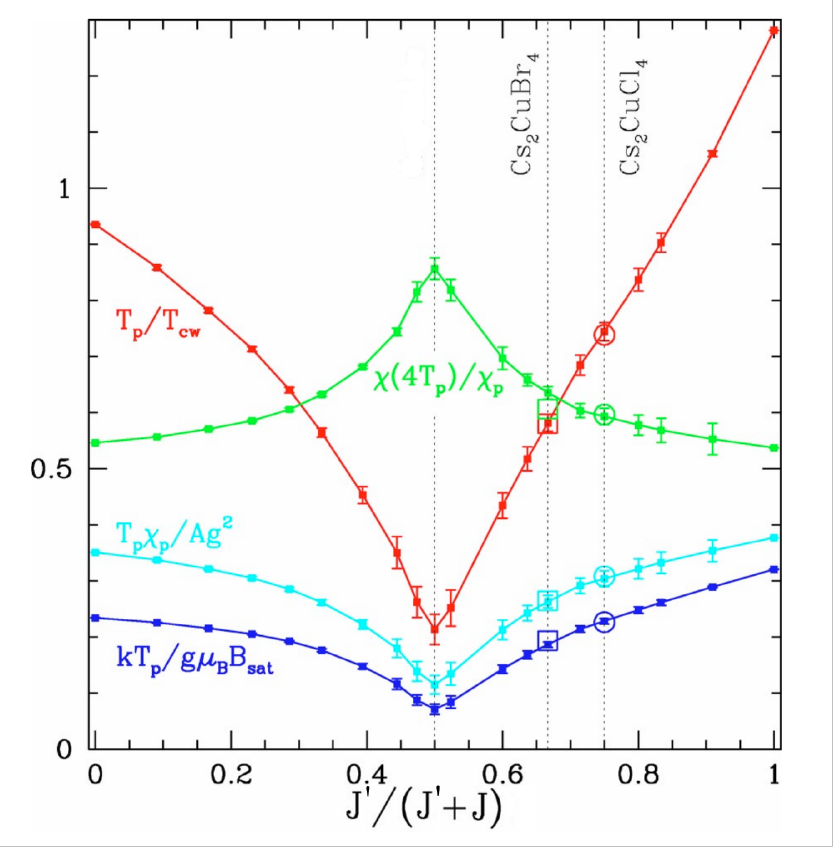}
\caption{
Effect of frustration on the 
temperature dependence of the magnetic susceptibility $\chi(T)$
 for the Heisenberg model on an anisotropic triangular
 lattice \cite{ZhengPRB05}.
  The variation of  key parameters is shown as a function of the ratio $J'/(J+J')$.
 $T_p$ is the temperature at which the susceptibility is a maximum, with a value $\chi_p \equiv \chi(T_p)$. $T_{cw}$ is the Curie-Weiss temperature which can be extracted from the high-temperature dependence of the susceptibility.
$B_{sat}$ is the magnetic saturation field and $Ag^2$ is the Curie-Weiss constant.
All quantities were calculated by a high-temperature series expansion.
All of the quantities plotted have extreme values for the isotropic triangular lattice, suggesting that in some sense it is the most frustrated.
[Modified from  \cite{ZhengPRB05}. Copyright (2005) by the American Physical Society.]
}
\label{fig:susc}
\end{centering}
\end{figure}

Figure \ref{fig:susc} shows results that might be the basis of some alternative measures of frustration.  The sensitivity of the temperature dependence of the susceptibility to the ratio $J'/J$ has been used to estimate this ratio for specific materials \cite{ZhengPRB05}.

In some sense then, the temperature 
$T_p$ at which the susceptibility has a maximum and the magnitude of that susceptibility is a measure of the amount of frustration. This is consistent with 
some intuition (or is it just prejudice?) that for the anistropic
triangular lattice  the frustration is largest
for the isotropic case.  
These measures of frustration are not dependent on dimensionality and so do not have the same problems discussed above that the ratio $f$ does.
On the other hand, these measures reflect short-range interactions rather than
the tendency for the system to fail to magnetically order.

Another issue that needs to be clarified is how one might 
distinguish quantum and classical frustration. 
In general the nearest neighbour spin correlation
 $f_s \equiv \langle 
 \hat{{\bf S}}_i \cdot \hat{{\bf S}}_j
 \rangle $ will be reduced by frustration.
Entanglement measures from quantum information theory can be
used to distinguish truly quantum from classical correlations.
For a spin rotationally invariant state (i.e. a total spin singlet
state) $f_s$ is related to a
measure of entanglement between two spins in a mixed state, known as the
concurrence $C$ by \cite{ChoPRA06}
\begin{equation}
C={\rm max}\{0,-2f_s-1/2\}.
\end{equation}
Hence, there is maximal entanglement ($C=1$) when the two
spins are in a singlet state and are not entangled
with the rest of the spins in the system.
Once the spin correlations decrease to $f_s=1/4$ there is
no entanglement between the two spins.

\subsubsection{Geometric frustration of kinetic energy }

In a non-interacting electron model we are aware of
only two proposed  quantitative measures of the geometrical frustration of the
kinetic energy. Both  are based on the observation
that, for frustrated lattices with $t>0$, an electron at the bottom of the band does not gain
the full lattice kinetic energy, while a hole at the top of the band
does. Barford and Kim  \cite{BarfordPRB91} 
suggested that for tight binding models a measure of the
frustration is then $\Delta = |\epsilon_k^{max}| -
|\epsilon_k^{min}|$, where $\epsilon_k^{max}$ and $\epsilon_k^{min}$
are the energies (relative to the energy of the system with no
electrons) of the top and bottom of the band respectively. This
frustration increases the density of states for positive energies
for $t>0$ (negative energies for $t<0$) which represents an
increased degeneracy and enhances the many-body effects when
the Fermi energy is in this regime. 

Together with Merino, we previously argued \cite{MerinoPRB06} that a simpler
measure of the kinetic energy frustration is $W/2z|t|$, where $W$ is
the bandwidth and $z$ is the coordination number of the lattice. The
smaller this ratio, the stronger the frustration is, while for an
unfrustrated lattice $W/2z|t|=1$. But, for example, on the triangular
lattice kinetic energy
frustration 
leads to a bandwidth, $W=9 |t|$, instead of
$12 |t|$ as one might na\"ively predict from $W=2z|t|$ since $z=6$.


We argued that geometrical frustration of the kinetic energy
is a key concept for understanding the properties of the
Hubbard model on the triangular lattice. In particular, it leads to
particle-hole asymmetry which enhances many-body effects for
electron (hole) doped $t>0$ ($t<0$) lattices.

It should be noted that geometrical frustration of the kinetic
energy is a strictly quantum mechanical effect arising from quantum
interference. This interference arises from hopping around
triangular plaquettes which will have an amplitude proportional to
$t^3$, which clearly changes sign when $t$ changes sign. In contrast
on the, unfrustrated, square lattice the smallest possible
plaquette is the square and the associated
amplitude for hopping around a square is
independent of the sign of $t$ as it is proportional to $t^4$.
Barford and Kim noted that the phase collected by
hopping around a frustrated cluster may be exactly cancelled by the
Aharonov-Bohm phase associated with hopping around the cluster for a
particular choice of applied magnetic field \cite{BarfordPRB91}.
 Thus a magnetic field
may be used to lift the effects of kinetic energy frustration. The
quantum mechanical nature of kinetic energy frustration is in
distinct contrast to geometrical frustration in antiferromagnets
which can occur for purely classical spins.

\section{Toy models to illustrate the interplay of frustration and quantum fluctuations}

We now consider some model Hamiltonians on just four lattice sites.
The same Hamiltonians on an infinite lattice are relevant
to the organic charge transfer salts and will be discussed
 in Sections \ref{sec:heisenberg} and \ref{sec:hubbard}.
 Although such  small systems are
far from the thermodynamic limit, these models can illustrate some of the essential physics associated with the interplay of strong electronic correlations, frustration, and quantum fluctuations. These toy models illustrate the
quantum numbers of  important low-lying quantum states,
the dominant short-range correlations, and how frustration changes the competition between these states.
Furthermore, understanding these small clusters is a pre-requisite for
cluster extensions of dynamical mean-field theory \cite{FerreroPRB09} and 
rotationally invariant slave boson mean-field theory \cite{LechermannPRB07} 
which describes band selective and momentum space selective Mott transitions. Insight
can also be gained by considering two, three, and four coupled Anderson 
impurities \cite{FerreroJPCM07}.
Small clusters are also the basis of the contractor renormalisation (CORE) method which has been used to study the doped Hubbard model
 \cite{AltmannPRB02} and frustrated spin models  \cite{BergPRL03}.

A similar approach of just considering four sites has been taken before
when considering the ground state of a Heisenberg model on a depleted lattice 
which is a model for CaV$_4$O$_9$ \cite{UedaPRL96}.
The authors first considered a single plaquette with frustration, albeit
along both diagonals (see also Section 3 in \cite{ValkovJETP06}).
Dai and Whangbo \cite{DaiJCP04}
 considered the Heisenberg model on a triangle and a tetrahedra.
Similar four site Heisenberg Hamiltonians
have also been discussed in the context of mixed valence
metallic clusters of particular interest to chemists \cite{AugustNJC05}.

\subsection{Four site Heisenberg model}

The four site Heisenberg model illustrates that frustration can 
lead to energy level crossings and consequently to
changes in the quantum numbers of the ground state and lowest
lying excited state. 

\begin{figure}\begin{centering}
\includegraphics[width=8cm]{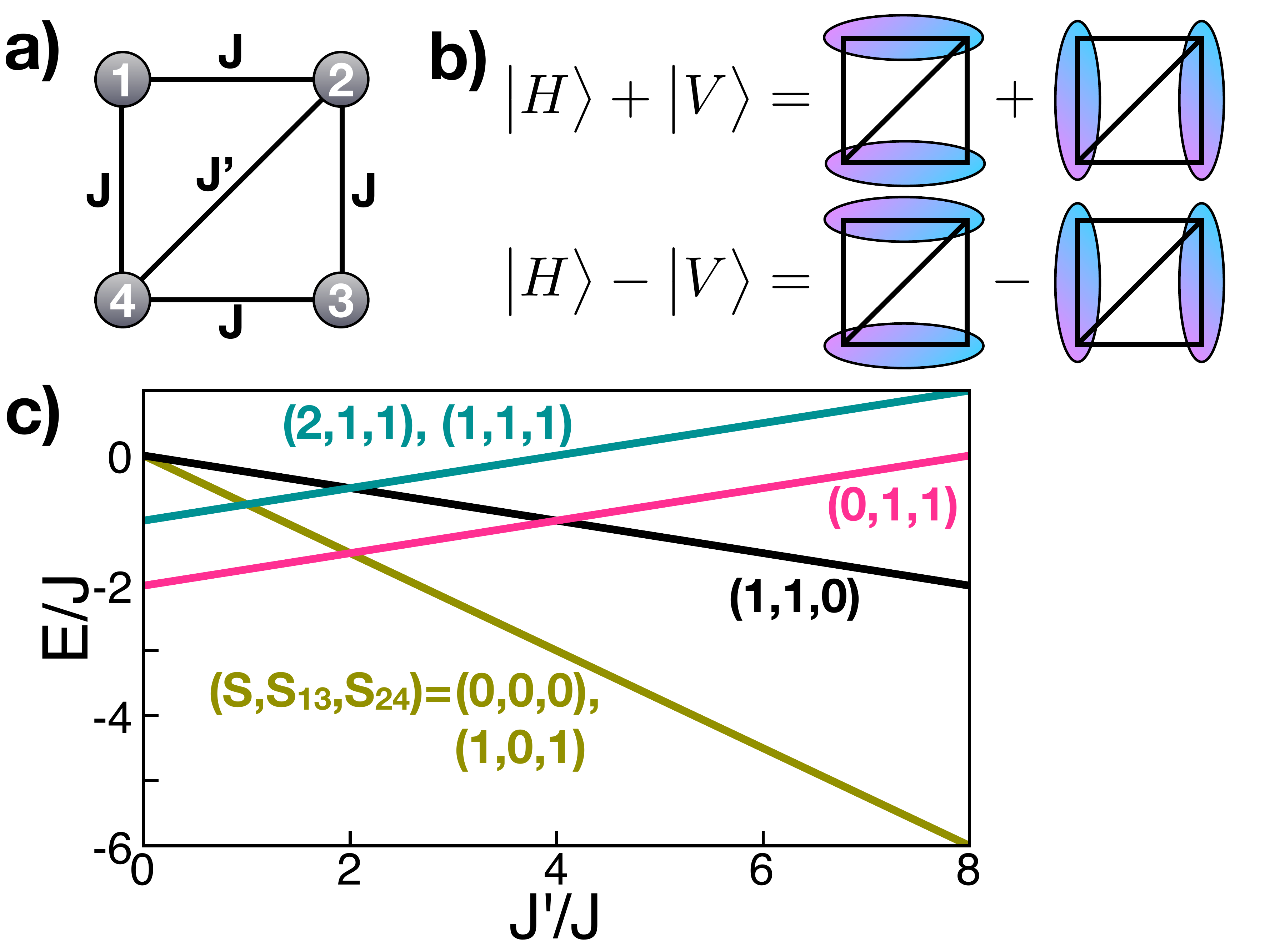}
\caption{
Eigenstates and eigenvalues of a
frustrated Heisenberg model on a single plaquette.
(a) The exchange interactions in the model.
(b) The two resonating valence bond states which span
all the singlet states (compare equations 
(\ref{vb000}) and (\ref{vb011})).
(c) Dependence of the energy eigenvalues as a function
of the diagonal interaction $J'/J$.
Note that the quantum
numbers of the lowest lying excited state change when 
$J'=J$ and the ground state changes when $J'=2J$.
Furthermore, the two resonating valence bond states become
degenerate at $J'=2J$.
}
\label{fig:plaquette}
\end{centering}
\end{figure}

The Hamiltonian is (see Figure \ref{fig:plaquette}(a))
\begin{eqnarray}
\hat{\cal H}&=&
 J \left(
\hat{{\bf S}}_1 \cdot \hat{{\bf S}}_2
+\hat{{\bf S}}_2 \cdot \hat{{\bf S}}_3
+\hat{{\bf S}}_3 \cdot \hat{{\bf S}}_4
+\hat{{\bf S}}_4 \cdot \hat{{\bf S}}_1
\right) \nonumber \\
&&+J' \hat{{\bf S}}_1 \cdot \hat{{\bf S}}_3.
\label{h4site}
\end{eqnarray}
It is helpful to introduce the total spin along each of the diagonals, 
$ \hat{{\bf S}}_{13} = \hat{{\bf S}}_1 +\hat{{\bf S}}_3$
and
$ \hat{{\bf S}}_{24} = \hat{{\bf S}}_2 +\hat{{\bf S}}_4$,
and note that these operators commute with each other and with
the Hamiltonian. The total spin of all four sites can be written in terms of these operators:
$ \hat{{\bf S}} =  \hat{{\bf S}}_{13} + \hat{{\bf S}}_{24}$.  Thus, the
total spin
$S$, and the total spin along each of the two diagonals, 
$S_{13}$ and $S_{24}$ are good quantum numbers.   
The term in (\ref{h4site}) associated with $J$ can be rewritten as
$J/2(\hat{{\bf S}}^2 - \hat{{\bf S}}_{13}^2 - \hat{{\bf S}}_{24}^2)$.
Hence, the energy eigenvalues are
\begin{eqnarray}
E(S,S_{13},S_{24}) &=&
\frac{1}{2}J S(S+1)
+\frac{1}{2}(J'-J) S_{13}(S_{13}+1)
\nonumber \\
&&-\frac{1}{2}J S_{24} (S_{24}+1)
-\frac{3}{4}J'.
\label{e4site}
\end{eqnarray}
Figure \ref{fig:plaquette} (c) shows a plot of these energy eigenvalues as a function
of $J'/J$. We note that  the quantum numbers of the lowest
 lying excited state change when $J'=J$ and $J'=4J$, and that the ground state changes when $J'=2J$.

The two singlet states can also be written as linear combinations of
two orthogonal valence bond states, denote $|H\rangle$ and $|V\rangle$,
which descibe a pair of singlets along the horizontal and 
vertical directions, respectively (see Figure \ref{fig:plaquette} (b)).
The state with quantum numbers $(S,S_{13},S_{24})=(0,0,0)$ is
\begin{equation}
|0,0,0\rangle = \frac{1}{\sqrt{2}}\left(|H\rangle-|V\rangle\right)
\label{vb000}
\end{equation}
and the state with $(S,S_{13},S_{24})=(0,1,1)$ state is
\begin{equation}
|0,1,1\rangle =  \frac{1}{\sqrt{2}}\left(|H\rangle+|V\rangle\right).
\label{vb011}
\end{equation}
Both of these singlet states
are resonating valence bond states (see Figure \ref{fig:plaquette} (b)).

The Hamiltonian has $C_{2v}$ with the $C_2$ axes along each diagonal (and out of the plane).
The two singlet states above have $A_1$ and $A_2$ symmetry, respectively.
However, if $J'=0$ there is $C_{4v}$ symmetry and 
the $(0,0,0)$ and $(0,1,1)$ states have $A_1$ and $B_1$ symmetry, respectively.
The latter, which is the ground state, connects naturally to the $B_1$
symmetry of a $d_{x^2-y^2}$ superconducting order parameter
on the square lattice.


It is possible to relate the two singlet  
states to the physical states of a $Z_2$ gauge field on a single
plaquette (see Section 3.2 of \cite{AletPA06}).
The gauge flux operator on the plaquette
$F_p$ flips the bonds between horizontal and vertical.
The RVB states (\ref{vb000}) and (\ref{vb011}) 
are eigenstates of $F_p$ with eigenvalues $\pm 1$.

\subsubsection{Effect of a ring exchange iteraction}

Consider adding to Hamiltonian (\ref{h4site}) the term
\begin{eqnarray}
\label{ringexch}
\hat{\cal H}_\square&=&J_\square(\hat P_{1234}+\hat P_{4321}) \\
&=&J_\square(
\hat P_{12} \hat P_{34}
+\hat P_{14} \hat P_{23}
-\hat P_{13} \hat P_{24}
+\hat P_{13} + \hat P_{24} -1 )
  \nonumber
\end{eqnarray}
where
$J_\square$ describes the ring-exchange interaction around a single plaquette,
the operator
$ \hat P_{12} = 2 \hat{\bf S}_1\cdot\hat{\bf S}_2 +1/2 $ permutes spins 1 and 2, and $\hat P_{1234}$ is the permutation operator around the plaquette \cite{ThoulessPPS65,slreviews}.

Intuitively, 
\begin{equation}
\hat{\cal H}_\square |H\rangle =2J_\square |V\rangle
\ \ \ \ \ 
\hat{\cal H}_\square |V\rangle =2J_\square |H\rangle
\end{equation}
Hence, the RVB states  (\ref{vb000}) and (\ref{vb011}) 
are eigenstates of the ring-exchange Hamiltonian    
with eigenvalues $-2J_\square$ and $2J_\square$, respectively. 
Hence, ring exchange has a similar effect to the diagonal interaction in
that it stabilises the state $|0,0,0\rangle$.

\subsection{Four site Hubbard model}

A comprehensive study of the $t'=0$ model (which has $C_{4v}$ symmetry)
has been given by Schumann \cite{SchumannAP02}.      
The analysis is simplified by exploiting this SU(2) symmetry
associated with particle-hole symmetry \cite{NocePRB96}.
In particular, the Hamiltonian matrix then decomposes
into blocks of dimension 3 or less.
Schumann has also solved the model on a tetrahedron and a triangle \cite{SchumannAP08}.
When $t' \neq 0$ the SU(2) symmetry is broken, but 
it may be that the SU(2) quantum numbers are still
useful to define a basis set in which to diagonalize the 
Hamiltonian and see the effect of $t' \neq 0$.

Freericks, Falicov, and Rokshar studied    an eight site 
Hubbard model with a next-nearest-neigbour hopping $t'$
and periodic boundary conditions \cite{FreericksPRB91}.
The model is invariant under a 128-element cluster 
permutation group.
For $t'=t/2$ the model is equivalent to an eight site triangular lattice cluster
or a face-centred-cubic cluster.
They found that at half filling the symmetry of 
the ground state changed
as a function of both $t'/t$ and $U/t$
[see Figure 3 in \cite{FreericksPRB91}]. 

Falicov and Victora \cite{FalicovPRB84}
 showed that the Hubbard model on a tetrahedron [which
has $T_d$ symmetry] with 
four electrons has a singlet ground state
with E  symmetry.\footnote{We use Mulliken notation, Falicov and Victora use Bethe notation and label this representation $\Gamma_3$, see \cite{Lax74} for details.}
Later Falicov and Proetta \cite{FalicovPRB93}
 also showed that an RVB state with complex pairing amplitude (and which thus breaks time-reveral symmetry)
and which they state has E  symmetry\footnote{Or $\Gamma_{12}$ in the Bouckaert, Smoluchoski, Wigner notation \cite{Lax74} that Falicov and Proetta use.} is within 
0.3\% of the exact ground state energy for $U=10t$.

More work is required to use the above results to 
extract insights about the role
of frustration. An important question is whether 
results on four sites can be related to a 
simple picture of how $d_{x^2-y^2}$ Cooper pairing 
emerges on the square lattice due to antiferromagnetic
interactions \cite{TrugmanPM96}. If so, does 
this pairing symmetry change with frustration,
as it does for the infinite lattice, at the mean-field RVB level
 \cite{PowellPRL07}?

%% file: ET.tex
\section{\kX}\label{ET}
\label{sec:ET}

\begin{figure}
\begin{centering}
\includegraphics[width=6cm]{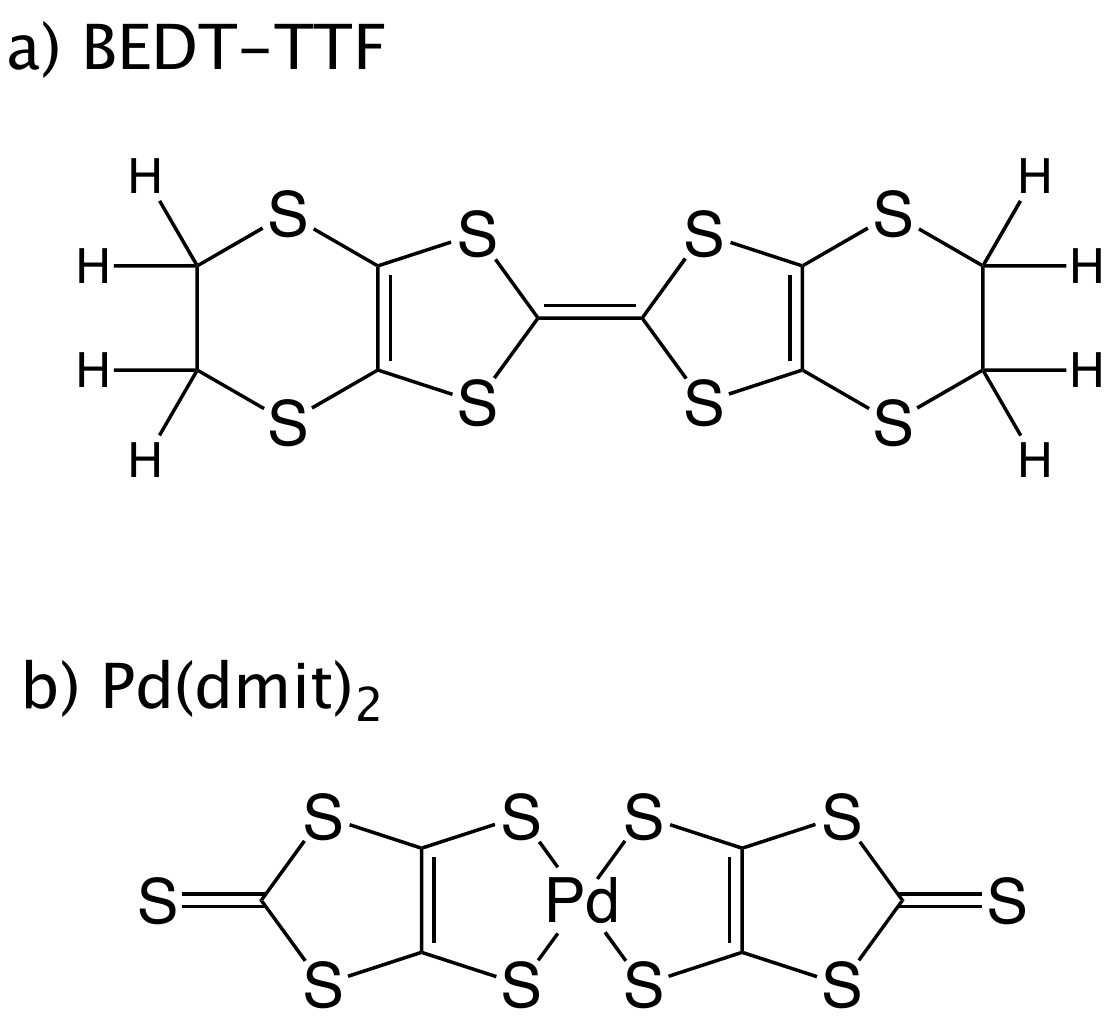}
\caption{The molecules BEDT-TTF and Pd(dmit)$_2$, which form charge transfer salts with frustrated lattices and in which the electrons are strongly correlated.
BEDT-TTF denotes bis(ethylene\-dithio)tetra\-thia\-fulvalene, an is an electron
donor.
Pd(dmit)$_2$ is an electron acceptor where
dmit is 1,3-dithiol-2-thione-4,5-dithiolate.
}
\label{fig:molecules}
\end{centering}
\end{figure}

An important class of model systems for quantum frustration is the organic charge transfer salts based on the molecule
 bis(ethylene\-dithio)tetra\-thia\-fulvalene
 (also known as BEDT-TTF or ET; shown in Fig. \ref{fig:molecules}a).
These salts have been extensively studied and          show a wide range of behaviours including, antiferromagnetism, spin liquids, (unconventional) superconductivity, Mott transitions, incoherent (or `bad') metals, charge ordering and Fermi liquid behaviour. In this section we focus on the aspects most relevant to the quantum frustration in these materials, a number of other reviews focusing on different aspects of these materials are also available elsewhere \cite{PowellJPCM06,Ishiguro1998,SingletonCP02,Lang2003,WosnitzaJLTP07,SeoJPSJ06} and
in the November 2004 issue of {\it Chemical Reviews}.
 Further, although a number of crystallographic phases are observed in the BEDT-TTF salts, we will limit ourselves to the $\kappa$ phase, which is by far the most widely studied and, in which, the most profound effects of frustration have been found.

\begin{figure}
\begin{centering}
\includegraphics[width=8cm]{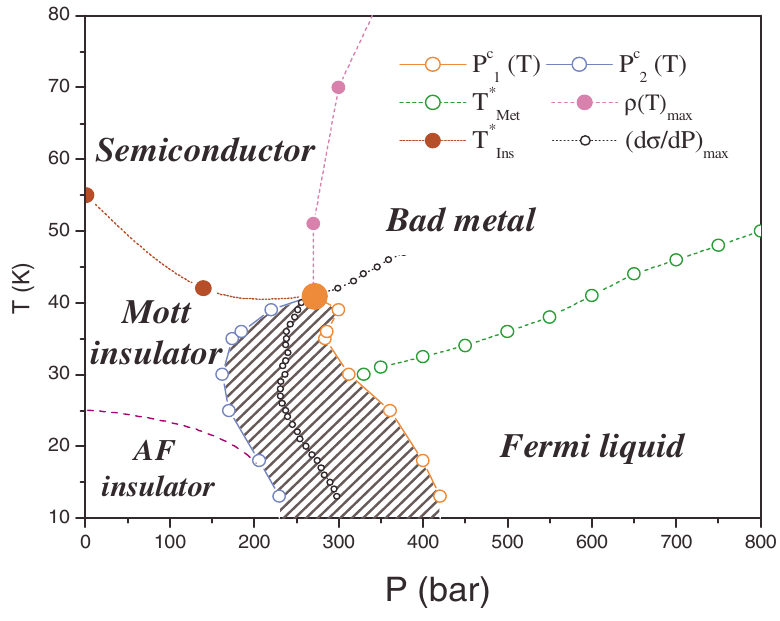}
\caption{Pressure-temperature phase diagram of \Cln. At low temperatures 
it undergoes a first-order Mott transition from an antiferromagnetic 
(AF) insulator (section \ref{ET-ins}) to a metal when hydrostatic pressure is applied (section \ref{ET-Mott}). As the temperature is raised the line of first order transitions ends in a critical point with novel critical exponents (section \ref{sect:exponents}). In the insulating phase raising the temperature destroys the antiferromagnetic order.  At the very lowest temperatures the metallic state becomes superconducting (section \ref{ET-super}). As the temperature is raised superconductivity gives way to a metal with coherent intralayer
 charge transport (section \ref{ET:FL}) and a pseudogap (section \ref{sect:pg}). Further, raising the temperature results in a loss of coherence in the intralayer
transport. This incoherent metallic phase is referred to as a `bad-metal' (section \ref{ET-metal}).  From \cite{LimelettePRL03}.
[Copyright (2003) by the American Physical Society.]
}
\label{fig:phase-diagram-kCl}
\end{centering}
\end{figure}

\begin{figure}
\begin{centering}
\includegraphics[width=7cm]{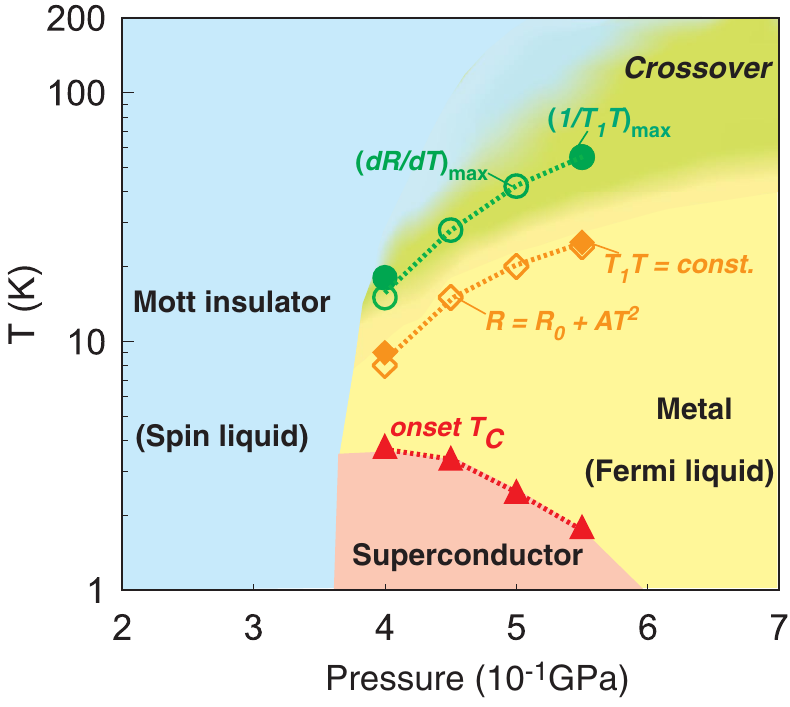}
\caption{Pressure-temperature phase diagram of \CNn. This is similar to that of \Cl (Fig. \ref{fig:phase-diagram-kCl}), but has important differences. Most importantly the Mott insulating phase does not show any signs of long range magnetic order down to 20 mK (the lowest temperature studied; see section \ref{ET-ins} and Fig. \ref{fig:CN-Cl-NMR}). Thus, \CN is believed to be a spin liquid at ambient and low pressures. Further, there is no evidence of a pseudogap in \CN (see section \ref{pseudo-CN}). These differences are believed to result from the greater geometrical frustration in \CN (cf. Table \ref{tab:t-DFT}, Eq. (\ref{eq:aniso}) and Fig. \ref{fig:ET-struct}). From \cite{KurosakiPRL05}.
[Copyright (2005) by the American Physical Society.]
}
\label{fig:phase-diagram-kCN}
\end{centering}
\end{figure}

The experimentally observed phase diagrams of two \kET salts (\Cl and \CNn) are shown in Figs. \ref{fig:phase-diagram-kCl} and \ref{fig:phase-diagram-kCN}. One should note how similar these two phase diagrams are (except for the magnetic order, or lack thereof, observed in the Mott insulating phase). Two important parameters are the strength of the electronic correlations and the degree of frustration. These parameters are determined by the choice of anion, $X$, in \kX and the applied hydrostatic pressure. Below we will focus on four of the most widely studied materials: \NCS and \Brn, which superconduct below  $\sim$10~K at ambient pressure; \Cln, which is an antiferromagnetic Mott insulator at ambient pressure; and \CNn, which appears to be a spin liquid at ambient pressure. Both \CN and \Cl undergo Mott transitions to superconducting states under modest pressures (a few 100 bar).


%

\subsection{Crystal and electronic structure}\label{ET-struct}


\kX salts form crystals with alternating layers of 
the  electron donors BEDT-TTF and electron acceptors ,$X$, leading
to  a quasi-two-dimensional (q2D) band structure. Charge is transferred from organic (\ETn) layer to the  anion ($X$) layer; for monovalent anions, which we consider here, one electron is transferred from each dimer [\ETtwo unit] to each anion formula unit. Band structure calculations \cite{KandpalPRL09,NakamuraJPSJ09} predict that the anion layer is insulating, but that the dimer layers are half-filled. Hence, these calculations predict that the organic layers are metallic, in contrast to the rich phase diagram observed (Figs. \ref{fig:phase-diagram-kCl} and \ref{fig:phase-diagram-kCN}).

\begin{figure}
\begin{centering}
\includegraphics[width=7cm]{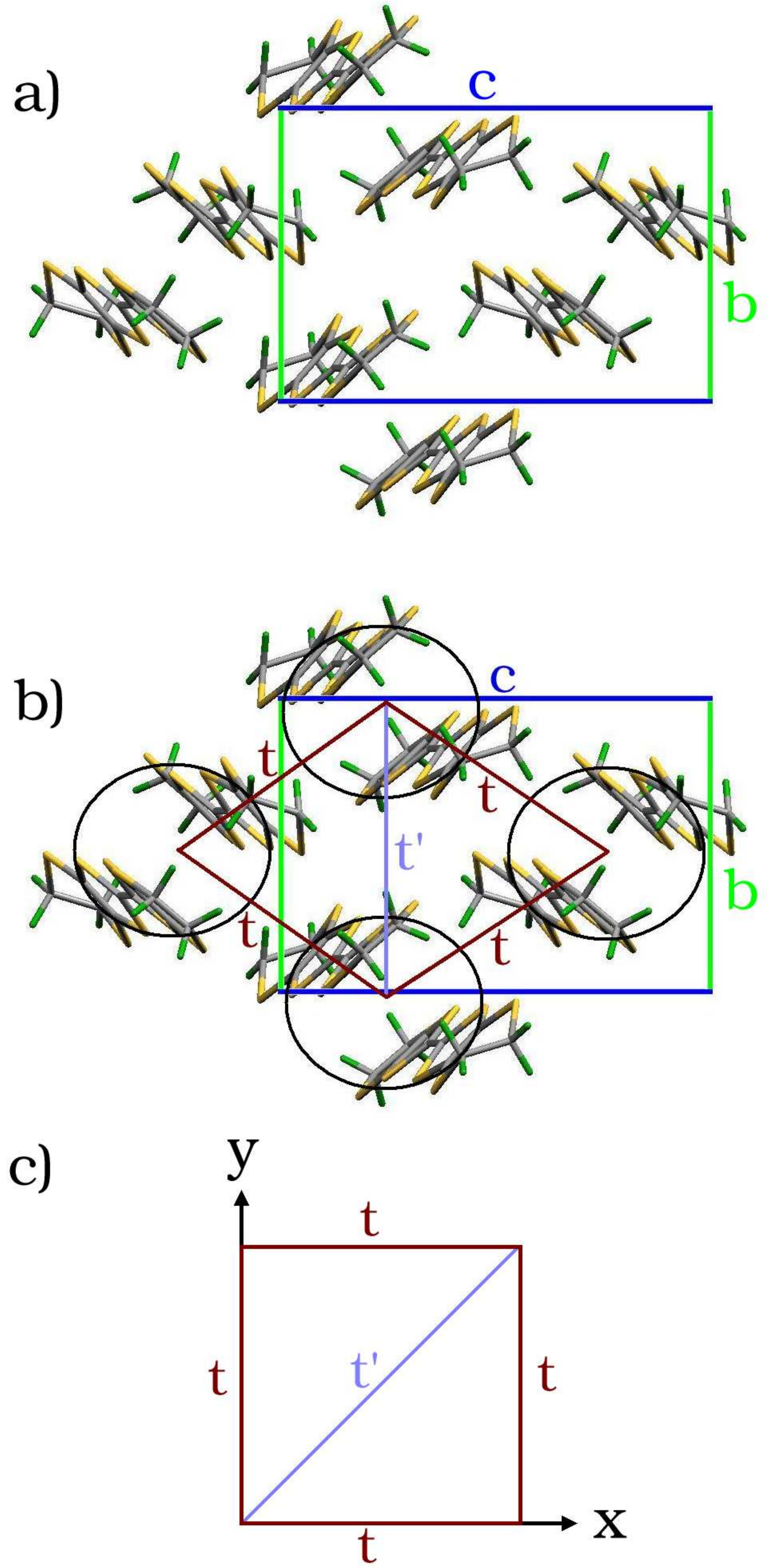}
\caption{Tight binding model of the electronic
band-structure of a $\kappa$-type \ET salt. Panel (a) shows a cross section of the crystal structure of \NCS in the organic layer. In panel (b) the black circles mark the dimers, within which the hopping integral is large and which serve as a `site' in lattice models of the band structure. Lines indicate the inter-dimer hopping integrals in both panels (b) and (c), which are topologically equivalent. [Taken from \cite{PowellJPCM06b}].}
\label{fig:ET-struct}
\end{centering}
\end{figure}

The $\kappa$ phase salts of \ET are strongly dimerised, that is the molecules stack in pairs within the crystal, cf. Fig. \ref{fig:ET-struct}. The frontier molecular orbitals of the \ET molecule are $\pi$ orbitals, i.e., they have nodes in the plane of the molecule, cf. Fig. \ref{fig:ET-HOMO}.   Thus, these orbitals overlap with the equivalent orbitals on the other molecule in the dimer, cf. Fig. \ref{fig:ET-HOMO}, more than they overlap with the orbitals of any other \ET molecule. This, combined with the greater physical proximity of the two molecules within the dimer, means that the amplitude for an electron to hop between two molecules within the same dimer has a much larger magnitude than the amplitude for hopping between molecules in different dimers. This suggests that the interdimer hopping might be integrated out of an effective low energy Hamiltonian \cite{KinoJPSJ96,McKenzieCCMP98}.

\subsubsection{Dimer model of the band structure of \kETX}
\label{sec:dimer}

The dimer model described above is the simplest, and most widely studied, model of the electronic structure for the $\kappa$-(\ETn)$_2X$
 salts and leads to the Hubbard model on an anisotropic lattice 
at half filling \cite{McKenzieCCMP98,PowellJPCM06}. The Hamiltonian of this model is 
\begin{eqnarray}
\hat{\cal H}=-t\sum_{\langle ij\rangle\sigma}\hat c_{i\sigma}^\dagger\hat c_{j\sigma}-t'\sum_{[ij]\sigma}\hat c_{i\sigma}^\dagger\hat c_{j\sigma}+U\sum_i\hat c_{i\uparrow}^\dagger\hat c_{i\uparrow}\hat c_{i\downarrow}^\dagger\hat c_{i\downarrow},
\label{eq:aniso}
\end{eqnarray}
where $\hat c_{i\sigma}^{(\dagger)}$ destroys (creates) an electron with spin $\sigma$ on site (dimer) $i$, $t$ and $t'$ are the hopping amplitudes between neighbouring dimers in the directions indicated in Fig. \ref{fig:ET-struct}, and $U$ is the effective Coulomb repulsion between two electrons on the same site (dimer). This model is, up to an overall scale factor, governed by two dimensionless ratios: $t'/t$, which sets the strength of the frustration in system and $U/W$, which determines the strength of electronic interactions. Here, $W$ is the bandwidth, which is determined by the values of $t$ and $t'$. These two ratios can be manipulated experimentally by hydrostatic pressure,\footnote{It has often been emphasised \cite{KanodaPC97} that increased hydrostatic pressures correspond to decreased correlation (decreased $U/W$), but it has become increasingly clear \cite{KandpalPRL09,CaulfieldJPCM94,PrattPB10} that pressure may also induce changes in the frustration, $t'/t$.} $P$, or by studying materials with different anions, $X$. Varying the anions is often referred to as chemical pressure, as both degrees of freedom lead to changes in the lattice constants. However, it appears that chemical pressure causes larger variations in $t'/t$ than hydrostatic pressure does. We will limit our discussions to monovalent anions,\footnote{See \cite{MoriCR04} for a discussion of anions with valencies other than one.} in which case we have $\kappa$-(\ETn)$_2^+X^-$, i.e., there is, on average, one hole per dimer and the appropriate dimer Hubbard model is half filled.

The anisotropic triangular lattice model extrapolates continuously between three widely studied lattice models. For $t'=0$ it is just the square lattice. For $t'=t$ we recover the (isotropic) triangular lattice. And for $t\rightarrow0$ one has quasi-one-dimensional chains with  weak zig-zag  interchain hopping. Thus, this model can be used to systematically
explore the effects of frustration in strongly correlated systems and would be of significant theoretical interest even without the experimental realisations of the model in organic charge transfer salts.

In order to make a direct comparison between theory and experiment one would like to know what parameters of the anisotropic triangular lattice (i.e., what values of $t$, $t'$ and $U$) represent specific materials. Significant effort has therefore been expended to estimate these parameters from electronic structure calculations. The first studies of the electronic structure of $\kappa$-\ET salts where limited, by the computational power available at the time, to extended H\"uckel theory \cite{Williams1992}. This is a semi-empirical, i.e. experimentally parameterised, tight-binding model and ignores the role
of the anions and electronic correlations.
 However, modern computing power means that density functional theory (DFT) calculations are no longer prohibitively expensive and several DFT studies have appeared recently.

\begin{table*}
\caption{Values of $t'/t$ of selected \kET salts, calculated from  Density
Functional Theory (DFT). 
H\"uckel theory gives systematically smaller values of  0.75, 0.84, and 1.06,
for these salts, respectively. Estimates of this parameter in  the metallic phase can also be made from quantum oscillation experiments
for \NCS  \cite{PrattPB10}, giving  $t'/t=0.6$ in agreement with the
DFT calculated value.}
\begin{center}
\begin{tabular}{c|c|c}
Material & $t'/t$& Reference\\\hline
\Cl & 0.4 & \cite{KandpalPRL09}\\
\NCS & 0.6 & \cite{KandpalPRL09,NakamuraJPSJ09}\\
\CN & 0.8 & \cite{KandpalPRL09,NakamuraJPSJ09}\\
\end{tabular}
\end{center}
\label{tab:t-DFT}
\end{table*}%

The large unit cells and complex anions of the $\kappa$ phase materials, meant that the first DFT studies of \ET salts focused on other crystallographic phases \cite{LeePRB03,FrenchJPCS04,YamaguchiP03,KublerSSC87,KasowskiIC90,KinoJPSJ06,MiyazakiKinoPRB03,MiyazakiPRB06}.
However, two groups have recently reported parameterisations of the tight-binding part of the Hamiltonian from DFT\footnote{Both GGA and LDA functionals give similar results.} calculations \cite{KandpalPRL09,NakamuraJPSJ09}. Both groups find that the frustration parameter, $t'/t$, is significantly smaller
than
 was previously thought on the basis of extended  H\"uckel calculations (summarised in Table \ref{tab:t-DFT}). Note that the frustration is least in \Cln, which has an antiferromagnetically ordered ground state and greatest in \CNn, which has a spin liquid ground state. However, even \CN is quite far from the isotropic triangular lattice ($t'=t$), which has been taken as the basis of a number of theories of \CN (discussed in Section \ref{sec:ring}) on the basis of  H\"uckel calculations \cite{KomatsuJPSJ96}.

\begin{figure}\begin{centering}
\includegraphics[width=9cm]{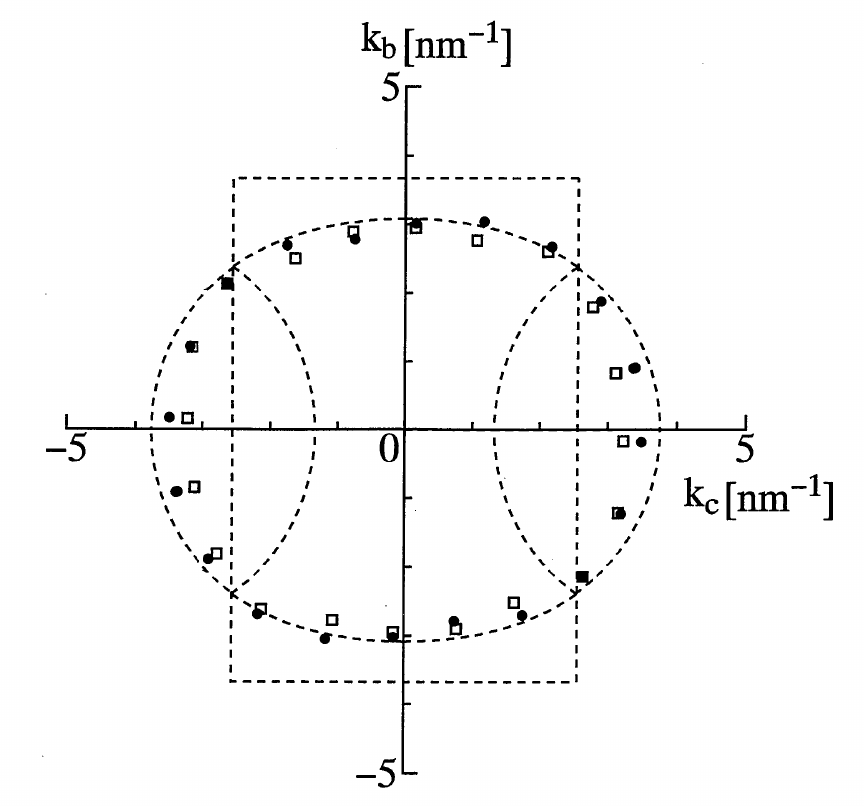}
\caption{
The Fermi surface of
$\kappa$-(BEDT-TTF)$_2$Cu$_2$(CN)$_3$ in the metallic phase
at pressures of 2.1 kbar (open squares) and 7.0 kbar (closed circles), as determined by
Angle-Dependent Magneto-Resistance Oscillations (AMRO) \cite{OhmichiJPSJ97}.
The dashed elliptical 
curves are the results of H\"uckel calculations.
}
\label{fig:amro}
\end{centering}
\end{figure}

The anisotropic triangular lattice has one site per unit cell. However, the $\kappa$-phase organics have two dimers per unit cell. This halves the Brillouin zone and causes the Fermi surface to be split into two sheets \cite{MerinoPRB00,PowellJPCM06}. Thus, two orbits are observed in quantum oscillations experiments. The lower frequency orbit, known as the $\alpha$ pocket corresponds to a hole like orbit. A higher frequency oscillation, known as the $\beta$ orbit, is only observed at higher fields and corresponds to the magnetic breakdown orbit around the Fermi surface of the dimer per unit cell model. The ratio of the areas of these orbits is strongly dependent on $t'/t$ \cite{PrattPB10}. Thus, estimates of $t'/t$ can be made from quantum oscillation or angle-dependent 
magnetoresistance (AMRO) experiments, 
which allow one to map out the Fermi surface \cite{kartsovnik}.
 In \NCS at ambient pressure this yields $t'/t=0.7$ \cite{CaulfieldJPCM94},
in reasonable agreement with the calculated value. In \CN at 7.6 kbar one finds that $t'/t=1.1$ \cite{OhmichiPRB09,PrattPB10}, which is rather larger than the ambient pressure value calculated from DFT. At 0.75 GPa 
the DFT calculations give $t'/t=0.75$ \cite{KandpalPRL09}, which is significantly smaller than the experimental estimate. AMRO experiments give a picture of  the Fermi surface
 that is qualitatively consistent with the calculated Fermi surface \cite{OhmichiJPSJ97}
(see Figure \ref{fig:amro}). However, the value of $t'/t$ has not been estimated from these measurements.

The area of the Fermi surface can be also determined by quantum oscillations.
For a wide range of organic charge transfer salts the area is
found to be consistent with Luttinger's theorem and the hypothesis
that these materials are always at half filling \cite{PowellJPCM04}.
This may put significant constraints on theories that 
the metal-insulator transition involves ``self-doping" \cite{BaskaranPRL03}.

The hopping between layers is much weaker than that within the layers. This can be measured in two separate ways: from AMROs \cite{Moses99,WosnitzaPRB02,SingletonPRL02,WosnitzaJPI96} or from a comparision of how disorder affects the superconducting critical temperature and the residual resistivity \cite{PowellPRB04}. Both methods find that the interlayer hopping integral, $t_\perp$ is a few tens of $\mu eV$ in the \kETXn, but that $t_\perp$ is an order of magnitude larger in the $\beta$ phase ET salts. DFT calculations \cite{LeePRB03} find that interlayer dispersion in $\beta$-(\ETn)$_2$I$_3$ is $\sim9$ meV. However, experimental estimates for the closely related material, $\beta$-(\ETn)$_2$IBr$_2$, yield $t_\perp\sim0.3$ meV \cite{PowellPRB04,WosnitzaJPI96}. 
However, one should note that the value of $t_\perp$ represents a very sensitive test of theory due its small absolute value and the small overlap of the atomic orbitals at the large distances involved in interlayer hopping.

\subsubsection{The Hubbard $U$}

There is a considerable literature that discusses the calculation of the Hubbard $U$ in a molecular crystals. Notable systems for which this problem has been tackled include the alkali doped fullerides \cite{Gunnarsson2004}, oligo-acene and thiopenes \cite{BrocksPRL04}, and the organic conductor tetrathiafulvalene-tetracyanoquinodimethane (TTF-TCNQ) \cite{CortesEPJB07}. These authors have proceeded by identifying  two separate contributions to the Hubbard $U$:
\begin{eqnarray}
U=U^{(v)}-\delta U^{(p)},
\end{eqnarray}
where $U^{(v)}$ is the contribution from the molecule (or cluster) in vacuum, and $\delta U^{(p)}$ is the reduction in the effective $U$ when the molecule is placed in the polarisable environment of the crystal.

One might think that $U^{(v)}$ is straightforward to calculate once one has a set
of  suitably localised orbitals as it is just the Coulomb repulsion between two holes (or electrons) in the same orbital:
\begin{equation}
F_0 = \int d^3 {\bf r}_1 \int d^3 {\bf r}_2  \frac{\rho_\uparrow({\bf r}_1) \rho_\downarrow({\bf r}_2)}{\left|{\bf r}_1-{\bf r}_2\right|}, \label{eqn:F0}
\end{equation}
where $\rho_\sigma({\bf r})$ the density of spin $\sigma$ electrons at the position $\bf r$ in the relevant orbital. However, this is incorrect. When one moves from a full band structure to the relevant one (or few) band model this interaction is significantly renormalised \cite{Gunnarsson2004,FreedACR83,IwataJCP92,ScrivenJCP09,Powell09}. Indeed, DFT calculations for a single \ET molecule find that the renormalised $U^{(v)}$ is about 50\% smaller than $F_0$ \cite{ScrivenJCP09}.

The first attempts to calculate $U^{(v)}$ from electronic structure calculations were also based on the extended  
H\"uckel method. It was noted \cite{KinoJPSJ96} that if one models the dimer as a two site Hubbard model where each site represents a monomer then in the limit $U_m^{(v)}\rightarrow\infty$, where $U_m^{(v)}$ is the effective Coulomb repulsion between two holes on the same monomer, one finds that $U^{(v)}\rightarrow2|t_{b_1}|$, where $t_{b_1}$ is the intra-dimer hopping integral. Whence, calculations of $t_{b_1}$ from the extended  H\"uckel approximation yield estimates of $U^{(v)}$ ranging between 0.14 eV \cite{RahalACB97} and 2.1 eV \cite{SimonovJMC05}. Note that this range of Hubbard $U$s is not caused just by changes in anions, but also the difference between different groups, who often find differences of more than a factor of two for the same material [for an extended discussion see \cite{ScrivenPRB09}]. More recently DFT has also been used to calculate $t_{b_1}$ and hence $U^{(v)}$ \cite{KandpalPRL09,NakamuraJPSJ09} - again there is a factor of two
 difference between the two different groups.


A better method of calculating $U^{(v)}$ is to note that 
\begin{eqnarray}
U=E_0(+2)+E_0(0)-2E_0(+1),
\end{eqnarray}
where $E_0(q)$ is the ground state energy of the molecule or cluster with charge $q$. This can be understood as $U$ is the energy required to activate the charge disproportionation reaction 2(\ETn)$_2^+\rightarrow$ (\ETn)$_2^0$ + (\ETn)$_2^{2+}$. Equivalently, $U$ is the difference in the chemical potentials for electrons and holes on the molecule or cluster. Calculations of this type for isolated \ET monomers show that $U_m^{(v)}$ is essentially the same for all monomers in the geometries in which they are found experimentally regardless of the anion in the salt,  the crystal polymorph, or the temperature or pressure at which the crystal structure was measured  \cite{ScrivenJCP09}. Remarkably, the same result holds for isolated dimers, consistent with the experimental finding that the dimer is a conserved structural motif in both the $\kappa$ and $\beta$ polymorphs \cite{ScrivenPRB09}.

Further, comparison of DFT calculations for monomers with those for dimers reveals that the approximation $U^{(v)}\rightarrow2|t_{b_1}|$ is \emph{incorrect} \cite{ScrivenPRB09}. This is because the effective Coulomb interaction between two holes on different monomers within the same dimer, $V_m^{(v)}$, is also large. Indeed, Scriven \etal
 found that $U_m^{(v)}\sim V_m^{(v)}\gg t_{b_1}$, in which case $U^{(v)}\simeq \frac12(U_m^{(v)} + V_m^{(v)})$, which is in reasonable agreement with their directly calculated value of $U^{(v)}$.

To date there are no calculations of $\delta U^{(p)}$ for \ET salts. This problem is greatly complicated for \ET relative to the other molecular crystals previously studied \cite{Gunnarsson2004,TsiperPRB03,BrocksPRL04,CortesEPJB07} because of the (often) polymeric anions and the fact that the intermolecular spacing is small compared to the size of the molecule. Therefore, Nakamura \etal \cite{NakamuraJPSJ09} have calculated $U$ directly from DFT band structure calculations by explicitly integrating out high energy interband excitations to leave an effective one band model. Interestingly, in order for the value of $U$ to converge Nakamura \etal had to include over 350 bands - corresponding to including excitations up to 16 eV above the Fermi level! Nakamura \etal find that the value of $U$ in the Mott insulator \CN (0.83 eV) is remarkably similar to that in the ambient pressure superconductor \NCS (0.85 eV). However, they find that $t=55$ meV for \CN and $t=65-70$ meV for \NCSn, yielding $U/t = 15.5$ for \CN and  $U/t=12.0-12.8$ for \NCSn, consistent with the experimental finding that former material is a Mott insulator that undergoes a Mott transition under moderate pressure and the later is an ambient pressure superconductor.

However, these values are much larger than those found from comparisons of DMFT calculations to optical conductivity measurements and on $\kappa$-(ET)$_2$Cu[N(CN)$_2$]Br$_x$Cl$_{1-x}$, which suggest that $U = 0.3$ eV \cite{MerinoPRL08}. These measurements are discussed in more detail in section \ref{ET-Mott}.

\subsubsection{The \ETtwo dimer}\label{ETdimer}

\begin{figure}
\begin{centering}
\includegraphics[width=8cm]{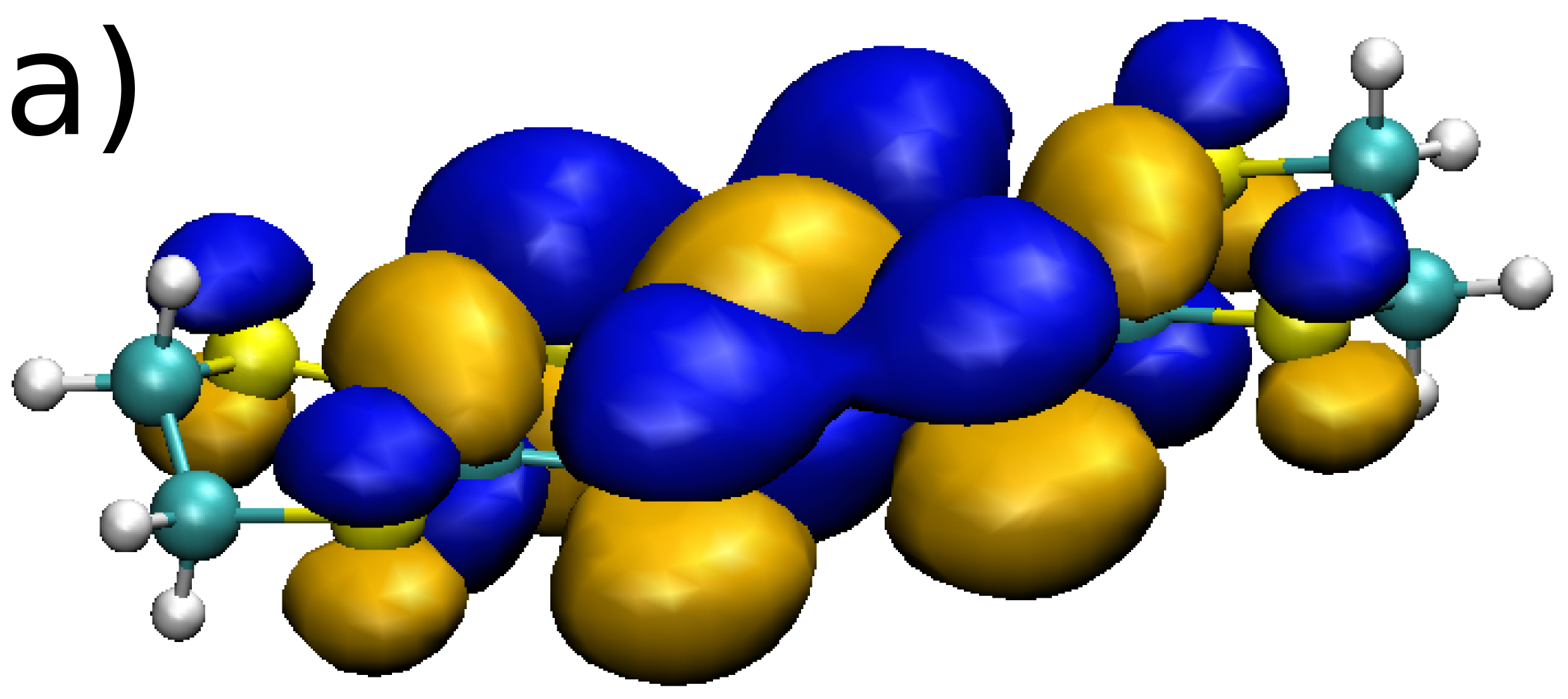}
\includegraphics[width=9cm]{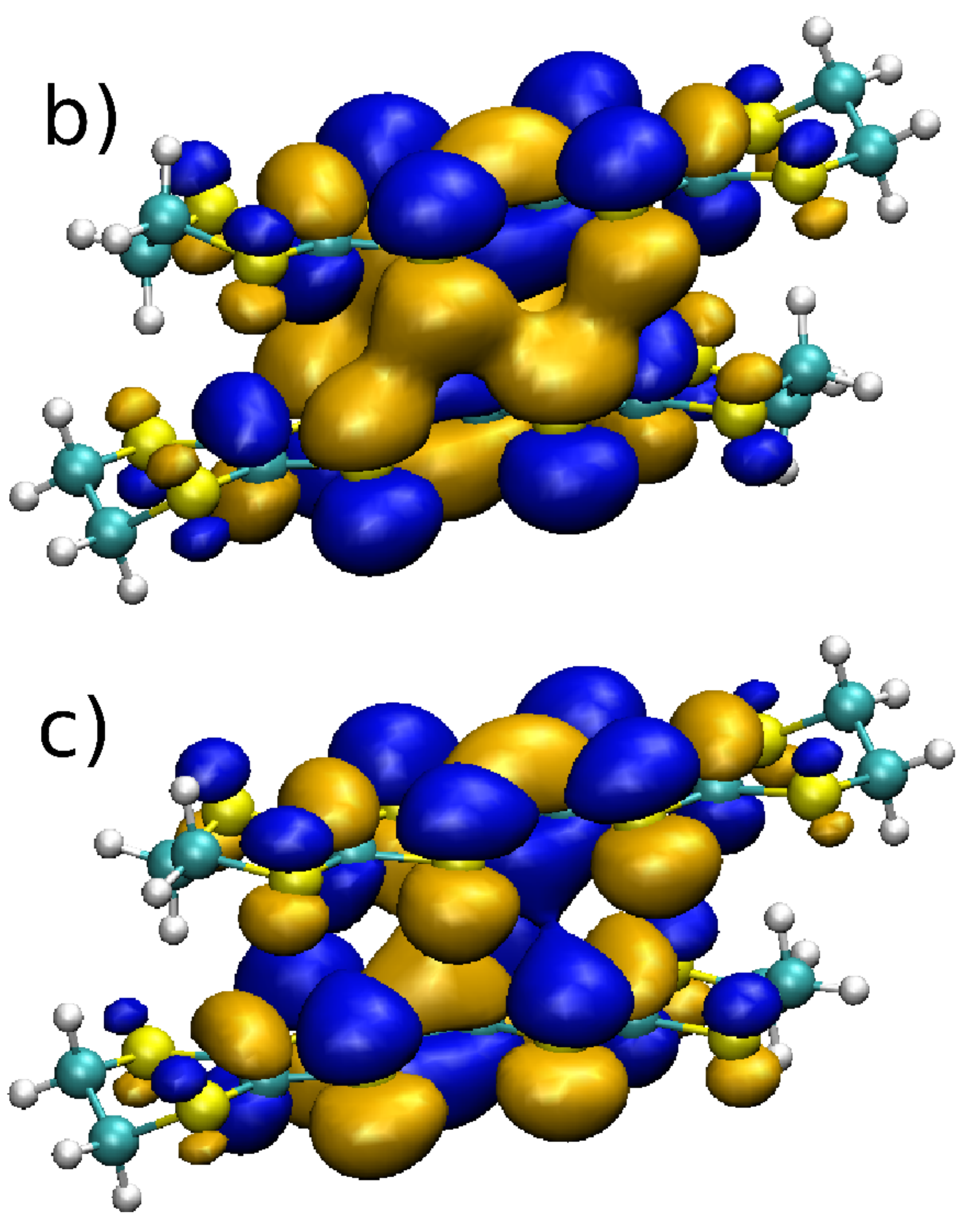}
\caption{The highest occupied molecular orbital (HOMO) of (a) an \ET molecule in the geometry found in \Cln, (b) an (BEDT-TTF)$_2^{2+}$ dimer in the geometry found in \CN and (c) a neutral (BEDT-TTF)$_2$ dimer in the geometry found in \CNn. It is clear from these plots that the HOMO of the neutral dimer is the antibonding combination of the two monomer HOMOs, whereas the HOMO of the (double) cation dimer is the bonding combination of the two monomer HOMOs. Thus, the \ETtwo dimer is held together by a `covalent bond' between the two monomers rather than bonds between any two particular atoms.  [Modified from \cite{ScrivenJCP09} and  \cite{ScrivenPRB09}].}
\label{fig:ET-HOMO}
\end{centering}
\end{figure}

Significant insight can be gained from comparing \ETtwo with the hydrogen molecule. In the molecular orbital (Hartree-Fock) picture \cite{Fulde95} the ground state wavefunction of H$_2$ is
\begin{eqnarray}
|\Psi\rangle = \frac12\left(|\phi_{1\uparrow}\rangle + |\phi_{2\uparrow}\rangle \right)\otimes\left(|\phi_{1\downarrow}\rangle + |\phi_{2\downarrow}\rangle \right),
\end{eqnarray}
where $|\phi_{i\sigma}\rangle$ is a basis function for an electron with spin $\sigma$ centred on atom $i$. This provides the simplest model of the chemical bond, which results from the stabilisation of the bonding combination of atomic orbitals, and implies an increased electronic density between the two atoms. If one includes electronic interactions the picture is somewhat complicated as the wavefunction becomes correlated. These correlations can be described in the two site Hubbard model, which is a good model for the hydrogen molecule, where each atom is treated as a site \cite{Powell09}. If one compares the electronic density in the HOMO of a single \ET molecule  (Fig. \ref{fig:ET-HOMO}a) with that of the HOMO of the \ETtwo dimer (Fig. \ref{fig:ET-HOMO}b), it is clear that the \ETtwo dimer wavefunction is close to being an antibonding combination of molecular wavefunctions, whereas the HOMO-1 (Fig. \ref{fig:ET-HOMO}c) is close to being a bonding combination of molecular wavefunctions. In the charge transfer salt there are, on average,  two electrons in the HOMO-1, but only one in the HOMO. Therefore, the net effect is bonding.

Electronic correlations also play an important role in the \ETtwo dimer. But, as in the case of H$_2$,  the two site Hubbard model, where each site is now an \ET molecule, provides a good description of the electronic correlations in the \ETtwo dimer \cite{PowellJPCM06,ScrivenPRB09}. This shows that the physics of the \ETtwo dimer is remarkably similar to that of the hydrogen molecule. Therefore, we can understand the \ETtwo dimer as being held together by a `covalent bond' not between any two atoms, but between the two \ET molecules themselves. As one expects this  `intermolecular covalent bond' to be strong compared to the interactions between dimers, this provides a natural explanation of the conservation of the dimer motif across different materials. 


\subsection{Insulating phases}\label{ET-ins}

Both \Cl and \CN are insulators at ambient pressure \cite{Ishiguro1998}, and undergo a metal-insulator transition under the application of hydrostatic pressure (which we will discuss in section \ref{ET-Mott}). This can be understood straightforwardly, in terms of the half-filled Hubbard model, introduced in section \ref{ET-struct}, as a Mott insulator phase \cite{McKenzieCCMP98,KanodaPC97}. However, despite these similarities in the charge sector, the spin degrees of freedom in the two materials behave very differently.


\subsubsection{Antiferromagnetic and spin liquid phases}

Shimizu \etal \cite{ShimizuPRL03} measured and compared bulk spin susceptibilities of \Cl and \CNn. Both materials are described by the Hubbard model on the anisotropic triangular lattice, cf. Fig. \ref{fig:ET-struct}.
In the Mott insulating phase the spin degrees of freedom are
described by a Heisenberg model (Section \ref{sec:heisenberg}) with exchange constants, $J\sim250$ K. However, \CN is significantly more frustrated than \Cl 
 (as expected by electronic structure calculations, cf. Table \ref{tab:t-DFT}). \Cl shows a clear magnetic phase transition at $\sim$27 K. This is an antiferromagnetic transition \cite{MiyagawaPRL95,KanodaPC97} and is only visible in the bulk spin susceptibility because there is a small canting of the magnetic moment \cite{MiyagawaPRL02}, which gives rise to a weak ferromagnetic moment \cite{WelpPRL92}. In contrast, no such phase transition is visible in the susceptibility of \CNn. Analyses \cite{ShimizuPRL03,ZhengPRB05} of the high temperature magnetic susceptibility  show that in both materials the effective Heisenberg exchange is $J\sim250$ K. 
Therefore, the absence of a phase transition in \CN down to 32 mK (the lowest temperature studied and four
orders of magnitude smaller than $J$)  led Shimizu \etal to propose that \CN is a spin liquid.

\begin{figure*}
\begin{centering}
\includegraphics[height=9cm]{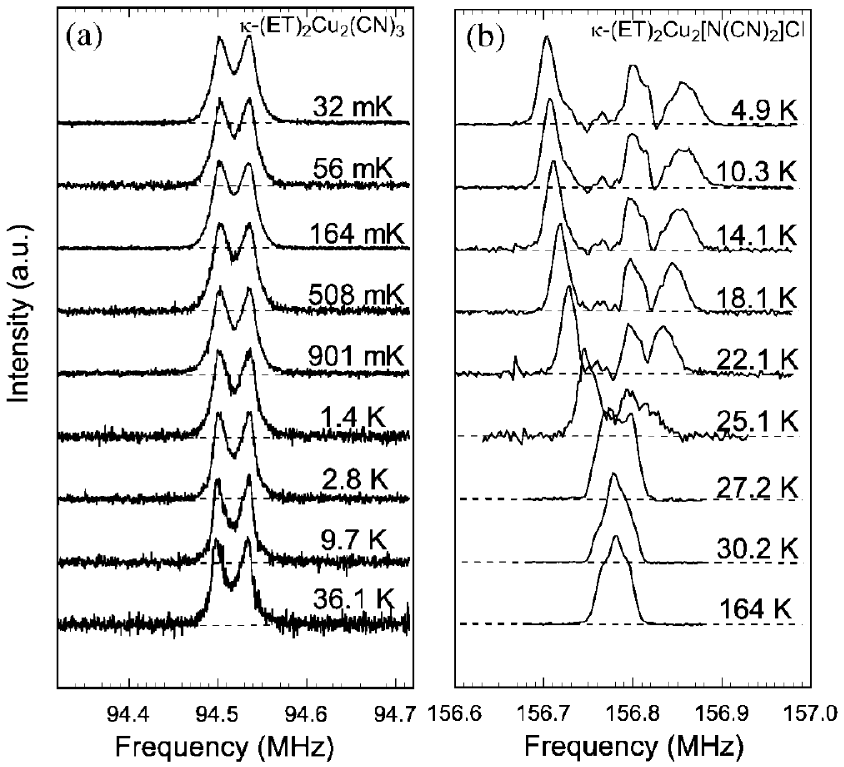}\includegraphics[height=9cm]{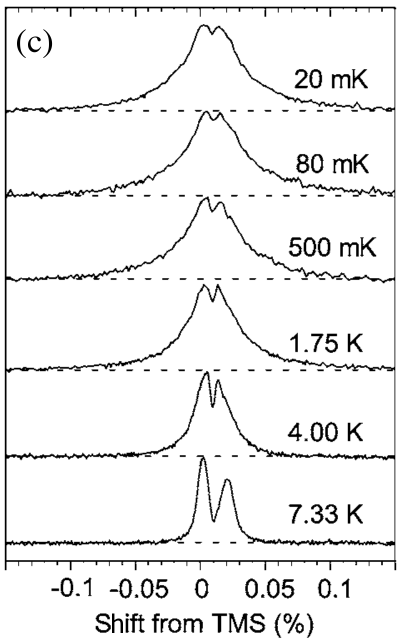}
\caption{The low temperature, ambient pressure $^1$H-NMR absorption spectra of (a) \CN  and (b) \Cln, and (c) the $^{13}$C-NMR absorption spectra of \CNn. The antiferromagnetic phase transition at $\sim27$ K is clear seen in \Cln. In contrast, no major changes occur with temperature in \CNn, consistent with the absence of long range magnetic order. However, the spectra do broaden as $T$ is lowered in \CNn. This broadening is seen even more dramatically in the $^{13}$C-NMR spectrum of \CN (panel c). Again, no signs of long range magnetic order are seen down to 20 mK, a temperature that is four orders of magnitude smaller than the antiferromagnetic exchange energy, $J\approx250$ K \cite{ShimizuPRL03,ZhengPRB05}.  [Panels (a) and (b) were taken from \cite{ShimizuPRL03} and panel (c) was modified from \cite{ShimizuPRB06}.]
[Copyright (2003,2006) by the American Physical Society.]
}
\label{fig:CN-Cl-NMR}
\end{centering}
\end{figure*}

\begin{figure}
\includegraphics[width=9cm]{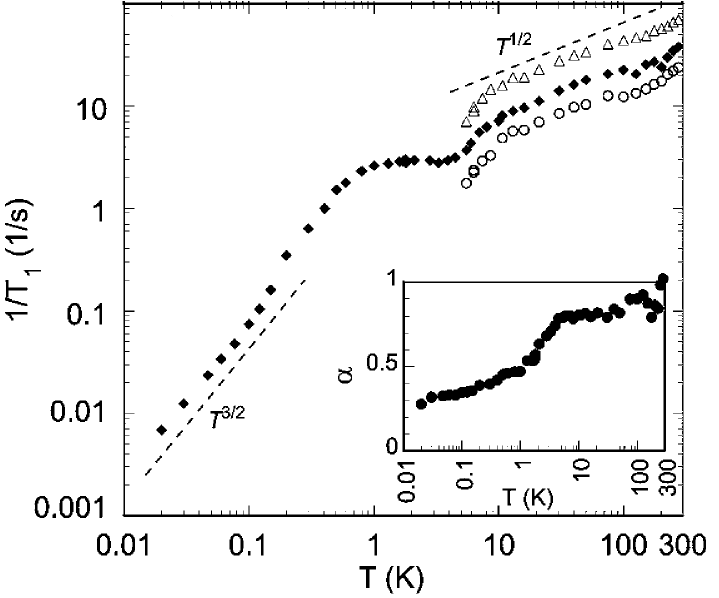}               
\caption{Temperature dependence of the 
  the $^{13}$C-NMR relaxation rate $1/T_1$
for the spin liquid material \CN at ambient pressure \cite{ShimizuPRB06}.
Between 20 mK and 1 K $1/T_1 \sim T^{3/2}$, which suggests that
the spin triplet excitation spectrum is gapless.
Such a power law dependence is consistent with quantum critical behaviour.
The inset shows the temperature dependence of the exponent $\alpha$
associated with the stretched exponential time dynamics of
the spin relaxation.
[Copyright (2006) by the American Physical Society.]
}
\label{fig:shimNMR}
\end{figure}



The form of the temperature dependence of the susceptibility turns out
to be quite sensitive to the amount of frustration \cite{ZhengPRB05}
(cf. Figure \ref{fig:susc}).
The values of both $J$ and $J'$ can be estimated by comparing
the observed temperature dependence of the uniform
magnetic susceptibility   with high
temperature series expansions  (above about $J/4$).
For \CN they agree for $J \simeq 200$  K and $J' \simeq J$. 
In Section \ref{sec:ring}
we discuss the possible effects of
ring exchange. Consequently, it is desirable to know
how they may modify the temperature dependence of the susceptibility
and the values of the exchange interaction estimated from the experimental data.

Further, evidence for the absence of magnetic ordering in \CN comes from its NMR spectrum. Fig. \ref{fig:CN-Cl-NMR} compares the $^1$H NMR absorption spectrum of \CN with that of \Cln. Shimizu \etal reported that ``the difference of the spectra between the two salts at high temperatures is explained by the difference in the orientation of ET molecules against the applied field and does not matter.'' In \Cl (Fig. \ref{fig:CN-Cl-NMR}b) they observe clear changes in the NMR spectrum below $T_C\sim27$ K. These multiple peaks are caused by the distinct crystal environments for the $^1$H atoms due to the antiferromagnetic ordering. In contrast, no quantitative changes are observed in
the spectrum of \CN down to 32~mK, the lowest temperature studied (Fig. \ref{fig:CN-Cl-NMR}a), consistent with an absence of long-range magnetic ordering. 

No evidence of long range magnetic order is observed in the $^{13}$C-NMR spectra of \CN down to 20 mK (the lowest temperature studied) (Fig. \ref{fig:CN-Cl-NMR}c). This is important because these experiments were carried out on samples where the $^{13}$C is one of the atoms involved in the central C=C double bond. The electron density is much higher for this atom (cf. Fig. \ref{fig:ET-HOMO}) than
 for the H atoms, which are on the terminal ethylene groups (cf. Fig. \ref{fig:molecules}). Therefore the $^{13}$C spectra demonstrate that the absence of long range order is genuine and not an artefact caused by low electronic density on the H atoms. We stress that 20~mK is four orders of magnitude smaller than the exchange coupling, which suggests that \CN may well be a true spin liquid.

The observed temperature dependence of the NMR relaxation rates for
$\kappa$-(BEDT-TTF)$_2$Cu$_2$(CN)$_3$ are also
inconsistent with this material
having a magnetically ordered ground state.
The observed \cite{ShimizuPRB06} decrease of the NMR relaxation rate, $1/T_1$, and the spin echo rate, $1/T_2$, with decreasing temperature for
$\kappa$-(BEDT-TTF)$_2$Cu$_2$(CN)$_3$ is distinctly
different from that expected for a material with a magnetically ordered ground state.
For such materials at low temperatures, both $1/T_1$ and $1/T_2$
should  increase   rapidly with decreasing temperature,
 rather than decreasing (since $1/T_1T \sim \xi(T)^2$).
The increase is seen in \Cl, above the antiferromagnetic ordering temperature.
 For materials described by the antiferromagnetic Heisenberg model on a square lattice [La$_2$CuO$_4$] \cite{sandvik2}
 and a chain [Sr$_2$CuO$_3$] \cite{takigawa},
both relaxation rates do {\it increase} monotonically
as the temperature decreases. 

 It is noteworthy that 
for $\kappa$-(BEDT-TTF)$_2$Cu$_2$(CN)$_3$ 
at low temperatures, from 1 K down to 20 mK, it was found \cite{ShimizuPRB06}
that $1/T_1 \sim T^{3/2}$ and $1/T_2 \sim $ constant
(see Figure   \ref{fig:shimNMR}).
As discussed in Section \ref{sec:qcdecon}
this is similar to  
that expected in the quantum critical regime, (\ref{qcrit}),
with the critical exponent $\eta \sim 1$, expected for
a non-linear sigma model with deconfined spinons.
However, the observed temperature dependence of $T_1$ and $T_2$
would lead to two different values for the exponent $\eta$.

However, caution is required in interpreting the data in Fig. \ref{fig:shimNMR}. The observed relaxation rate is not well described by a single exponential. Shimizu \etal extracted $T_1$ from a fit to a stretched exponential, the exponent, $\alpha$, is plotted in the inset to Fig. \ref{fig:shimNMR}. This could be indicative of multiple relaxation rates or some other complex phenomena that has not yet been adequately explained.


There is a way to check that  the NMR relaxation is actually
due to spin fluctuations and not another physical mechanism.
The magnitude of the relaxation rate at high
temperatures can be used to provide an independent estimate of $J$.             
Data 
for $\kappa$-(BEDT-TTF)$_2$Cu$_2$(CN)$_3$ at ambient pressure
 \cite{kawamoto:kcn32006} 
gives, for the outer $^{13}$C site,
$1/T_1\simeq 10-30$/sec in the range 100-300 K.
From the $K-\chi$ plot a value of
$A=0.07$ T/$\mu_B$ is deduced for the outer site \cite{ShimizuPRL03}.
Using the above values in the expression (\ref{t1-highT})
gives $J\simeq 200-600$ K, consistent with the value $J=250$ 
deduced from the temperature dependence of the
uniform magnetic susceptibility \cite{ShimizuPRL03,ZhengPRB05}.

However, Shimizu \etal did observe a slight broadening of the $^1$H NMR spectrum of \CN as the temperature is \emph{lowered}. They observed 
an even more dramatic broadening  in the $^{13}$C NMR (Fig. \ref{fig:CN-Cl-NMR}c) \cite{ShimizuPRB06}.   This is somewhat counterintuitive and has provoked some theoretical interest, discussed below.
Spin echo $^{13}$C experiments show that the broadening is inhomogenous ($T_2^*$) rather than an increase in the homogeneous $T_2$. Similar broadenings are also seen in EtMe$_3$Sb[Pd(dmit)$_2$]$_2$ and EtMe$_3$P[Pd(dmit)$_2$]$_2$, cf. Section \ref{sect:dmit-spin-liquid} and Fig. \ref{fig:ItouPRB08}, which could hint that this is a rather general phenomenon.
It was also observed that a magnetic field induces  spatially non-uniform
local moments \cite{ShimizuPRB06}. Motrunich has given an
 interpretation of this observation in terms
of spin liquid physics \cite{motrunich2}: the fluctuating
gauge field associated with the spinons leads to
the nuclear spins "seeing" a distribution in local magnetic fields. 

 Several model calculations  have been performed to 
attempt to explain the large broadening by taking into account the role of disorder \cite{GregorPRB09}.
 They found that they could only explain the experimental data
for temperatures above about 5 K, if there is much larger disorder than expected and that it is strongly temperature dependent.
This is in contrast to previous work
where  comparable calculations for a kagome antiferromagnet 
could explain experimental data for ZnCu$_3$(OH)$_6$Cl$_2$ \cite{GregorPRB08}.
Gregor and Motrunich mention that it is hard to estimate the strength of the disorder and the role of temperature dependent screening.
It is desirable to connect this work to recent estimates of the strength of
disorder in the \kETX materials \cite{ScrivenPRB09}.

\subsubsection{Is the spin liquid in \CN gapped?}

Key questions about a spin liquid are: is it gapped and what are
 the nature of the low lying excitations? In particular, are there deconfined spinons? Two experiments have recently tried to address these questions in \CNn, one by measuring the specific heat  \cite{YamashitaNP08}, the other by measuring the thermal conductivity \cite{YamashitaNP09}. However, as will now discuss, the two groups reached contradictory conclusions on the basis of these different measurements.

S. Yamashita \etal \cite{YamashitaNP08} concluded that there are gapless fermionic excitations, i.e., deconfined spinons, in \CN on the basis of their specific heat, $C_p$, measurements. A plot of $C_p/T$ against $T^2$ 
is linear in the range $\sim$0.75-2.5 K, implying that $C_p=\gamma T+\beta T^3$, with $\gamma=20\pm5$ mJ K$^{-2}$ mol$^{-1}$. Moving to lower temperatures complicates heat capacity measurements as there is a significant Schottky anomaly. Nevertheless, the data in the temperature range 0.075 - 3 K fits well to the form $C_p=\alpha/T^2+\gamma T+\beta T^3$ with $\gamma=12$  mJ K$^{-2}$ mol$^{-1}$.  
One expects a large linear term in the heat capacity if there are gapless fermionic excitations. Indeed, the values of $\gamma$ estimated by S. Yamashita \etal are the same order of magnitude as those found in the metallic phases of \kX salts. Further, comparing this value with the previous measurements of the bulk magnetic susceptibility \cite{ShimizuPRL03} gives a Sommerfeld-Wilson ratio, $R_W=(\pi^2k_B^2/\mu^2) (\chi_0/\gamma)$, of order unity \cite{YamashitaNP08}, which is what one would expect if the same fermions were responsible for both the linear term in the specific heat and the susceptibility \cite{leelee2}.
In contrast, other organic charge transfer salts which undergo magnetic
ordering were found to have no such linear term but to have a specific heat capacity that was quadratic in temperature.

In a discussion of these results Ramirez \cite{RamirezNP08} pointed out that S. Yamashita \etaln's data is fit equally well by $C_p=\alpha/T^2+\gamma_{2/3} T^{2/3}+\beta T^3$. This is consistent with the predictions for spinons coupled to a $U(1)$ gauge field \cite{MontrunichPRB05} (as discussed in Section \ref{sec:gauge}. Ramirez was also concerned that the entropy associated with the $\gamma$ term estimated by S. Yamashita \etal is only about $\frac{R\ln2}{40}$,
which is only a small fraction of the total spin entropy. However,  
it is not clear to us that this should be a point of concern since for at
temperatures of order $J/5$ the entropy of a Heisenberg antiferromagnet
is already much less than the high temperature value due to short-range 
spin correlations \cite{ElstnerJAP94}.

\begin{figure}
\begin{centering}
\includegraphics[width=7cm]{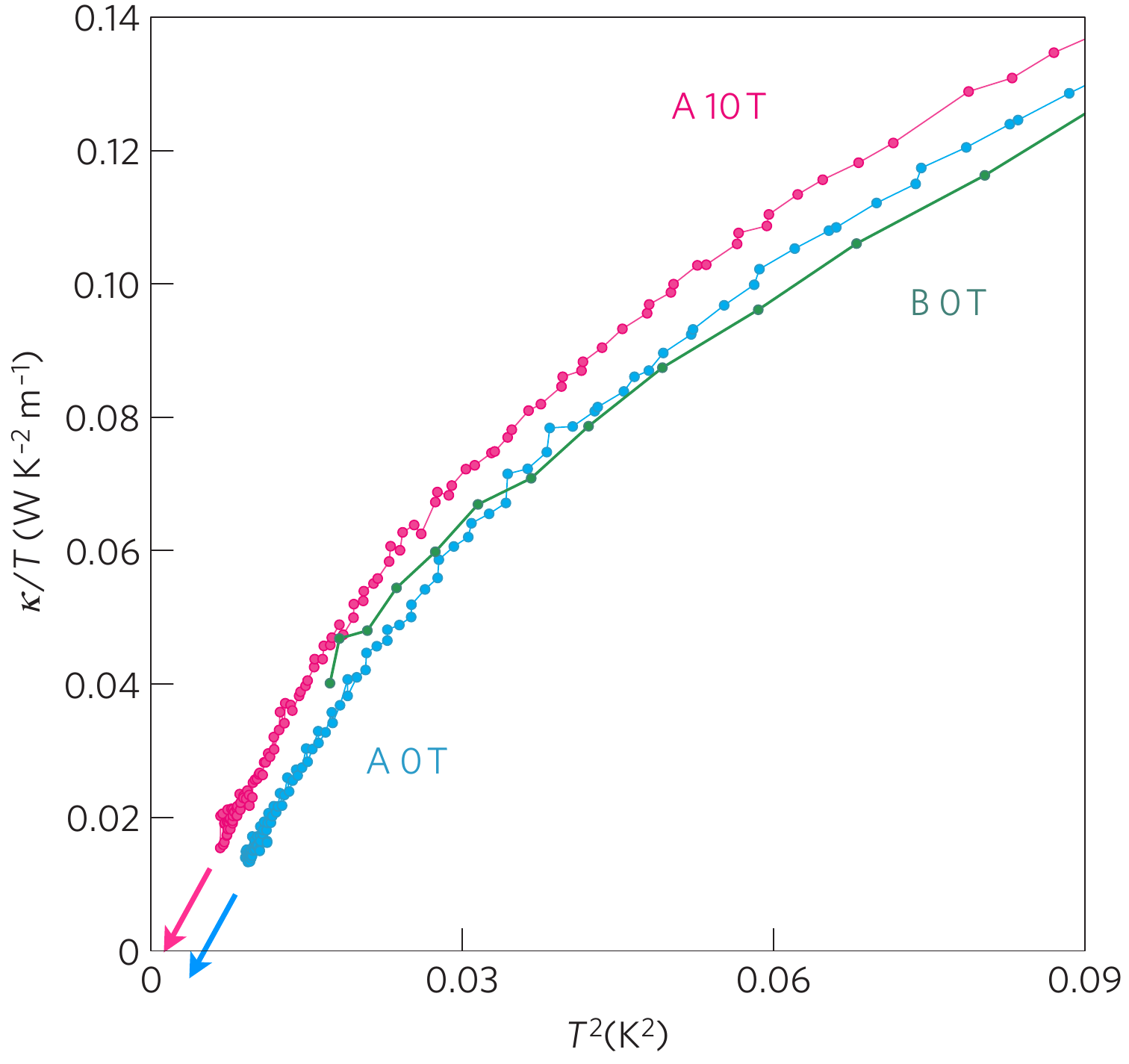}
\caption{The thermal conductivity, $\kappa$, of \CN measured
 in two samples (A and B) and in different magnetic fields. As with the heat capacity, simple arguments suggest that, at low temperatures $\kappa/T=\alpha+\beta T^2+\dots$. Clearly this is not what is observed. These data suggest that $\alpha\simeq0$ (a simple extrapolation gives $\alpha<0$, which is  unphysical). This suggests that the spin liquid state of \CN is gapped. However, finite temperatures do not lead to a quadratic increase in $\kappa/T$, suggesting that the low-lying excitations may be more complex that simply magnons and phonons. [Modified from \cite{YamashitaNP09}.]}
\label{fig:YamashitaNP09f3}
\end{centering}
\end{figure}

In contrast to the specific heat results
described above M. Yamashita \etal \cite{YamashitaNP09} concluded, on the basis of thermal conductivity measurements, that the spin liquid state in \CN is fully gapped. As with the heat capacity, one expects that for a simple metal the thermal conductivity is given by $\kappa=\alpha T+\beta T^3+\dots$ \cite{Ziman60}, with the  fermions  giving rise to the linear term and bosons, typically phonons, giving rise to the cubic term. Note particularly that, as $\kappa$ is only sensitive to itinerant excitations, one does not need to subtract a Schottky term. Fig. \ref{fig:YamashitaNP09f3} shows M. Yamashita \etaln's data in the temperature range 0.08-0.3 K plotted as $\kappa/T$ against $T^2$. One immediately notices that the data does not lie on a straight line, which suggests that it is not dominated by phonons and therefore that M. Yamashita \etal did resolve the contribution from magnetic excitations. Further support for this assertion comes from the field dependence of the data, which one would not expect if the heat transport was dominated  by phonons. However, more importantly, one should notice that an extrapolation of the data to $T=0$ will not give a significant $\kappa/T$ (indeed the simplest extrapolation, indicated by the arrows in the figure, gives $\kappa/T<0$, which is unphysical). Therefore, M. Yamashita \etal concluded that $\kappa/T$ vanishes at $T=0$ K. If correct this would imply that the spin liquid state of \CN  is gapped. 

M. Yamashita \etal also attempted to quantitatively analyse the very lowest temperature part of their data. One complication in this exercise was that they were unable to directly determine what fraction of the measured thermal conductivity is due to magnetic excitations and what fraction, $\kappa_{ph}=\beta T^3$, is due to phonons. M. Yamashita \etal found that $\kappa(T)$ cannot be well described by a power law even if $\kappa_{ph}$ is large enough to represent three quarters of the measured $\kappa$ at $T=100$ mK, which seems a rather generous upper bound given their arguments (described above) that the phonons do not dominate the thermal conductivity. This suggests that the gap does not have nodes, which
would give a small but non-zero intercept.

M. Yamashita \etal also made an Arrhenius-plot of their data. A reasonable fit was found for a value of $\beta$ that implies
that about one quarter of the thermal conductivity at 100 mK is due to phonons. This fit yields a gap of 0.46 (0.38) K in zero field (10 T). However, as M. Yamashita \etal stress, one should be cautious about taking this precise value too seriously as that fit was limited to less than a decade of temperature (0.08-0.5 K) due to the low energy scales involved and current limitations in cryogenic technology. Nevertheless, this analysis does show that, if there is a gap, it is 2-3 orders of magnitude smaller than the exchange energy $J\sim250$ K.

Clearly, an important question is why these two experiments
(specific heat and thermal conductivity)
 lead to such different conclusions. M. Yamashita \etal \cite{YamashitaNP09} argued that this disagreement  results from an incorrect subtraction of the Schottky term in the heat capacity. However, this is unlikely to be the full story because the Schottky term only dominates the heat capacity below $\sim0.2$ K. One point of interest is that the value of $\gamma$ extracted from the heat capacity measured between $\sim$0.75-2.5 K (in which no Schottky anomaly is evident) is almost twice that found from the fit of the data taken between 0.075 and 3 K. The gap estimated by M. Yamashita \etal is small compared to 0.75 K, so one would expect there to be high densities of thermally excited fermions in the higher temperature range. Indeed, significant densities of thermally excited fermions would remain over most of the lower temperature range. 

\subsubsection{The 6 K anomaly}

One thing both groups \cite{YamashitaNP08,YamashitaNP09}
do agree on is that something interesting happens at temperatures around 6 K.
S. Yamashita \etal found a broad `hump' when they replot their data as $C_pT^{-3}$ against $T$ (in this plot the phonon term should just appear as a constant offset, while the Schottky term is not relevant at these relatively high temperatures). They also present a provocative plot of $\Delta C_p/T$ against $T$, where $\Delta C_p$ is the difference between the heat capacities of \CN and \NCSn. However, \NCS becomes superconducting at $\sim10$ K, so its heat capacity is changing rapidly in the relevant temperature range. This makes it difficult to distinguish which  of the changes in $\Delta C_p$ are due to \CNn. Note that the estimation of $\gamma$, discussed above, is from lower temperature data. However, the `hump' appears as  a change in slope of $C_p/T$ versus $T^2$ around 6 K, so it is not clear that whatever causes this effect can be neglected in the estimation of $\gamma$. S. Yamashita \etal also found that the heat capacity is remarkably insensitive to magnetic fields (they studied fields up to 8 T).  

The anomaly in the thermal conductivity is, however, very clear cut. A hump is immediately obvious in the plot of $\kappa$ against $T$, Fig. 1 of  \cite{YamashitaNP09}, which begins at $\sim6$ K and reachs a broad maximum at $\sim4$ K. Clear anomalies have also been reported  in the NMR spin-lattice relaxation rate, $1/T_1$, \cite{ShimizuPRL03} and the uniaxial expansion coefficients \cite{MannaPRL10} in this temperature range.

A number of theoretical explanations have been proposed for the 6~K anomaly including: pairing of spinons \cite{LeePRL07}, the formation of visons (vortices in a $Z_2$ spin liquid) \cite{QiPRL09},  spin-chirality ordering \cite{BaskaranPRL89} and exciton condensation \cite{QiPRB08}. These theories will be discussed in section \ref{sec:quasiparticle}. 

%
%

\subsection{Mott metal-insulator transition}\label{ET-Mott}

In the cuprates\footnote{As we only refer to the cuprates here for pedagogical reasons we will neglect subtleties relating to the role of the oxygen p-levels and the distinction between charge transfer and Mott insulators \cite{ZaanenPRL85}.} the metal-insulator transition is driven by chemically doping charge carriers into the copper-oxygen plane of the insulating parent compound \cite{LeeRMP06}. This is sometimes referred to as the `band-filling controlled Mott transition'. However, in the organics the `parent' insulating compound can be driven metallic by decreasing $U/W$ the ratio of the Hubbard $U$ to the bandwidth, $W$. This is often referred to as the `bandwidth controlled Mott transition' (cf. Figure \ref{fig:motttrans})
There are several ways to drive the bandwidth controlled Mott transition in the \kX salts: 

\begin{enumerate}
\item \emph{Hydrostatic pressure.} This is a beautiful realisation of Mott's original proposal \cite{MottPRS49} of how to drive a Mott insulator metallic. Because they form rather soft crystals, only moderate pressure (sometimes
as small as a few hundred bars) are required to drive very significant changes, including the Mott transition, in organic charge transfer salts. 

\item \emph{Uniaxial stress.} This seems a particularly promising approach as it holds out the prospect of also tuning the frustration, i.e., $t'/t$. However, this method has not yet been widely applied to the \kX salts. For a recent review see \cite{KagoshimaCR04}.

\item `\emph{Chemical pressure}'. Changes in the anion have a significant effect on the unit cell parameters - particularly in systems with polymeric anions. Thus, tuning the chemistry of the anion is remarkably similar to applying a pressure. For example, \Cl is an antiferromagnetic Mott insulator, but the isostructural \Br is a metal, which superconducts at low temperatures. 
A particularly elegant form of chemical pressure is to alloy the anions Cu\-[N\-(CN)$_2$]\-Br and Cu\-[N\-(CN)$_2$]\-Cl to form crystals of \ClBrn. We stress that as both anions are monovalent this \emph{does not dope} the organic layer away from half filling. 

\item \emph{Deuteration of the cation.} Each \ET molecule contains eight hydrogen atoms, cf. Fig. \ref{fig:molecules}. \Br is extremely close to Mott transition and crystals containing the fully deuterated molecule are antiferromagnetic insulators \cite{TaniguchiPRB03}. Crystals of partially deuterated \ET molecules, which can be made uniformly deuterated throughout the entire crystal, sit at different positions spanning the first order Mott transition \cite{TaniguchiPRB03} and the macroscopic coexistence of the metallic and insulating phases can be seen in these crystals \cite{SasakiJPSJ05}. No detailed explanation of how this
deuteration effect operates has been presented to date. 
Presumably, deuteration weakens
the hydrogen bonding interaction between the cation and anions 
because of the different quantum zero point motion (cf. \cite{HayashiJCP06}).
\end{enumerate}
 Although the Mott transition in organics is commonly called `bandwidth controlled' we stress that really the important quantity is the ratio $U/W$, and as $U$ is significantly renormalised by interdimer (as well as intramolecular and intradimer) processes. It has been suggested that both hydrostatic and chemical pressure may also result in variations in $U$ \cite{ScrivenPRB09,NakamuraJPSJ09}. Further, the ratio $t'/t$ also has an important impact on wether the ground state is metallic or insulating \cite{PowellPRL07}.
 
\subsubsection{Critical exponents of the Mott transition}\label{sect:exponents}

Much attention has focused on the Mott transition from the antiferromagnetic state to a correlated metal and superconductor. Indeed, the phase diagram of this transition in \Cl has been mapped out in considerable detail \cite{PowellJPCM06,LimelettePRL03,KagawaN05,FaltermeierPRB07,KagawaNP09}. 

Theoretical arguments predict that the Mott transition belongs to the Ising universality class \cite{CastellaniPRL79,KotliarPRL00}. These can be understood on the basis of an analogy between the Mott transition and the lattice gas \cite{CastellaniPRL79}. Here one views the metallic phase as a liquid of doubly occupied and vacant sites [corresponding to (BEDT-TTF)$_2^0$ and (BEDT-TTF)$_2^{2+}$] moving on a background of singly occupied [(BEDT-TTF)$_2^{+}$] sites. The Mott insulating phase is then simply the gaseous phase of this model. Indeed, a formal basis for this analogy can be given within the dynamical mean-field approximation \cite{KotliarPRL00}. In this theory the Mott critical point is described by a scalar (Ising) order parameter, which couples to the singular part of the double occupancy. Experimental support for this theory have come from measurements of the critical exponents associated with the metal-insulator transition in (V$_{0.989}$Cr$_{0.011}$)$_2$O$_3$ that suggest that this transition belongs to the 3D Ising  universality class \cite{LimeletteS03}.


It was therefore surprising when a novel set of critical exponents ($\beta\approx1$, $\gamma\approx1$, $\delta\approx2$) were reported for the metal-insulator transition in \Cl \cite{KagawaN05,KagawaNP09}. Indeed, the critical exponents found by Kagawa \etal from measurements of the conductivity \cite{KagawaN05} are far from those of the Ising model in either two- or three-dimensions. Nevertheless, the Widom scaling relation, $\gamma=\beta(\delta-1)$, is obeyed and, when appropriately scaled, the data collapses onto two curves (one for data above the critical temperature, the other for data below  the critical temperature). Kagawa \etal \cite{KagawaNP09} have also reported the same order parameter exponent, $\beta$, from NMR measurements. This is interesting as NMR probes the magnetic degrees of freedom, whereas the conductivity probes the charge degrees of freedom.

 A number of theories have been proposed to try and explain these results. Imada \etal \cite{ImadaJPSJ05,ImadaPRB05,MisawaJPSJ06} predicted exponents close to those observed by Kagawa \etal in a theory based on the proximity of the first order Mott transiton to a quantum critical point as the critical end point is moved to $T=0$. Alternatively, Papanikolaou \etal \cite{PapanikolaouPRL08} have argued that the experiments are indicative of an Ising universality class. Papanikolaou \etal showed that the conductivity is not only sensative to the order parameter, but can also depend on other singular variables, particularly the energy density. In the regime where the energy density dominates they found that  $\beta=1$, $\gamma=7/8$ and $\delta=15/8$, consistent with the Widom scaling relation and in reasonable agrement with the experimental results. However, this theory does not explain finding that $\beta=1$ from an NMR experiment \cite{KagawaNP09}.

Bartosch \etal have recently shown that a scaling theory based on the Ising universality class can describe
the observed temperature dependence of the  thermal expansion near
the critical point for the fully deuterated \Br
  material \cite{BartoschPRL10}. It would therefore be interesting to know what Imada \etaln's theory predicts for these experiments.


\subsubsection{Optical conductivity}\label{sect:opt-cond}

Faltermeier \etal \cite{FaltermeierPRB07} have studied the evolution of the reflectivity and optical conductivity spectra as \BrCl as it is driven through the metal-insulator transition by increasing the Br density, $x$. At low temperatures, three important features can be identified in these spectra: a Drude peak and two broad peaks that are fit well by Lorentzians at around 2200 cm$^{-1}$ and 3200 cm$^{-1}$. 

The Drude peak is absent in the pure Cl and low Br density compounds:
 as expected for the Drude peak arises from
Fermi liquid quasiparticle excitations, which are absent in the Mott insulating phase \cite{KotliarPT04}. As the Br density is increased the system is driven metallic by chemical pressure and the Drude peak appears - it can be seen rather weakly for $x=0.73$. Increasing $x$ further increases the width of and 
the spectral weight under the Drude peak.

The optical spectrum of the Hubbard model at half filling and near the Mott transition is only expected to show two main features: the Drude peak in the metallic phase and a single broad peak centred on $\sim U$ and of width $\sim W$ \cite{MerinoPRB00,KotliarPT04}. This broad peak corresponds to excitations that change the number of doubly occupied sites (and therefore change the number of vacant sites so as to ensure charge conservation). Therefore, Faltermeier \etaln's observation of two broad Lorentzians requires explanation.

Faltermeier \etal \cite{FaltermeierPRB07} argued that the lower frequency Lorentzian is the peak predicted by the dimer Hubbard model of \BrCln.  If correct, this assignment would yield a estimate of $U=0.27$ eV, which is significantly smaller than that found from downfolding DFT calculations \cite{NakamuraJPSJ09} (cf. section \ref{ET-struct}). Further, Faltermeier \etal argued that the higher frequency feature is due to intra-dimer transitions.

However, Werner and Millis \cite{Werner10} have recently made significant advances in dealing with dynamical screening near the Mott transition via DMFT. Their calculations of the spectral function differ significantly from those with only a static $U$. For simple models of the frequency dependence of the effective on-site Coulomb repulsion they find two broad peaks in the spectral function at finite frequencies for a broad range of parameters. At high screening frequencies these two peaks appear to correspond to a screened-$U$ band and a bare-$U$ band. However, it is important to stress that Werner and Millis conclude that ``peak positions in the spectral functions do not provide quantitative estimates of either the screened or unscreened $U$ values.'' This may explain why Faltermeier \etaln's $U$ value is significantly smaller than theoretical estimates. While Werner and Millis did not carry out explicit calculations of the optical conductivity, one can anticipate that dynamical screening will change the optical conductivity significantly from what is expected for a static $U$. Indeed, on the basis of Werner and Millis's calculations one would expect to find an additional broad peak at finite frequency in the optical conductivity, precisely as is seen experimentally. This suggests a possible reinterpretation of Faltermeier \etaln's data. An interesting question is whether this theory is capable of accounting for the observed changes in vibrational frequencies that are naturally explained by Faltermeier \etaln's theory.

\begin{figure}\begin{centering}
\includegraphics[width=9cm]{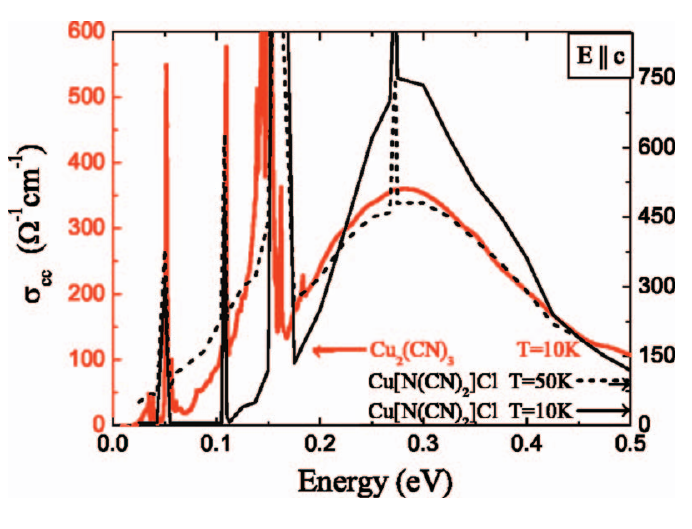}
\caption{
Comparison of the frequency dependent conductivity in two Mott insulators,
\Cl 
and  $\kappa$-(BEDT-TTF)$_2$Cu$_2$(CN)$_3$ \cite{KezsmarkiPRB06}.
The former has a ground state with  Neel antiferromagnetic order and clearly
has an energy gap of about 0.1 eV. 
In contrast, the latter compound may have a spin liquid ground state
and has a much smaller energy gap.
The sharp peaks are due to intramolecular vibrational modes and not the electronic degrees of freedom.
[Copyright (2006) by the American Physical Society].
}
\label{fig:optcond}
\end{centering}
\end{figure}

The frequency dependence of the conductivity, 
shown in Figure \ref{fig:optcond},
suggests that the charge gap in  \CN is smaller than \Cl
\cite{KezsmarkiPRB06}.
Indeed, it has been suggested that in the former compound 
there is  a charge gap, but that the optical conductivity 
has a power law dependence at low frequencies \cite{NgPRL07}.
Note, that the charge gap (the energy cost of adding
an electron or hole; a signature of a Mott insulator)
is  a different physical quantity from the optical gap
(the energy required to produce a charge neutral, spin singlet excitation,
with a non-zero transition dipole moment)
and so it is possible, at least in principle, that the former is non-zero
and the latter is zero.
However, the relative size of the energy gaps also presents a puzzle
because one can also argue that \Cl
 is closer to the metallic phase than the other compound.
Since \Cl
requires a smaller pressure to destroy the Mott insulating phase
(300 bar versus 4 kbar).

Motivated by these experimental results,
Ng and Lee calculated the frequency dependence of the optical
conductivity in a Mott
insulating state which is a spin liquid with a spinon Fermi surface
and coupled to a fluctuating U(1) gauge theory \cite{NgPRL07}.
(This theory is discussed further in Section \ref{sec:gauge}.)
They find that there is a power-law frequency dependence at low
frequencies due to the conductivity of the spinons. The spinons
are charge neutral and so do not couple directly to an electromagnetic field.
However, they couple indirectly because the external field induces an
internal gauge field in order to maintain the constraints associated
with the slave-rotor representation of the electrons.

Deducing whether the experimental data shown in Figure \ref{fig:optcond}
does imply zero optical gap for \CN 
could be made more rigorous by subtracting the vibrational 
contributions. A robust   procedure now exists for this and
has been applied in the analysis of the optical conductivity
of alloys of \Cl
 and \Br \cite{FaltermeierPRB07,MerinoPRL08,DummPRB09}.

\subsubsection{The spin liquid to metal transition}

\CN also undergoes a Mott transition under hydrostatic pressures $\sim0.35$ GPa \cite{KurosakiPRL05}. What little is known experimentally about how the spin liquid insulator to metal transition differs from the antiferromagnetic-insulator--metal transition comes mainly from the pioneering work of Kurosaki \etal \cite{KurosakiPRL05}. They reported measurements of the resistivity and NMR, but did not examine the critical end-point closely. Kurosaki \etal  observed two NMR spin-lattice relaxation rates, $1/T_1$ at 0.35 GPa, i.e., close to the of  metal-insulator transition. This suggests that the metal-insulator transition in \CN is first order and that the two rates are caused by the coexistence of the insulating and metallic phase. Note that the metal-insulator transitions in \Cl and  (as a function of deuteration in) \Br are also first order. Kurosaki \etal also found that pressure (up to 0.8 GPa) does not induce any significant changes in the $^1$H NMR spectrum at 1.4 K. This shows that, at least at this temperature, pressure does not induce long range magnetic ordering and hence, one presumes, the spin liquid state remains right up until the first order Mott transition.

\subsubsection{Reentrance of the Mott transition - explanation from undergraduate thermodynamics}
\label{sec:undergrad}

One interesting difference between the Mott transitions in \Cl and \CN is the shape of the first order line in the ($P$-$T$) phase diagram \cite{KagawaPRL04,KurosakiPRL05}. For the pressure driven metal-insulator transition the Clausius-Clapeyron relation is 
\begin{eqnarray}
\frac{dT}{dP}=\frac{\Delta V}{\Delta S}, \label{eqn:CC}
\end{eqnarray} 
where $\Delta V=V_\textrm{ins}-V_\textrm{met}$ and $\Delta S=S_\textrm{ins}-S_\textrm{met}$.
As the metal is the high pressure phase one presumes\footnote{However, some care should be exercised with this assumption. For example, famously for the ice-water transition $\Delta V<0$.} that $\Delta V>0$. Therefore, the sign of $dT/dP$ is determined by the sign of $\Delta S$. In \CN  $dT/dP>0$ along the entire phase transition,  cf. Fig. \ref{fig:phase-diagram-kCN}, \cite{KurosakiPRL05}. In contrast, the phase transition in \Cl is reentrant,  cf. Fig. \ref{fig:phase-diagram-kCl}, \cite{KagawaPRL04}, i.e.,  $dT/dP$ changes sign along the phase boundary. Therefore, at certain pressures ($\sim25$ MPa) isobaric cooling results in first a insulator-to-metal transition (at $T\sim35$ K) followed by a metal-to-insulator transition (at $T\sim20$ K). The change in sign of $dT/dP$ occurs in the region of the phase diagram where antiferromagnetism is observed.

Fermi statistics imply that the entropy of the electrons in a metal varies linearly with temperature.\footnote{As $S(T)=\int_0^T\frac{C_v}{T}dT$ and $C_v=\gamma T$ for a gas of fermions.} In the antiferromagnetic Mott insulator phase the entropy is dominated by the spin degrees of freedom. One expects the entropy to be carried by spin waves in the magnetically ordered phase, thus $S(T)\sim T^\alpha$. For a quasi-two dimensional material, such as \Cln, one expects that  $\alpha>1$.\footnote{In the antiferromagnetically order states one finds $\alpha=2$ in two dimensions, e.g., on the square lattice antiferromagnet, and $\alpha=3$ in three dimensions. As \Cl is quasi-two dimensional an intermediate behaviour may also be possible.} Therefore, at low temperatures the entropy of the antiferromagnetically ordered state is proportional to $T^\alpha$. At low enough temperatures this will always be less than the entropy of a Fermi liquid, which is proportional to temperature. 

In a paramagnetic insulator, the entropy becomes
independent of temperature at high temperatures.\footnote{For
a paramagnetic insulator with $T \gg J$ 
the  entropy per spin is $S/N=k_B\ln2$.} The phase diagrams of \CN \cite{KurosakiPRL05}, \Cl \cite{LefebvrePRL00,KagawaPRL04} and V$_2$O$_3$ \cite{McWhanPRB73,LimeletteS03} demonstrate that $\Delta S>0$ for the paramagnetic insulator-phase transition in all three of these materials.  
In principle, this argument could be quantitatively tested from measurements of the heat capacity; however, performing such measurements under pressure is extremely challenging. 

For the triangular lattice Heisenberg model the
entropy is much larger than that of the square
lattice model \cite{ElstnerJAP94,BernuPRB01}.
For example, a value of $0.2R$ is obtained at temperatures of 
$0.5J$ and $0.15J$ respectively.
For both models  the entropy (and specific heat) are 
quadratic in temperature at low temperatures.
However, the coefficient of proportionality is twenty times
larger for the triangular lattice than the square lattice \cite{BernuPRB01}.

\begin{figure*}
\begin{centering}
\includegraphics[width=16cm]{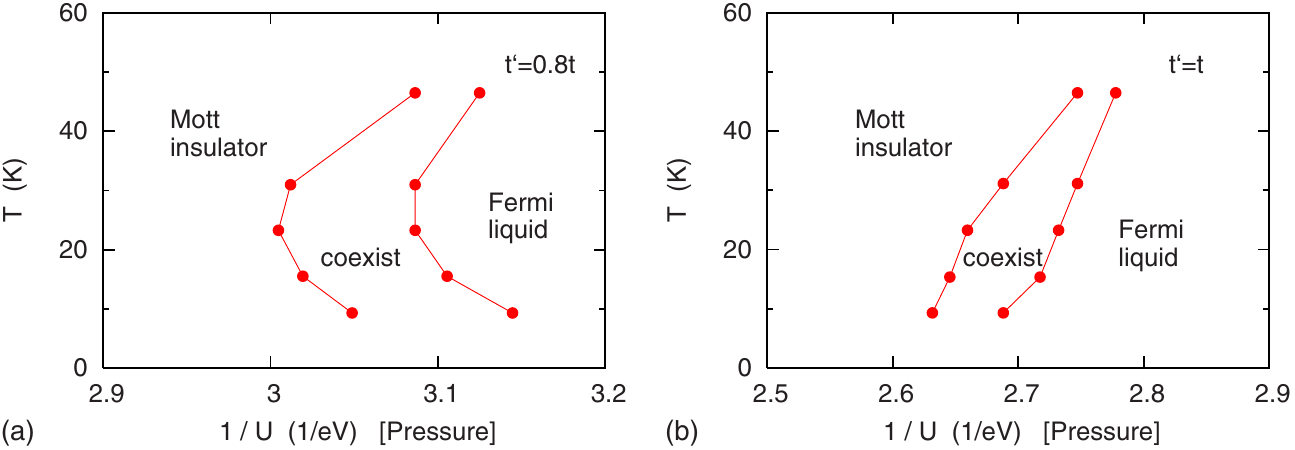}
\caption{Phase diagram at finite temperature from cluster dynamical mean-field
theory (CDMFT)
of the Hubbard model on the anisotropic triangular lattice
at half filling \cite{LiebschPRB09}.
As $U/t$ increases there is a first order phase transition
from a metallic to a Mott insulating phase.
This first order line ends at a critical point.
(a) and (b) are for $t'/t=0.8$ and 1, respectively. For  $t'=0.8t$ a reentrant Mott transition is found. This can be understood from the Clausius-Clapeyron equation (\ref{eqn:CC}) as showing that, at low temperatures, the insulating phase has lower entropy than the metallic phase. This would be expected if the insulating phase were magnetically ordered (see text) and can even be caused by the short range antiferromagnetic correlations associated in incipient magnetic ordering, as is the case here. At high temperatures the reverse is true, consistent with a simple paramagnetic metal. For $t'=t$ the insulating phase has higher entropy at all temperatures, consistent with a spin liquid ground state. The phase diagrams shown in panels (a) and (b) are consistent with those of  \Cl (Fig. \ref{fig:phase-diagram-kCl}) and \CN (Fig. \ref{fig:phase-diagram-kCN}) respectively. [From \cite{LiebschPRB09}.]
[Copyright (2009) by the American Physical Society].
}
\label{fig:LiebschPRB09}
\end{centering}
\end{figure*}

It is interesting to compare the experimental phase diagrams of \CN and \Cl with the cluster dynamical mean field
theory (CDMFT) calculations of 
Liebsch \etal \cite{LiebschPRB09} for the phase diagram of the Hubbard model on an (an)isotropic triangular lattice, Fig. \ref{fig:LiebschPRB09}.  Similar results were obtained independently for
$t'=0.8t$ by  different group {\it et al.} \cite{OhashiPRL08}.
They find a first order Mott transition as $U/W$ is decreased.
However, they found interesting differences in the phase diagrams as the frustration, $t'/t$, is varied. 
For the isotropic triangular lattice ($t'=t$) the line of first order transitions always has a positive slope. It follows from the Clausius-Clapeyron equation (\ref{eqn:CC}) that the insulating state has a larger entropy than the metallic state, even at low temperatures. For $t'=0.8t$ the slope of the phase boundary becomes negative at low temperatures, indicating that the metallic state has greater entropy at low temperatures. 
This is in semi-quantitative agreement with the observed temperature-pressure phase diagram of a range of organic charge transfer salts if we associate \CN with $t'=t$ and \Cl with $t'=0.8t$. However, the parameterisation of the tight-binding model from DFT (cf. section \ref{ET-struct}) suggest that $t'/t$ is actually rather smaller for both materials (cf. Table \ref{tab:t-DFT}).

It is interesting to note that Liebsch \etal did not allow for long range antiferromagnetic order in their calculations. Thus, short-range magnetic fluctuations must be sufficient to account for decreased entropy in the insulating state. The parameterisations of the tight binding model for \CN (see section \ref{sec:dimer}) would put this material in the regime where Liebsch find a reentrant phase transition. However, even
 more sophisticated calculations, going beyond the three site cluster Liebsch \etal studied, may result in a shift in the parameter regime in which reentrance is observed. Therefore, the lack of reentrance in the in the phase diagram of \CN is consistent with a spin liquid ground state.

\subsection{Magnetic frustration  in the normal state}\label{ET-metal}

A striking feature of the normal state is that the resistivity \cite{KurosakiPRL05,LimelettePRL03,AnalytisPRL06} Hall coefficient  \cite{SushkoSM97,TanatarPRB97,MurataSSC90} and thermopower \cite{YuPRB91,BuravovJPI92,DemishevJETP98} all vary non-monotonically with temperature \cite{MerinoPRB00}. This is in marked contrast to what is found in weakly correlated metals \cite{Ashcroft76}, where these quantities have a monotonic temperature dependence. 

Further, at high temperatures the conductivity 
is less than the Mott-Ioffe-Regal limit \cite{KurosakiPRL05,LimelettePRL03,AnalytisPRL06}, which would mean that, in a Drude
picture, electrons are scattering more frequently than they hop from site to site \cite{MerinoPRB00,GunnarssonRMP03}. In weakly correlated metals this only found as one approaches the Anderson transition in extremely disordered systems \cite{Phillips03}. Whereas, the organics are remarkably clean systems \cite{kartsovnik,AnalytisPRL06,SingletonRPP00}. 

A third difference between \kETX and weakly correlated metals is that no Drude peak is observed in the optical conductivity above
a relatively low temperature ($T\gtrsim40$ K) \cite{DresselPRB94,TamuraJPSJ91,EldridgeSSC91,KornelsenSSC89,MerinoPRL08,FaltermeierPRB07}. 

However, at low temperatures the mean free path returns below the Mott-Ioffe-Regal limit \cite{KurosakiPRL05,LimelettePRL03,AnalytisPRL06} and a Drude peak is seen in the optical conductivity \cite{DresselPRB94,TamuraJPSJ91,EldridgeSSC91,KornelsenSSC89,MerinoPRL08,FaltermeierPRB07}. 
This phenomenology is observed for both \CN  \cite{KurosakiPRL05} and its more weakly frustrated brethren.

\subsubsection{Dynamical mean-field theory (DMFT)}\label{sec:DMFT}

DMFT \cite{GeorgesRMP96,KotliarPT04} provides both a fundamental explanation \cite{MerinoPRB00} and an accurate description  \cite{LimelettePRL03,MerinoPRL08} of these phenomena. 
Merino and McKenzie \cite{MerinoPRB00} found that, even without taking account of the details of the band structure, the basic features of the $\kappa$ phase organics, described above, are captured by DMFT. Furthermore,
these features are seen in a broad range of other strongly
correlated electron materials such 
as transition metal oxides. In particular, DMFT predicts an
 incoherent or `bad' metal at high temperatures. In this regime there are no quasiparticles (and hence no Drude peak). Incoherence implies that momentum is not a good quantum number, i.e., that electrons frequently scatter off one another, and gives rise a mean free path less than
a lattice constant \cite{MerinoPRB00,GunnarssonRMP03}. 

Below a characteristic temperature, $T_{coh}$, DMFT predicts a Fermi liquid.  Hence, the resistivity drops below the Mott-Ioffe-Regal limit and the Drude peak returns. However, the Fermi liquid is strongly correlated and the effective mass is almost an order of magnitude larger than the band mass. The change from the bad metal to the Fermi liquid is a crossover rather than a phase transition. 
In DMFT it is this crossover that is largely responsible for the nonmonotonicity of many response functions, including the resistivity, the thermopower and the Hall coefficient.

The success of DMFT in describing this broad range of experiments is, initially, rather puzzling. DMFT is exact in infinite dimensions or for an infinite co-ordination number. Hence, one expects DMFT to be a good approximation in the limit of large dimensions, but
the $\kappa$-phase organics are quasi-two-dimensional. However, it has recently been argued \cite{MerinoPRB06} that DMFT is a much better approximation for frustrated systems than unfrustrated systems as frustration suppresses long range correlations.
 The applicability of DMFT to low-dimensional
systems with large frustration is consistent with the fact that a Curie-Weiss law holds down to a much
lower temperature  for
frustrated magnetic models than for unfrustrated
models \cite{RamirezARMS94,SchifferPRB97,ZhengPRB05}. 
Deviations from Curie-Weiss behavior result from spatially dependent
correlations. Hence, the DMFT treatment of the Hubbard
model on frustrated lattices is expected to be a good approximation down to
much lower temperatures than it is for unfrustrated models. 

Furthermore, in
the ``bad metal'' region magnetic properties such as the uniform
susceptibility and spin relaxation rate, can be described by the
Heisenberg model because the electrons are essentially localized due
to the proximity to the Mott insulating phase. This means that the
susceptibility follows a Curie-Weiss form down to temperatures
much less than the exchange energy $J$. The spin
correlation length of the antiferromagnetic Heisenberg model
increases with temperature  much more slowly for the triangular
lattice than the square lattice \cite{ElstnerPRL93,ElstnerJAP94}. Specifically, at
$T=0.3J$ the spin correlation length is only one lattice constant
for the triangular lattice. In contrast, for the square lattice the
correlation length is about 50 lattice constants, at
$T=0.3J$ \cite{ElstnerPRL93,ElstnerJAP94}.

Dynamical cluster approximation (DCA) calculations provide
a means to systematically go beyond DMFT.
They show that for the isotropic
triangular lattice the solution is remarkably similar to that found from single site DMFT.
In particular, a quasiparticle peak appears at the Fermi energy.
However, if the frustration is released a pseudogap opens 
in the one-electron spectra as a result of short range
antiferromagnetic correlations \cite{ImaiPRB02}. We will delay more detailed discussion of these results until section \ref{pseudo-CN}.

A further hint that DMFT is a better approximation on the triangular
lattice than it is on the square lattice comes from the fact that one finds
that, at half filling, the Mott
transition occurs at $U_c\approx 15 |t|$ \cite{MerinoPRB06}. This 
can be compared with more sophisticated numerical treatments
which find that the Mott transition
takes place at $U\approx 6-8 |t|$ (see Section \ref{sec:heisenberg}).
 On the square lattice it is known
that perfect nesting of the Fermi surface
 means that the ground state is insulating 
for any finite $U$. However, DMFT predicts \cite{GeorgesRMP96} that
$U_c\gg|t|$ unless antiferromagnetism is included. Thus
(without including antiferromagnetism) DMFT gives a qualitatively
incorrect result for the (unfrustrated) square lattice, but a
qualitatively correct result for the (frustrated) triangular
lattice.

Hence, it appears that frustration plays an important role even in the normal state of the organic charge transfer salts. Counterintuitively, by suppressing long range spin correlations, frustration makes the normal state of the $\kappa$-phase organics easier to understand than would be the case without significant frustration. This may be taken as a major blessing if one compares the comparative simplicity of the normal state of the organics to the complexities of the `normal' state of the cuprates \cite{LeeRMP06}. Thence, an important question is: are the differences between the normal states of the organics and the cuprates intrinsic differences between the band-width controlled Mott transition and the band-filling controlled Mott transition, or are extrinsic effects responsible for the non-Fermi liquid effects observed in the cuprates?

\subsubsection{Fermi liquid regime}\label{ET:FL}

At low temperatures ($T<T_{coh}$) DMFT reduces to a local Fermi liquid theory and hence predicts that the temperature dependence of the resistivity is given by  $\rho(T)=\rho_0+AT^2$, where $\rho_0$ results from impurity scattering and the quadratic term results from electron-electron scattering. This temperature dependence is indeed seen experimentally in a range of organic charge transfer salts \cite{KurosakiPRL05,LimelettePRL03,StrackPRB05}. DMFT also predicts an enhancement of the effective mass over the band mass predicted by electronic structure calculations. This enhanced mass can be observed experimentally via the linear terms in the heat capacity, $C_v=\gamma T$. These two facts are not unrelated.

The Kadowaki-Woods ratio is defined as $A/\gamma^2$. This ratio is found to take the same value in a range of transition metals \cite{RicePRL68}. Twenty years later it was found empirically that in a range of heavy fermions the Kadowaki-Woods ratio takes the same value \cite{KadowakiSSC86}, albeit with several well known outliers. However, the ratio in the heavy fermions was found to be an order of magnitude larger than that in the transition metals. It was pointed out some time ago that the Kadowaki-Woods ratio is even larger in the organics \cite{DresselSM97,StrackPRB05}. 

This large Kadowaki-Woods ratio has recently been shown to be the consequence of the details of the band structure of the organics \cite{JackoNP09}. Indeed, Jacko \etal found that, quite generally, the Kadowaki-Woods ratio depends on the band structure of the material in question. Jacko \etal proposed a new ratio, closely related to the Kadowaki-Woods ratio, that takes these band structure effects into account. They found that this ratio takes the same, predicted, value in a wide range of transition metal, heavy fermion materials, transition metal oxides and organic charge transfer salts.
This new understanding of the Kadowaki-Woods ratio  shows that the mass enhancement measured by the specific heat and the quadratic term in the resistivity share the same physical origin. This strongly suggests that electron-electron scattering are responsible for both effects, which had been questioned in the organics \cite{StrackPRB05}. As this suggests that electron-electron interactions are the strongest forces immediately above $T_c$, it may also imply that these same interactions        are implicated in the mechanism of superconductivity in these materials.



\subsubsection{NMR and the pseudogap}\label{sect:pg}

Beyond the arguments above that frustration enhances localisation and thus emphasises the Mott physics captured by DMFT, the properties discussed above do not depend crucially on the frustration at play in organic charge transfer salts. Therefore, to better understand the role of frustration, it is desirable to experimentally probe the spin correlations in the metallic state. The most direct method would be inelastic neutron scattering. However, this requires large single crystals, which have never been grown for organic charge transfer salts \cite{PintschoviousEPL97,ToyotaSM97,TaniguchiJLTP06}. Therefore, the best remaining probe in nuclear magnetic resonance (NMR) spectroscopy. 

Two key properties measured in an NMR experiment are the Knight shift, $K_s$, which is the shift in the resonance frequency due the screening of the applied magnetic field by the conduction electrons in a metal, and the spin-lattice relaxation rate, $1/T_1$,  which is the characteristic time taken for spins flipped by a magnet field to return to their equilibrium distribution. In a metal, both of these quantities  are related to the dynamic spin susceptibility, $\chi({\bf q},\omega)=\chi'({\bf q},\omega)+i\chi''({\bf q},\omega)$, of the electrons. It may be surprising that the nuclear relaxation rate is a probe of electrons. However, this is because the total system (nuclei and their environment) must conserve energy and spin. Therefore, the nuclei can only relax by interacting with their environment. In a metal the low energy relaxation pathways are dominated by exchanging spin with the conduction electrons. Thus, one finds that \cite{MoriyaJPSJ63} 
\begin{eqnarray}
\frac{1}{T_1} &=& \lim_{\omega \to 0}\frac{2k_BT}{\gamma_e^2\hbar^4}
\sum_{\bf q} |A({\bf q})|^2\frac{\chi''({\bf q},\omega)}{\omega},
\label{t1t}
\end{eqnarray}
and
\begin{eqnarray}
K_s &=& \frac{|A({\bf 0})| \chi'({\bf0},0)}{\gamma_e\gamma_N
\hbar^2}, \label{eqn:Ks}
\end{eqnarray}
where $A({\bf q})$ is the hyperfine coupling between the nuclear and
electron spins, and $\gamma_N$ ($\gamma_e$) is the nuclear (electronic)
gyromagnetic ratio. Note that, because of the factor $T$ in
the expression (\ref{t1t}),  $1/T_1T$ often gives  more
direct access the temperature dependence
of the spin fluctuations than $T_1$ itself.

For non-interacting electrons one finds that 
\begin{eqnarray}
K_s\propto N(\epsilon_F)\label{eqn:KsNI}
\end{eqnarray}
 and 
\begin{eqnarray}
\frac{1}{T_1T}\propto N(\epsilon_F)^2,\label{eqn:T1TNI}
\end{eqnarray} 
where $N(\epsilon_F)$ is the density of states at the Fermi level, if the hyperfine coupling is constant in reciprocal space, which is strictly true if there is only one atom per unit cell, and is an approximation otherwise. Note that both $K_s$ and $1/T_1T$ are independent of temperature in this approximation. Further, taking the ratio $1/T_1TK_s^2$ removes the dependence on $N(\epsilon_F)$, which is  generally not known \emph{a priori} \cite{KorringaP50}. One finds that, for non-interacting electrons,
the dimensionless ratio
\begin{eqnarray}
\mathcal{K} \equiv \frac{\hbar}{4\pi
k_B}\left(\frac{\gamma_e}{\gamma_N}\right)^2 \frac{1}{T_1T K_s^2}=1.
\end{eqnarray}
$\mathcal{K}$ is known as the Korringa ratio. These three results: that $1/T_1T$ and $K_s$ are independent of $T$ and that $\mathcal{K}=1$ are collectively known as Korringa behaviour. Indeed, one can show \cite{YusufJPCM09} that these results hold for interacting systems provided vertex corrections to the dynamic spin susceptibility are negligible. This holds regardless of the form of the self energy, so long as it is consistent with Ward identities. However, magnetic fluctuations lead to vertex corrections to $\chi({\bf q},\omega)$ \cite{YusufJPCM09,Doniach98}. Therefore systems with strong magnetic fluctuations do not display Korringa behaviour. In particular,  $\mathcal{K}<1$ in systems with ferromagnetic fluctuations and $\mathcal{K}>1$ in systems with antiferromagnetic fluctuations \cite{DoniachJAP68}.

\begin{figure}
\begin{centering}
\includegraphics[width=7cm]{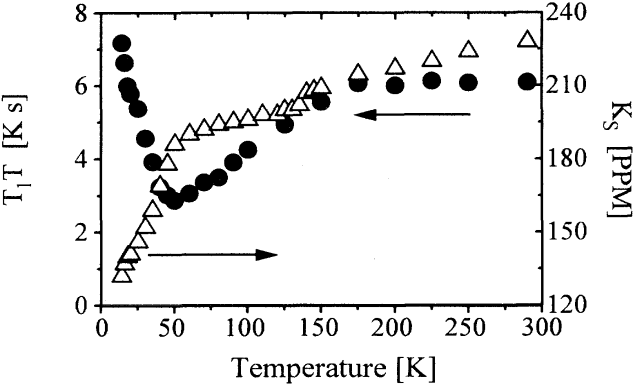}
\caption{NMR spectroscopy shows that there are strong spin fluctuations in \Br and other organic charge transfer salts. $1/T_1T$ shows a maximum at $\sim50$ K. This temperature corresponds with the temperature, $T_{coh}$, at which the crossover from the Fermi liquid to the bad metal is observed in the DC resistivity 
and the optical conductivity
. Indeed this correspondence between the maximum in $1/T_1T$ and $T_{coh}$ is found in a wide range of \ET salts \cite{PowellPRB09}.  The data above 50 K is well described  \cite{YusufPRB07} by a phenomenological spin fluctuation theory \cite{MoriyaAP00,MillisPRB90}. Below 50 K there is a sudden drop in both $1/T_1T$ and the Knight shift $K_s$, suggesting the a pseudogap opens \cite{PowellPRB09}.  [Modified from \cite{deSotoPRB95}.]
[Copyright (1995) by the American Physical Society].
}
\label{fig: deSoto}
\end{centering}
\end{figure}

There have been numerous studies of NMR in metallic organic charge transfer salts (for a review see \cite{KanodaCR04}). We begin by discussing investigations of the more weakly frustrated materials such as \Brn, \Cl and \NCS \cite{MayaffreEPL94,KawamotoPRL95,KawamotoPRB95,deSotoPRB95,ItayaPRL09}. These materials all show clear non-Korringa behaviours, cf. Fig. \ref{fig: deSoto}. As the temperature is lowered from room temperature, both $1/T_1T$ and $K_s$ rise to a maximum at a temperature we will denote as $T_{NMR}$. Below $T_{NMR}$ both $1/T_1T$ and $K_s$ decrease; both drop more rapidly below the superconducting critical temperature, $T_c$. For weakly frustrated compounds $T_{NMR}\simeq T_{coh}$ \cite{PowellPRB09,ItayaPRL09}, the coherence temperature marking the crossover from a Fermi liquid to a bad metal, for a range of anions  and pressures  to within experimental error. Measurements of the Korringa ratio \cite{deSotoPRB95,KawamotoPRB95,ItayaPRL09} find that $\mathcal{K}\gg1$ indicating that there are strong antiferromagnetic fluctuations.

For $T>T_{NMR}$ the experimental data is naturally explained \cite{YusufPRB07,PowellPRB09} by Moriya's self-consistent renormalised theory \cite{MoriyaAP00} in the phenomenological form used
by Millis, Monien and Pines in the context of the cuprates \cite{MillisPRB90}. In this model there are two contributions to the dynamic susceptibility, one arising from long wavelength background from Fermi liquid like excitations and a 
second contribution from spin fluctuations that is strongly peaked at some wavevector $\bf Q$ associated with the nascent magnetic order. In the limit of strong magnetic fluctuations this model predicts that  \cite{PowellPRB09}
\begin{eqnarray}
\frac{T_1T}{(T_1T)_{{NMR}}} =
\frac{T_{NMR}}{T_{NMR}+T_x}\left(\frac{T}{T_{NMR}}\right)
+ \frac{T_x}{T_{NMR}+T_x}, 
\label{eqn:scalingNMR}
\end{eqnarray}
where $(T_1T)_{{NMR}}$ is the value of $T_1T$ at $T=T_{NMR}$ and $T_x$ sets the scale for the temperature dependence of the spin correlation length. Plotting the experimental data as ${T_1T}/{(T_1T)_{NMR}}$ against $T/T_{NMR}$ indeed yields the straight line predicted above for $T>T_{NMR}$ \cite{PowellPRB09}. A more detailed analysis \cite{YusufPRB07} allows one to estimate the spin correlation length, $\xi(T)$. For example, $\xi(T_{NMR})\simeq3a$, where $a$ is the lattice constant, in \Brn. It has been shown \cite{DingPRL90} that, on the square lattice, the
antiferromagnetic Heisenberg model has a correlation length of order
$\xi(T)/a \sim 1$ for $T = J$ and of order $\xi(T)/a \sim
30$ for $T= 0.3 J$. 
On the other hand, for the antiferromagnetic Heisenberg model on the
isotropic triangular lattice, the correlation length is only of order
a lattice constant at $T= 0.3J$ \cite{ElstnerPRL93}.
Therefore a correlation length of $\sim3a$ is consistent with the intermediate value of $t'/t$ calculated for \Br (cf. section \ref{ET-struct}) placing this compound somewhere between the square and triangular lattices.

An important  question is: what causes the reduction in $1/T_1T$, $K_s$ and $\mathcal K$ for $T<T_{NMR}$?  Given its successes in describing many of the phenomena discussed in this section, one should first ask what DMFT predicts. 
It predicts a temperature dependence
that can be fit to the form (\ref{eqn:scalingNMR})
for all temperatures.
In the bad metal phase DMFT predicts that same behaviour as the spin fluctuation theory \cite{PruschkeAP95}. Therefore, for $T>T_{NMR}$, DMFT agrees with experiment qualitatively - although we are not aware of a specific quantitative comparison. However, DMFT also predicts
that $1/T_1$ increases monotonically with
temperature and in the Fermi liquid regime it
 predicts a constant Knight shift and a constant $1/T_1T$.
 None of these are seen experimentally. Therefore, although DMFT may provide an adequate description of the spin physics for $T>T_{NMR}$ some additional ingredient is required for $T<T_{NMR}$. As DMFT is a purely local (single site) theory this immediately suggests that some non-local correlations are important for understanding the spin correlations.

Two pictures have been proposed to try to explain the NMR below $T_{NMR}$: (i) the opening of a pseudogap \cite{YusufPRB07,PowellPRB09,MayaffreEPL94,MiyagawaPRL02} and (ii) a loss of spin correlations \cite{ItayaPRL09, deSotoPRB95,KawamotoPRB95,LefebvrePRL00}.

In the pseudogap scenario one assumes that non-local interactions cause a loss of spectral weight at the Fermi energy. This would lead to the suppression of both $1/T_1T$ and $K_s$, cf. Eqs. (\ref{eqn:KsNI}) and (\ref{eqn:T1TNI}).  
 In context of this hypothesis it is interesting to note that the fit of the data to Eq. \ref{eqn:scalingNMR} shows that $T_{NMR}\simeq T_x$, which suggests that the spin correlations play an important role in determining $T_{NMR}$. The interpretation of this result in this picture is then that the growing spin correlations cause a pseudogap to open as the temperature is lowered. 

If spin correlations were to decrease below $T_{NMR}$ this would clearly cause a reduction in $1/T_1T$. However, it is not clear that such a decay of spin correlations would also lead to a decrease in the Knight shift as $K_s$ is a measure of the ferromagnetic (${\bf q}=\bf0$) fluctuations, cf. Eq. \ref{eqn:Ks}. However, if a peak in the dynamic susceptibility at ${\bf q}\ne{\bf0}$ were sufficiently broad, antiferromagnetic spin correlations could contribute significantly to the Knight shift and lead to the observed behaviour. This picture also gives a natural explanation of why the Korringa ratio decreases below $T_{NMR}$: because $1/T_1T$ is more sensitive to antiferromagnetic spin fluctuations than $K_s$.

\subsubsection{There is no pseudogap in \CN}\label{pseudo-CN}
  
In contrast to the weakly frustrated materials there is no evidence for a pseudogap in \CN \cite{ShimizuPRB10}. In the metallic state 
Korringa-like behaviour is seen at low temperatures: both $1/T_1T$ and $K_s$ are constant. 

This is consistent with the finding from DCA calculations \cite{ImaiPRB02} that find no pseudogap on the isotropic triangular lattice. Yet when the frustration is reduced  a pseudogap caused by short range
antiferromagnetic correlations is found. Imai \etal find a pseudogap for $t'\lesssim0.6t$ from DCA calculations, using the non-crossing approximation (NCA) to solve the effective cluster problem. While one may have some concern over whether the accuracy of the NCA is sufficient for a quantitative comparison with experiment, this result seems to fit nicely with the experimental picture of no pseudogap in \CN and pseudogaps in \Cln, \Br and \NCS if one uses  the values of $t'/t$ calculated from DFT (Table \ref{tab:t-DFT}).   
  
\subsubsection{Other evidence for a pseudogap in the weakly frustrated materials}

Independent evidence for the suppression of
density of states at the Fermi level 
can come from the temperature dependence of the 
of electronic specific heat \cite{TimuskRPP99}. This
probes the density of excitations within $k_B T$ of the Fermi energy.
Any gap will suppress the density of states near the Fermi surface
which results in the depression of the specific heat coefficient
$\gamma$. Kanoda \cite{KanodaJPSJ06} compared $\gamma$ for several of
the \kX salts and found that in the region close to the Mott
transition, $\gamma$ is indeed reduced. One possible interpretation of
this behaviour is a pseudogap which becomes bigger as one approaches the
Mott transition. However, other interpretations are also possible, in
particular one needs to take care to account for the coexistence of
metallic and insulating phases; this is expected as the Mott transition
is first order in the organic charge transfer
salts \cite{KagawaN05,SasakiJPSJ05}. The existence of a pseudogap has also been
suggested in $\lambda$-(BEDT-TSF)$_2$GaCl$_4$ \cite{SuzukiJLTP06} from
microwave conductivity measurements. The reduction of the real part of the
conductivity $\sigma_1$ from the Drude conductivity
$\sigma_\mathrm{dc}$ and the steep upturn in the imaginary part of the
conductivity $\sigma_2$ have been interpreted in terms of preformed pairs
leading to a pseudogap in this material.

Scanning tunnelling microscopy (STM) has given important insights into the pseudogap phase of the high temperature superconductors \cite{FischerRMP07}. Therefore, it is natural to ask what can be seen via STM in the organics. This is complicated by the difficulty in obtaining high quality surfaces in the organics and these results should be treated with caution. However, Aria \etal \cite{AriaSSC00} did find evidence that at pseudogap opens below $T\sim45$ K in \NCSn. This temperature scale coincides  with $T_{NMR}$. Further the pseudogap is 
about five times larger than the superconducting gap. This is consistent with the observation that the pseudogap gap opens at a temperature about five times larger than the superconducting critical temperature. Further, the superconducting gap appears `on top' of the pseudogap. This `two gap' picture is similar to what is observed in the cuprates         \cite{BoyerN07,FischerRMP07}.

Clearly more work is required, from both a theoretical and experimental perspective, to resolve this issue.
The most obvious theoretical avenues are to study non local correlations in the \kX salts are the cluster extensions to DMFT such as CDMFT and the DCA. These include some off-site correlations, either in real (CDMFT) or reciprocal (DCA) space. However, there are significant technical challenges to overcome to accurately calculate the properties measured in NMR spectroscopies by these methods.

\subsubsection{Tests of the pseudogap hypothesis}

There are a number of key experiments needed to
resolve whether or not a pseudogap is present in the
paramagnetic metallic phase of \kXn. The pressure and magnetic field
dependences of the nuclear spin relaxation rate and Knight shift would
be valuable in determining the pseudogap phase boundary, estimating the
order of magnitude of the pseudogap, and addressing the issue how the
pseudogap is related to superconductivity. In the cuprates, there have
been several investigations of the magnetic field dependence of the
pseudogap seen in NMR experiments. For
Bi$_2$Sr$_{1.6}$La$_{0.4}$CuO$_6$ the nuclear spin relaxation rate does
not change with field up to 43 T \cite{ZhengPRL05}. However, since the pseudogap temperature $T^*
\sim 200$ K, one may require a larger field to reduce the pseudogap.
Similar results were found in YBa$_2$Cu$_4$O$_8$ \cite{ZhengPRL05}.
However, in YBa$_2$Cu$_3$O$_{7-\delta}$ [see especially Fig. 6 of 
\cite{MitrovicPRB02}] a field of order 10 T is enough to start to
close the pseudogap. Mitrovic \etal \cite{MitrovicPRB02} interpreted
this observation in terms of the suppression of `$d$-wave'
superconducting fluctuations.

The interlayer magnetoresistance of the cuprates has been used as 
a probe of the pseudogap.
\cite{MozorovPRL00,ShibauchiPRL01,KawakamiPRL05,ElbaumPRB04} Moreover, it has been argued that for
the field perpendicular to the layers (which means that the Zeeman effect
will dominate orbital magnetoresistance effects
due to the Lorentz force) the pseudogap is
closed at a field given by
\begin{equation}
H_{PG} \simeq \frac{k_B T^* }{\hbar \gamma_e},
\end{equation}
where  $\gamma_e$ is the gyromagnetic ratio of the electron. For the hole doped cuprates
this field is $\sim100$ T. In contrast, for the
electron-doped cuprates this field is of the order $\sim30$ T (and $T^*
\sim 30-40$ K), and so this is much more experimentally
accessible \cite{KawakamiPRL05}. The field and temperature dependence of the
interlayer resistance for several superconducting organic charge
transfer salts \cite{ZuoPRB99} is qualitatively similar to that for the
cuprates. In particular, for temperatures less than the zero-field
transition temperature and fields larger than the upper critical field,
negative magnetoresistance is observed for fields perpendicular to the
layers. A possible explanation is that, as in the cuprates, there is a
suppression of the density of states near the Fermi energy, and the
associated pseudogap decreases with increasing magnetic field.

Angle dependent magnetoresistance has proven to be a 
powerful probe of Fermi surface properties in the organic
charge transfer salts \cite{kartsovnik} and
more recently in the cuprates \cite{AbdelNP06,KennettPRB07}.
Recently, it has been shown that an anisotropic pseudogap
should produce distinct signatures in the interlayer magnetoresistance
when the magnetic field is rotated parallel to the layers \cite{SmithPRB09}.
This is a realistic and important experiment that should
be done on \Brn.

One could also study the pressure dependence of the linear coefficient
of heat capacity $\gamma$. Since $\gamma$ is proportional to the
density of states at the Fermi energy, a detailed mapping of
$\gamma(P)$ would be an important probe for the study the pseudogap.
Finally, measurements of the Hall effect have also led to important
insights into the pseudogap of the cuprates \cite{TimuskRPP99}, therefore,
perhaps, the time is ripe to revisit these experiments in the organic
charge transfer salts.

\subsubsection{The Nernst effect and vortex fluctuations above $T_c$}

\begin{figure}
\begin{centering}
\includegraphics[width=7cm]{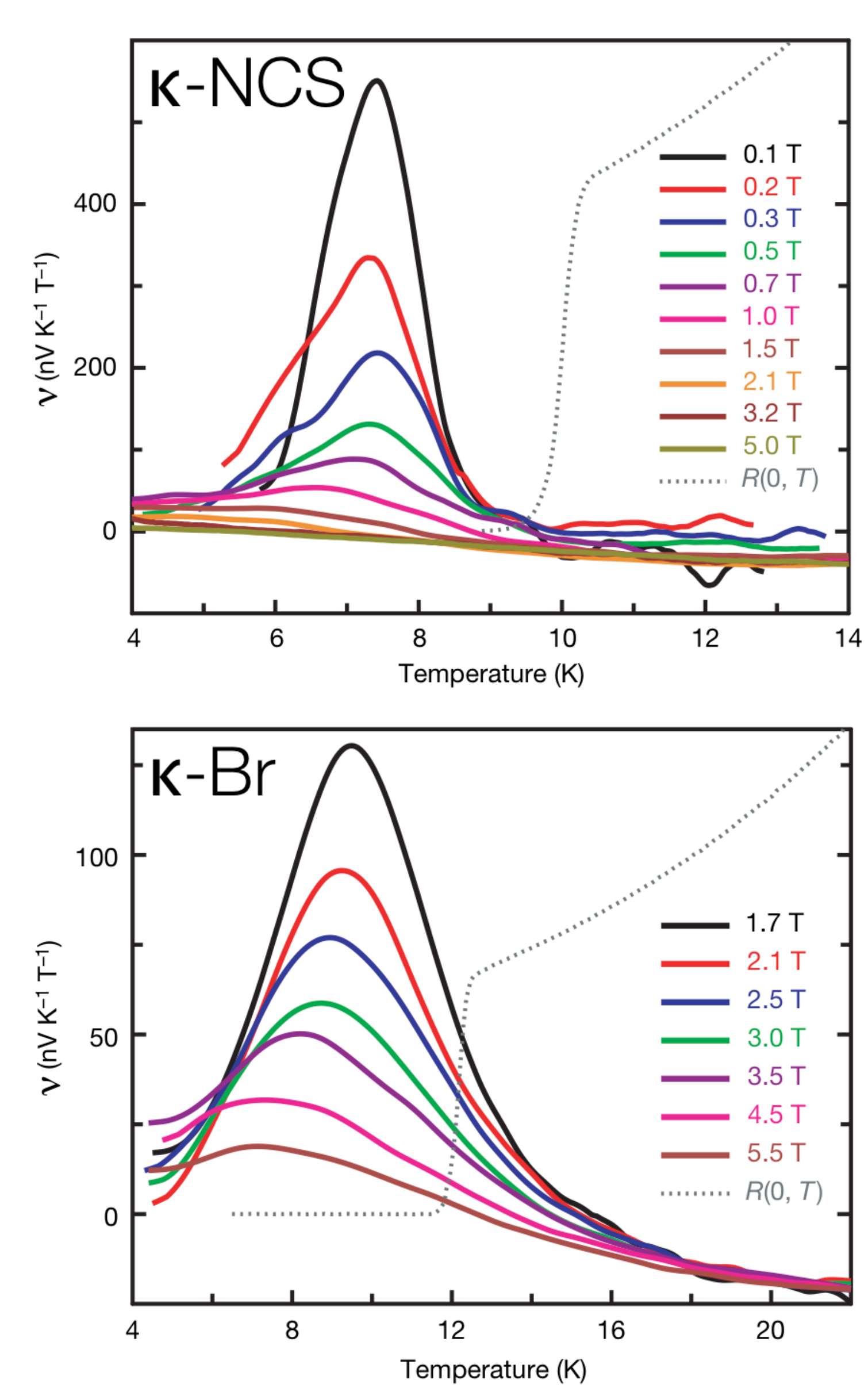}
\caption{Nernst coefficient, $\nu$, in \NCS (top) and \Br (bottom). In \NCS a large  Nernst coefficient is observed below the superconducting critical temperature, $T_c$, (for reference the temperature dependence of the zero field resistance is shown as a dotted line). Nam \etal \cite{NamN07} attributed this increase to the effect of vortices in the superconducting state, which are known to give a large positive contribution to $N$. In \Br the Nernst coefficient is large even above $T_c$, which Nam \etal interpreted as evidence of fluctuating superconductivity even above $T_c$. The contribution to the Nernst coefficient from quasiparticles can take either sign. Therefore, the negative Nernst coefficient at the highest temperatures in the lower panel presumably arises from quasiparticles.}
\label{fig:Nam}
\end{centering}
\end{figure}

A most interesting observation, which may be related to the pseudogap, is the large normal state Nernst effect in \Br \cite{NamN07}, shown in Fig. \ref{fig:Nam}. In \NCS the Nernst signal is of order the noise in the experiment for $T>T_c$. However, a large positive Nernst signal is observed just below $T_c$. It is extremely likely that this arises from the motion of superconducting vortices, which freeze out at lower temperatures as the vortex lattice forms. Nam \etal point out that  ``there is nothing unexpected in these observations.'' What was unexpected however, is that in \Br a Nernst signal is seen even for $T>T_c$. Nam \etal interpreted this as evidence of superconducting fluctuations that support vortices above $T_c$. 

A large normal state Nernst effect is also seen in the underdoped cuprates \cite{WangPRB06}. This has also often been interpreted as evidence for vortices above $T_c$. In part this was due to a misunderstanding of the ``Sondheimer cancellation''. Sondheimer \cite{SondheimerPRSA48} showed that for the dispersion characteristic of free fermions, $\epsilon_{\bf k}=\hbar^2|{\bf k}|^2/2m^*$, the normal state Nernst effect is small. This was often taken to be a general result and it was therefore assumed that a large Nernst signal was a definitive signal of vortices in the normal state. However, the Sondheimer cancellation turns out to be a special property of the free fermion dispersion relation \cite{BehniaJPCM09}.


   In the last few years it has become clear that there are several other effects that could give rise to large Nernst effects including an electronic nematic or Pomeranchuk phase \cite{Fradkin09,DaouN10,HacklPRB09}, stripes \cite{HacklPRB10}, a $d$-density wave \cite{KotetesPRL10} or even just the details of the band structure  \cite{BehniaJPCM09}. None of these effects have been yet been considered as possible explanations for Nam \etaln's results. Another interesting question, given that Nam \etal only see the normal state Nernst in \Brn, which is very close to the first-order Mott transition, is: would the coexistence of small amounts of the insulating phase with the metallic phase lead to an enhanced Nernst signal. 

An important consideration is that vortices can only give rise to a positive Nernst coefficient\footnote{In the convention employed by Nam \etal Note that two different sign conventions are used in the literature, which can be rather confusing. A clear discussion of this is given in \cite{BehniaJPCM09}.}. Yet, in the normal state of \Br Nam \etal report a \emph{negative} Nernst coefficient above $\sim15$ K. This seems to suggest that while the Nernst signal below $\sim15$ K may indeed be caused by vortices, the large normal state Nernst signal above $\sim15$ K arises from quasiparticles, which may give rise to a Nernst coefficient of either sign \cite{BehniaJPCM09}.

 An order of magnitude estimate of the quasiparticle contribution to the Nernst coefficient, $\nu$, can be made from \cite{BehniaJPCM09}
\begin{equation}
\frac{\nu}{T}=-\frac{\pi^2}{3}\frac{k_B}{e}\frac{\mu_c}{T_F},
\end{equation}
where $T_F=E_F/k_B$ is the Fermi temperature, and the carrier mobility, $\mu_c$, is given by
\begin{equation}
\mu_c=\frac{\tan\theta_H}B=\frac{e\tau}{m^*}=\frac{el}{\hbar k_F},
\end{equation}
where $\theta_H$ is the Hall angle, $\tau$ is the quasiparticle lifetime, and $l$ is the mean free path.
$k_F\sim\frac\pi{2a}$, where $a\sim 1$ nm is a lattice constant. As the temperature is raised towards the bad metal regime the $l\sim a$. Thus, $\mu_c\sim10^{-3}$ T$^{-1}$. $T_F\sim10^3$ K \cite{PowellPRB04}. Between 50 K and 20 K the resistivity decreases by an order of magnitude \cite{AnalytisPRL06} and hence the mean free path increases by an order of magnitude.  So, at 20 K, one expects that $\nu\sim10$ nV K$^{-1}$ T$^{-1}$. (One can also construct an estimate of this order of magnitude by extrapolating from the measured scattering rate \cite{PowellPRB04} in the $T\rightarrow0$ limit.) This is indeed the order of magnitude observed at $T\sim20$ K  in both \NCS and \Brn. 
Thus these two estimates suggest that the magnitude of the Nernst coefficient observed in the normal state of \NCS and \Br may be reasonable although given the multiband Fermi surface with both electron and hole sheets a more careful calculation is required to test this and to establish the sign of the quasiparticle contribution to the Nernst coefficient.


\subsection{The superconducting state}\label{ET-super}

\subsubsection{\CN}

Little is known experimentally about the effects of frustration on the superconducting state of the \kX salts. In particular, there have only been a very few studies of the superconducting state of \CNn. 
 However, this has not prevented significant interest in the effects of frustration on superconductivity from the theoretical community \cite{KondoJPSJ04,WatanabeJPSJ06,ClayPRL08,KyungPRL06,PowellPRL05,PowellPRL07,HuangPRB07,SahebsaraPRL06,GanPRB06,WrobelPRB07,LeePRL07,GalitskiPRL07}. 
Recently, Shimizu \etal \cite{ShimizuPRB10} have reported NMR experiments under pressure in the superconducting phase. They found that $1/T_1T\propto T^2$, which is consistent with line nodes on a three dimensional Fermi surface or point nodes on a two dimensional Fermi surface. Further, Shimizu \etal did not observe any signs of a Hebel-Slichter peak, which suggests that the pairing has a non-$s$-wave symmetry.  

Another, potentially important result is that Shimizu \etal only observed a very small reduction in the Knight shift of \CN below $T_c$. They suggested two possible explanations for this result. Firstly, it could be an experimental artefact due to radio frequency (rf) heating during their spin-echo experiments. Shimizu \etal were not able to rule this out as free induction decay experiments are complicated by the short $T_2^*$ in \CN and low power rf experiments were not sufficiently sensitive in the small crystals that are currently available. Therefore larger crystals are important to rule out this trivial explanation. However, the second, more interesting, explanation is that there is little change in $1/T_1T$ below $T_c$ because \CN is a triplet superconductor. There is a small drop in $K_s$ below $T_c$, which suggests that the pairing state is not purely equal spin pairing, like the A phase of $^3$He. However, many of the plethora of exotic phases that are available to $^3$He are ruled out \cite{PowellJPCM08} by the low symmetry of the \CN crystal. 

\begin{table}
\begin{center}
\begin{tabular}{c|cccccc|c}
$C_{6v}$ & $E$ & $C_2$ & $2C_3$ & $2C_6$ & $3\sigma_d$ & $3\sigma_v$ & states \\\hline
$A_1$ & 1 & 1 & 1 & 1 & 1 & 1 & $s$, $s_{x^2+y^2}$ \\
$A_2$ &  1 & 1 & 1 & 1 & -1 & -1 &  \\
$B_1$ & 1 & -1  & 1 & -1 & -1 & 1 &  \\
$B_2$ & 1 & -1 & 1 & -1 & 1 & -1 &  \\
$E_1$ & 2 & -2 & -1 & 1 & 0 & 0 &  \\
$E_2$ & 2 & 2 & -1 & -1 & 0 & 0 & $(d_{x^2-y^2},d_{xy})$ 
\end{tabular}
\caption{The character table of $C_{6v}$, which represents the point group
symmetry of the isotropic triangular lattice.}
\label{table:C6v}
\end{center}
\end{table}%

A wide range of theories that invoke a magnetic pairing mechanism give rise to $d_{x^2-y^2}$ pairing on the square lattice \cite{LeeRMP06,MonthouxN07}. More formally one should say that the superconducting order parameter transforms like the $B_1$ representation of $C_{4v}$, which is the point group symmetry of the square lattice. 
The isotopic triangular lattice (i.e., $t'=t$) has $C_{6v}$ symmetry, cf. Table \ref{table:C6v}. A $d_{x^2-y^2}$ order parameter would belong to the $E_2$ representation of $C_{6v}$. This is interesting because $E_2$ is a two-dimensional representation, which means that one naturally expects a two component order parameter, $(\eta_1,\eta_2)$, for which the Ginsburg-Landau free energy would be \cite{SigristRMP91,AnnettAP90}
\begin{eqnarray}
F_{e_2}&=&F_n+\alpha(T-T_c)(|\eta_1|^2+|\eta_2|^2) +
\beta_1(|\eta_1|^2+|\eta_2|^2)^2 \notag\\&& +
\beta_2(\eta_1^*\eta_2-\eta_1\eta_2^*)^2, \label{eq:GL}
\end{eqnarray}
where $F_n$ is the free energy of the normal state and $\alpha$, $\beta_1$ and $\beta_2$ are the parameters of the theory, which need to be determined from experiment or derived from a microscopic theory.

Eq. (\ref{eq:GL}) has three solutions: (i)
$\vec{\eta}=(1,0)$ or (ii) $\vec{\eta}=(0,1)$ for $\beta_2>0$ (the
degeneracy is lifted by sixth order terms
\cite{SigristRMP91,AnnettAP90}); (iii) $\vec{\eta}=(1,i)$ for
$\beta_2<0$. 
The two components of the order parameter can be associated with, say, the $d_{x^2-y^2}$ and the $d_{xy}$ pairing channels, which gives the physical interpretation of the theory. Solution (i) corresponds to $d_{x^2-y^2}$ pairing, solution (ii) corresponds to  $d_{xy}$ pairing, and solution (iii) corresponds to  $d_{x^2-y^2}+id_{xy}$ pairing, which we will refer to as the $d+id$ state. The $d+id$ state is therefore predicted for a large fraction of the possible parameter values in the theory, including the weak coupling solution \cite{SigristRMP91,AnnettAP90,PowellJPCM06b}. The $d+id$ state is also found in microscopic calculations for the, strong coupling, resonating valence bond (RVB) theory on the isotropic triangular lattice \cite{PowellPRL07}. In principle the broken time reversal symmetry of the $d+id$ state should be directly detectable via muon spin relaxation experiments \cite{SigristRMP91}. However, such experiments are yet to be performed on the superconducting state of \CNn.

\begin{figure}
\begin{centering}
\includegraphics[width=6cm]{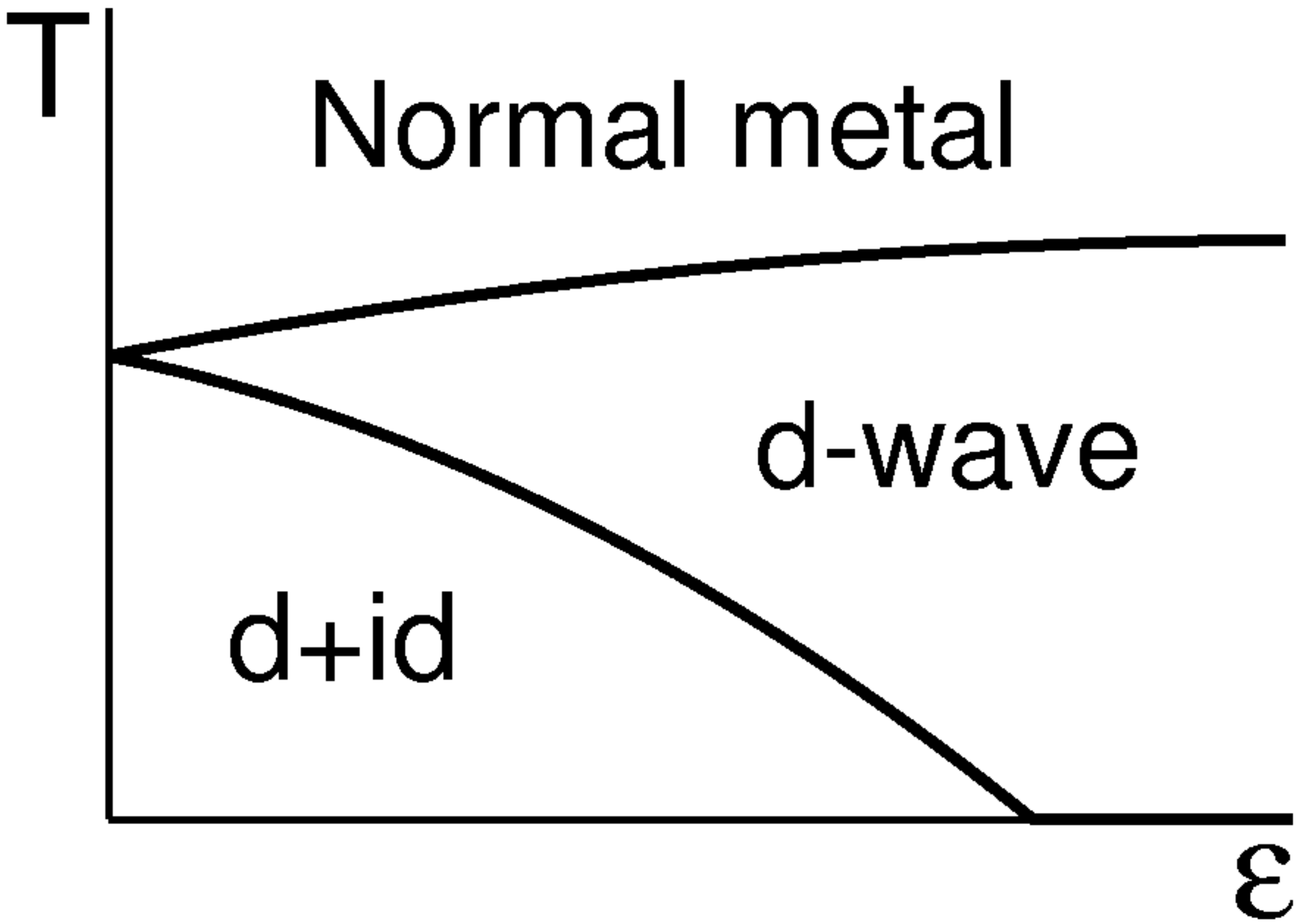}
\caption{Breaking the symmetry of the isotropic triangular lattice destroys the $d+id$ superconducting state in favour of a pure d-wave state. For weak symmetry breaking a double superconducting transition will occur. The symmetry breaking parameter $\epsilon \sim 1-t'/t$. From \cite{PowellJPCM06b}.
}
\label{fig:splitters}
\end{centering}
\end{figure}

However, \CN crystals actually have $C_{2h}$ symmetry rather than the $C_{6v}$ symmetry of the isotropic triangular lattice. Similarly, the anisotropic triangular lattice ($t'\ne t$) has $C_{2v}$ symmetry. It can be seen from Table \ref{tab:C2v} that for both $C_{2h}$ and $C_{2v}$ $d_{x^2-y^2}$ and $d_{xy}$ order parameters belong to different one dimensional representations. Thus, one does not expect a two-component order parameter generically. A simple way to understand what will happen near the isotropic case is to introduce a symmetry breaking perturbation into Eq. (\ref{eq:GL}) \cite{PowellJPCM06b}. This perturbation lifts the degeneracy and results in a double superconducting transition, see Fig. \ref{fig:splitters}. Physically such a perturbation corresponds to varying $t'/t$ away from unity, but because of its $C_{2h}$ crystal symmetry one always excepts this perturbation to always be present in \CNn. Therefore, if the superconducting transition of \CN breaks time reversal symmetry, this will be signified by a double superconducting transition, which would be visible to any number of thermodynamic probes. However, to date, no suitable experiments have been performed, presumably this is due, at least in part, to the difficulty in performing many of these measurements under pressure.

\begin{table}
\begin{center}
\begin{tabular}{c|cccc|c}
$C_{2v}$ & $E$ & $C_2$ & $\sigma_v$ & $\sigma_v'$ & states \\\hline
$A_1$ & 1 & 1 & 1 & 1 & $s$, $d_{xy}$ \\
$A_2$ &  1 & 1 & -1 & -1 & $d_{x^2-y^2}$ \\
$B_1$ & 1 & -1  & 1 & -1 &  \\
$B_2$ & 1 & -1 & -1 & -1 &  
\end{tabular} \\\vspace{10pt}
\begin{tabular}{c|cccc|c}
$C_{2h}$ & $E$ & $C_2$ & $\sigma_h$ & $i$ & states \\\hline
$A_g$ & 1 & 1 & 1 & 1 & $s$, $d_{xy}$ \\
$A_u$ &  1 & 1 & -1 & -1 &  \\
$B_g$ & 1 & -1  & -1 & 1 & $d_{x^2-y^2}$ \\
$B_u$ & 1 & -1 & -1 & 1 &  
\end{tabular}  
\end{center}
\caption{The character tables of $C_{2v}$ and $C_{2h}$. The anisotropic triangular lattice has $C_{2v}$ symmetry for $t'\ne t$ (cf. Fig. \ref{fig:ET-struct}). However, the point group symmetry of \CN is $C_{2h}$. Note that for the $C_{2v}$ point group we use the coordinate system defined in Fig. \ref{fig:ET-struct}c, in which the $C_2$ axis is along the $x+y$ direction, thus the indicated transformation properties of coordinate system are different from those found in many textbooks e.g. \cite{Tinkham92,Lax74}.}
\label{tab:C2v}
\end{table}%

%

\subsubsection{Weakly frustrated materials}

We have given an extended review of the superconducting states of the more weakly frustrated materials, such as \Br and \NCSn, somewhat recently \cite{PowellJPCM06}. We will not repeat that discussion here and will limit ourselves to highlighting the main issues and discuss some of the more recent results.

One key issue, that remains controversial, is the pairing symmetry. There is clear evidence from the suppression of the Knight shift below $T_c$ that the weakly frustrated \kX salts are singlet superconductors \cite{PowellJPCM06b}. However, no Hebel-Slichter peak is seen in $1/T_1T$ \cite{KawamotoPRB95,deSotoPRB95}, which suggests that the pairing state is not $s$-wave. Further, thermodynamic measurements down to the lowest temperatures \cite{TaylorPRL07} suggest that there are nodes in the gap. Given the low symmetry of crystals of organic charge transfers salts, this evidence suggests that a $d_{x^2-y^2}$-wave state is realised in these materials \cite{PowellJPCM06b}. This is also natural on theoretical grounds given the proximity to antiferromagnetic order in the more weakly frustrated compounds.  

\begin{figure}
\begin{centering}
\includegraphics[width=7cm]{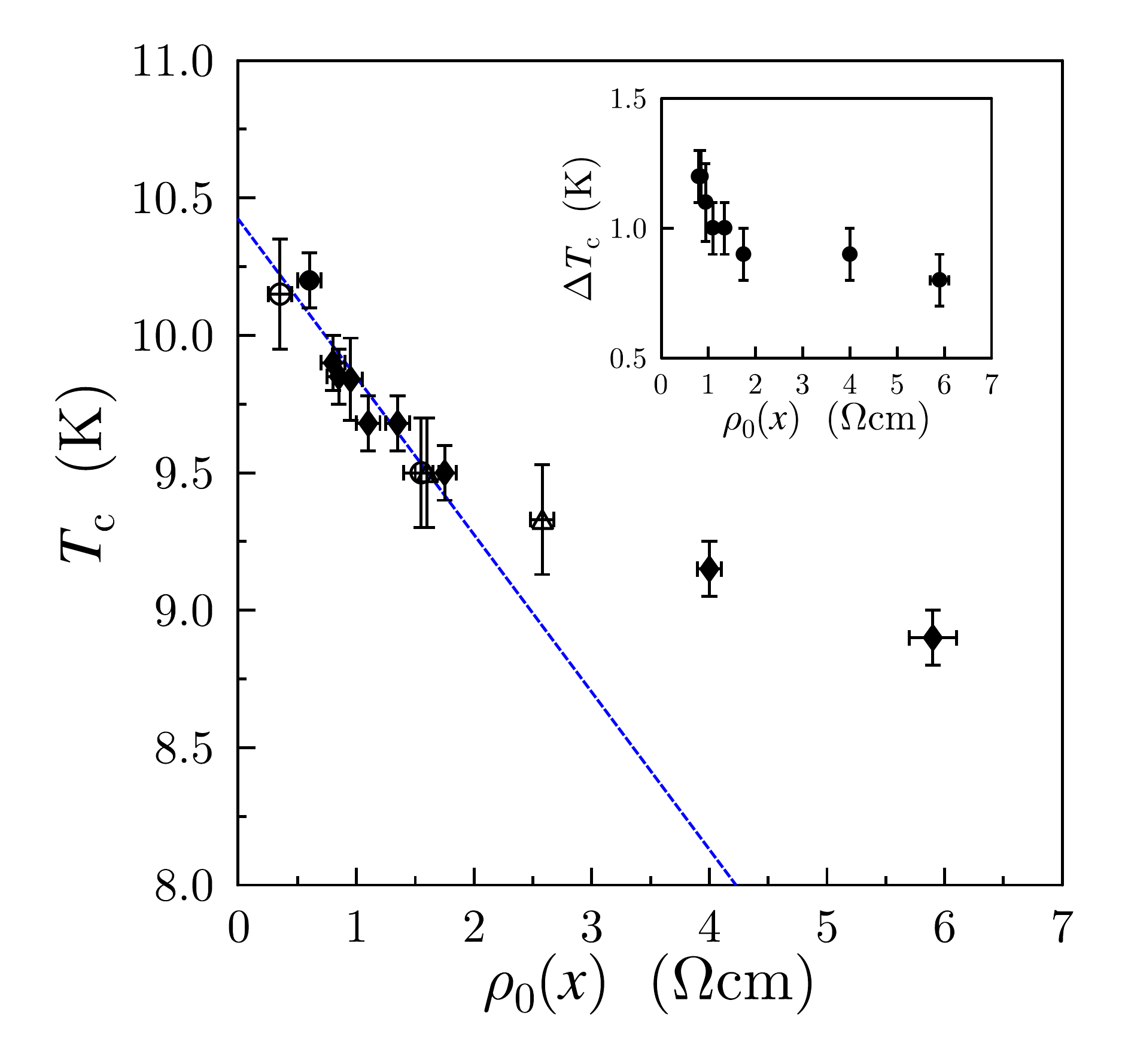}
\caption{Suppression of superconductivity by non-magnetic impurities. At low concentrations of impurities the data is well described by Eq. (\ref{eqn:AGlinear}) with a reasonable value of $t_\perp$ (line). But, for higher impurity concentrations strong deviations from the predictions of the Abrikosov-Gorkov formula, Eq.  (\ref{eqn:AG}) are observed.  From \cite{AnalytisPRL06}; impurities were introduced by x-ray and proton irradiation.
[Copyright (2006) by the American Physical Society].
}
\label{fig:Analytis}
\end{centering}
\end{figure}

In an unconventional superconductor (i.e., any superconductor in which the order parameter does not transform like the trivial representation) non-magnetic disorder suppresses the superconducting critical temperature in accordance with the Abrikosov-Gorkov forumla \cite{LarkinJETP65,Mineev99}:
\begin{eqnarray}
\ln \left(\frac{T_{c0}}{T_{c}} \right) = \psi\left( \frac{1}{2} +
\frac{\hbar}{4\pi k_BT_{c}}\frac{1}{\tau} \right) - \psi\left(
\frac{1}{2} \right), \label{eqn:AG}
\end{eqnarray}
where $T_{c0}$ is the critical temperature of the clean system, $1/\tau$ is the rate at which electrons scatter from impurities and $\psi(x)$ is the digamma function. Combining this result with the Fermi liquid expression for the interlayer conductivity it can be shown \cite{PowellPRB04} that, to leading order in $1/\tau$, the suppression in $T_c$ is given by 
\begin{eqnarray}
T_c = T_{c0} - \frac{e^2m^*ct_\perp^2}{4k_B\hbar^3} \rho_0,
 \label{eqn:AGlinear}
\end{eqnarray}
where $m^*$ is the effective mass, $t_\perp$ is the interlayer hopping amplitude and $\rho_0$ is the interlayer residual resistivity. 
For low impurity concentrations this linear behaviour is indeed observed in the \kET superconductors \cite{PowellPRB04,AnalytisPRL06}. Furthermore, the value of $t_\perp$ found from a fit of Eq. (\ref{eqn:AGlinear}) to this data yields excellent agreement with estimates of $t_\perp$ from other experimental techniques, such as angle-dependent magnetoresistance and quantum oscillations \cite{PowellPRB04,AnalytisPRL06}.   However,  for higher disorder concentrations the data does \emph{not} follow Eq. (\ref{eqn:AG}) \cite{AnalytisPRL06,SasakiA10}, cf. Fig \ref{fig:Analytis}. Until this deviation from the prediction of Eq. \ref{eqn:AG} is understood the question of the pairing symmetry cannot be considered to have been resolved.  

Another puzzle about the superconducting state is that, at low temperatures, the in-plane penetration depth, $\lambda$ has been found to vary as $\lambda\sim T^{3/2}$. As the penetration depth is proportional to the density of states with $\sim k_BT$ of the Fermi energy, one expects that  $\lambda\sim \exp{(-\Delta/k_BT)}$ for a fully gapped superconductor, $\lambda\sim T^2$ for a 3D superconductor with point nodes and $\lambda\sim T$ for a 3D superconductor with line nodes or a 2D superconductor with point nodes \cite{Annett90}. The observation of an intermediate power law has not yet received an adequate explanation.

Another interesting issue is the zero temperature superfluid stiffness, $\rho_s(0)\propto1/\lambda^2$, where $\lambda$ is the zero temperature penetration depth. In the underdoped cuprates it is found that $\rho_s(0)\propto T_c$, which is known as the Uemura relationship. A number of explanations have be advanced to explain this, but most boil down to the idea that underdoped cuprates become normal due to a loss of phase coherence as the temperature is raised \cite{EmeryN95}. This should be contrasted with the BCS theory where superconductors become normal at finite temperatures due to the suppression of pairing by the
entropy associated with quasi-particle excitations. Pratt \etal \cite{PrattP03,PrattPRL05} have found that in a wide range of \ET salts $T_c\propto1/\lambda^3\propto\rho_s(0)^{3/2}$. Furthermore, their results disagree, by orders of magnitude, with the prediction of both the BCS theory and Emery and Kivelson's theory of phase fluctuations \cite{PowellJPCM04}, which gives a good description of the phenomena observed in the cuprates. No theoretical explanation of Pratt \etaln's results has been given yet. Doing so remains a major challenge to theory and a major test for any proposed microscopic theory of superconductivity in these materials.

%% file: dmit.tex
\section{$\beta'$-$Z\textrm{[Pd(dmit)}_2 \textrm{]}_2$}
\label{sec:dmit}

\begin{figure}
\begin{centering}
\includegraphics[width=7cm]{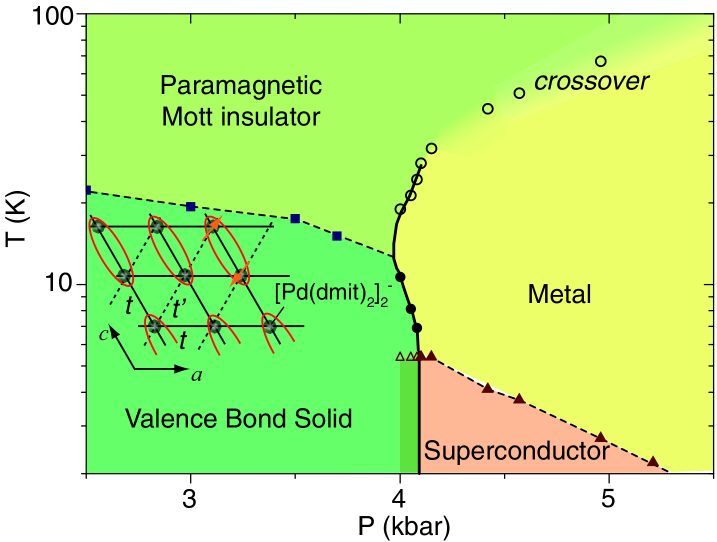}
\caption{Pressure-temperature phase diagram of EtMe$_3$P[Pd(dmit)$_2$]$_2$ (P-1). This is remarkably similar to the phase diagrams of \Cl (Fig. \ref{fig:phase-diagram-kCl}) and \CN (Fig. \ref{fig:phase-diagram-kCN}). However, the Mott insulator phase of P-1 shows valence bond crystalline order unlike the antiferromagnetism observed in \Cl and the spin liquid seen in \CNn. From \cite{ShimizuPRL07}.}
\label{fig:P-1}
\end{centering}
\end{figure}

\begin{figure}
\begin{centering}
\includegraphics[width=7cm]{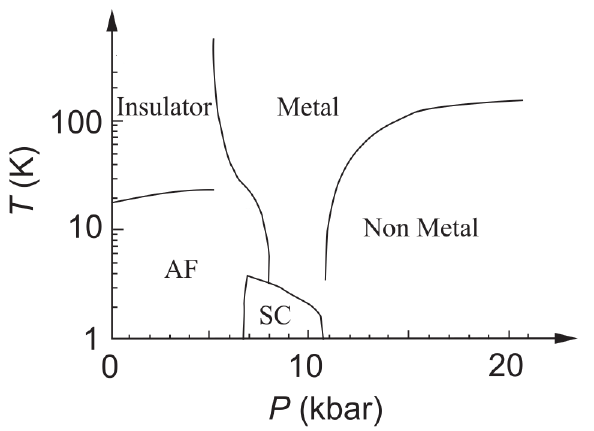}
\caption{Pressure-temperature phase diagram of Et$_2$Me$_2$P[Pd(dmit)$_2$]$_2$ (P-2). This has strong similarities to the phase diagrams of \Cl (Fig. \ref{fig:phase-diagram-kCl}), \CN (Fig. \ref{fig:phase-diagram-kCN}) and P-1 (Fig. \ref{fig:P-1}). An important difference is that P-2 shows long range antiferromagnetism, like \Cln. The  high pressure
non-metal  phase is associated with a change in crystal structure \cite{YamauraJPSJ04}.  Modified from \cite{YamauraJPSJ04}.}
\label{fig:P-2}
\end{centering}
\end{figure}

Organic charge transfer salts based on the \dmit molecule, shown in Fig. \ref{fig:molecules}b, are less well known, and have been less widely studied, than the \kX materials discussed above (section \ref{ET}). However, the salts of Pd(dmit)$_2$ show a fascinating range of behaviours, which we review in this section. We will see that the Pd(dmit)$_2$ salts have much in common with the \kETX salts, as is apparent from their similar phase diagrams (compare Figs. \ref{fig:P-1} and \ref{fig:P-2} with Figs. \ref{fig:phase-diagram-kCl} and \ref{fig:phase-diagram-kCN})

\begin{figure}
\begin{centering}
\includegraphics[width=8cm]{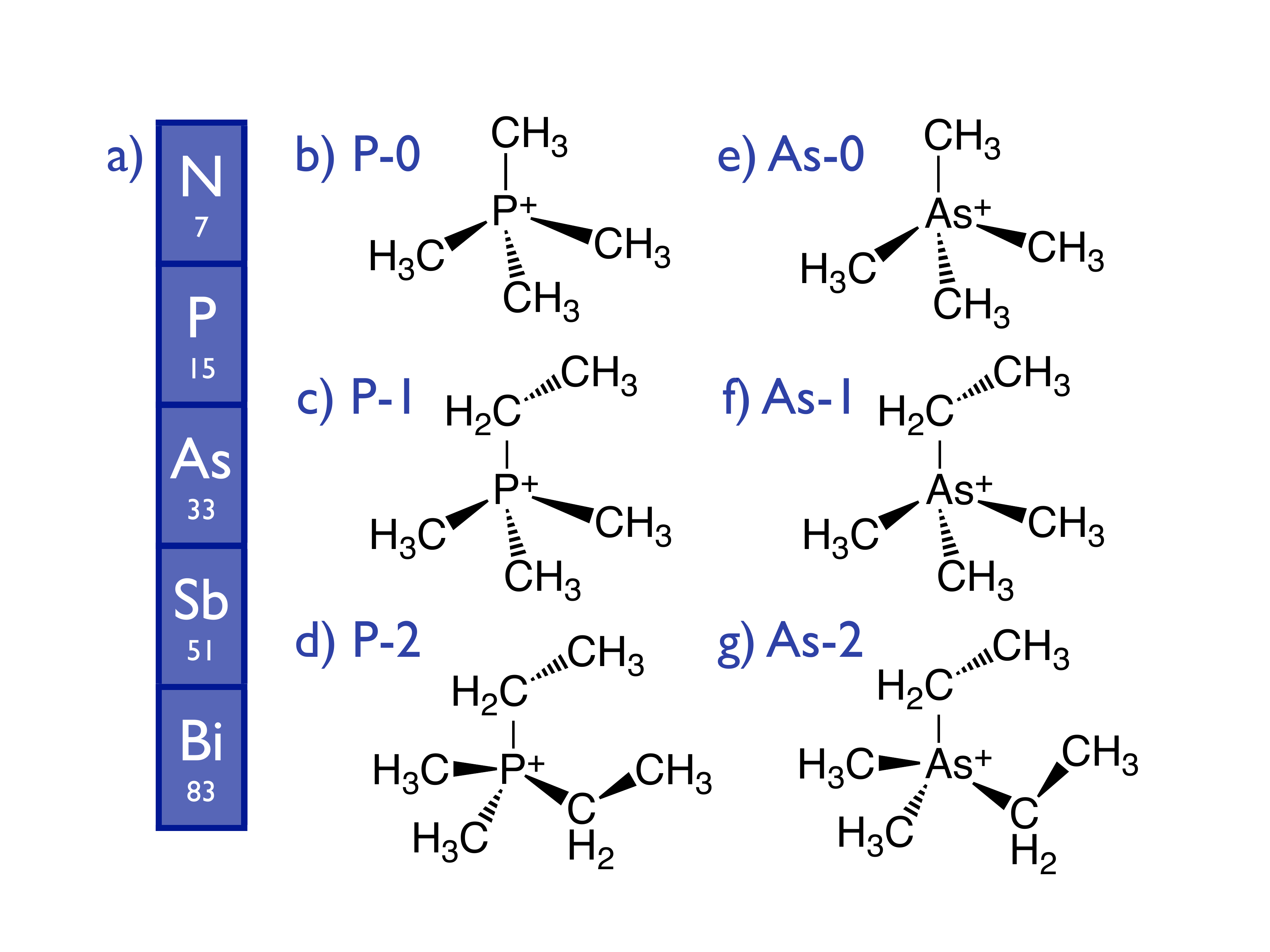}
\caption{Typical cations in charge transfer salts with Pd(dmit)$_2$ (shown in Fig. \ref{fig:molecules}b) are shown in panels (b)-(g). These cations share the $Pn$Me$_{4-n}$Et$_n$ motif, where $Pn$ is a group V element (pnictogen), panel (a), Me is a methyl group (CH$_3$) and Et is an ethyl group (C$_2$H$_5$). We use the shorthand $Pn$-$n$ for the salt $Pn$Me$_{4-n}$Et$_n$[Pd(dmit)$_2$]$_2$}
\label{fig:cations}
\end{centering}
\end{figure}

In this section we will discuss many materials with rather similar chemical formulae. To simplify our discussion we introduce the following nomenclature: Et$_n$Me$_{4-n}Pn$[Pd(dmit)$_2$]$_2$ will be written as $Pn$-$n$, where $Pn$ is a pnictogen, Et is the ethyl group, C$_2$H$_5$, and Me is the methyl group, CH$_3$, cf. Fig. \ref{fig:cations}.

\subsection{Crystal and electronic structure}\label{dmit-struct}

The \dmit molecule is shown in Fig \ref{fig:molecules}b. A number of other $M$(dmit)$_2$ molecules, where $M$ is a transition metal, can also form charge transfer salts.  An interesting example is Ni(dmit)$_2$, whose salts have quasi-one-dimensional properties; in contrast to the quasi-two-dimensional behaviour found in salts of \dmitn. These differences arise because of subtle changes in the molecular orbitals of the dimers of these two molecules as we will discuss below.

\begin{figure}
\begin{centering}
\includegraphics[width=7cm]{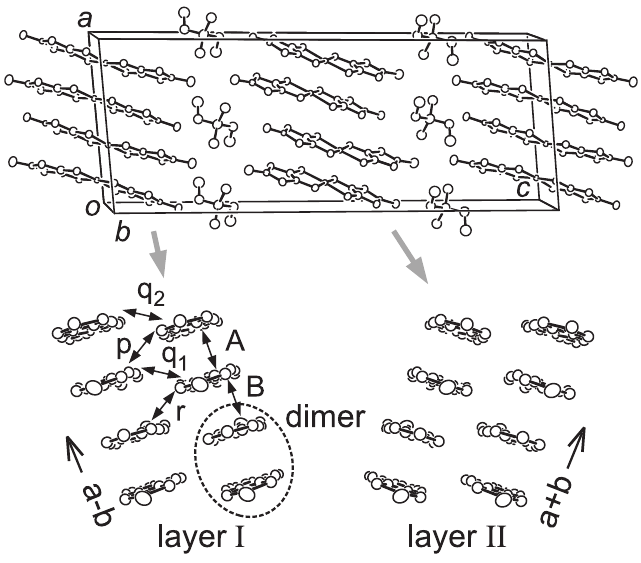}
\caption{The crystal structure of $\beta'$-Et$_2$Me$_2$P[Pd(dmit)$_2$]$_2$ (P-2). Note that there are two organic layers per unit cell. In the first layer the molecules stack along the ${\bf a}-{\bf b}$ direction, whereas in the second layer the molecular stacks are orientated along the ${\bf a}+{\bf b}$ direction. From \cite{YamauraJPSJ04}.}
\label{fig:crystal-bp}
\end{centering}
\end{figure}

Most of the crystals that we will discuss below take the so-called $\beta^\prime$ phase, shown in Fig. \ref{fig:crystal-bp}. An important feature of this structural motif is that the \dmit molecules are arranged in dimers. Electronic structure calculations show that the amplitude for hopping between two monomers within a dimer is much larger than the amplitude for hopping between two monomers in different dimers \cite{MiyazakiPRB99,CanadellCCR99}. This arises not only from the greater proximity of the two monomers in a dimer, but also because of
the shape of the relevant molecular orbitals, which have a large contribution from the $\pi$ orbitals. Thus, the face-to-face stacking within a dimer leads to a large overlap.
In Section \ref{ET-struct} we discussed a simple model for the electronic structure of a BEDT-TTF dimer. This model is based on a single molecular orbital (the HOMO) on each BEDT-TTF molecule. We will now discuss a similar model for the electronic structure of [$M$(dmit)$_2$]$_2$. However, the model for $M$(dmit)$_2$ differs from that for \ET in several important ways because more than one molecular orbital on each $M$(dmit)$_2$ molecule is involved. Our discussion is  based on extended H\"uckel calculations \cite{CanadellCCR99}, although we also note that  DFT calculations  \cite{MiyazakiPRB99} give the same qualitative picture.

%

\begin{figure}
\begin{centering}
\includegraphics[width=7cm]{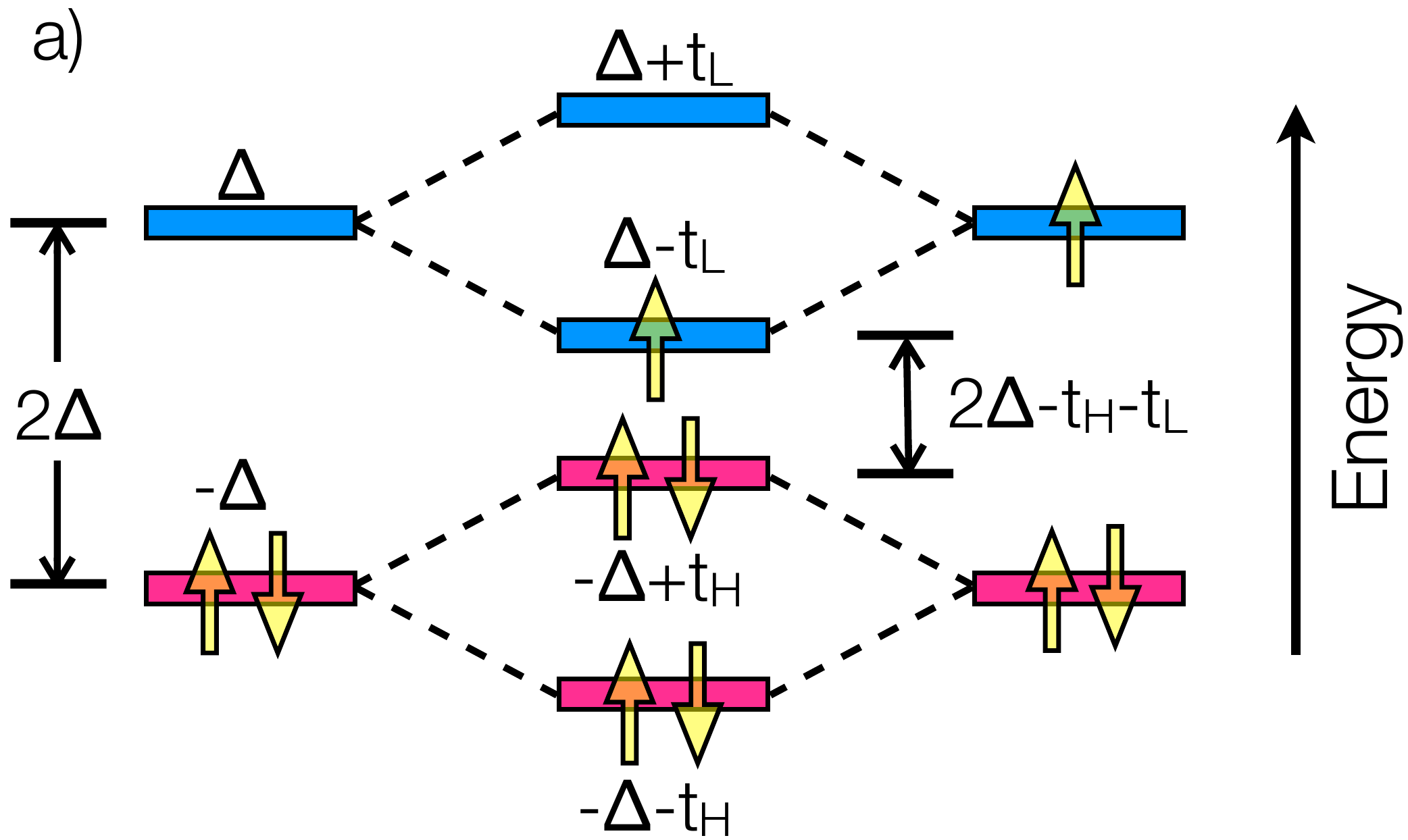}\\\vspace{1cm}
\includegraphics[width=7cm]{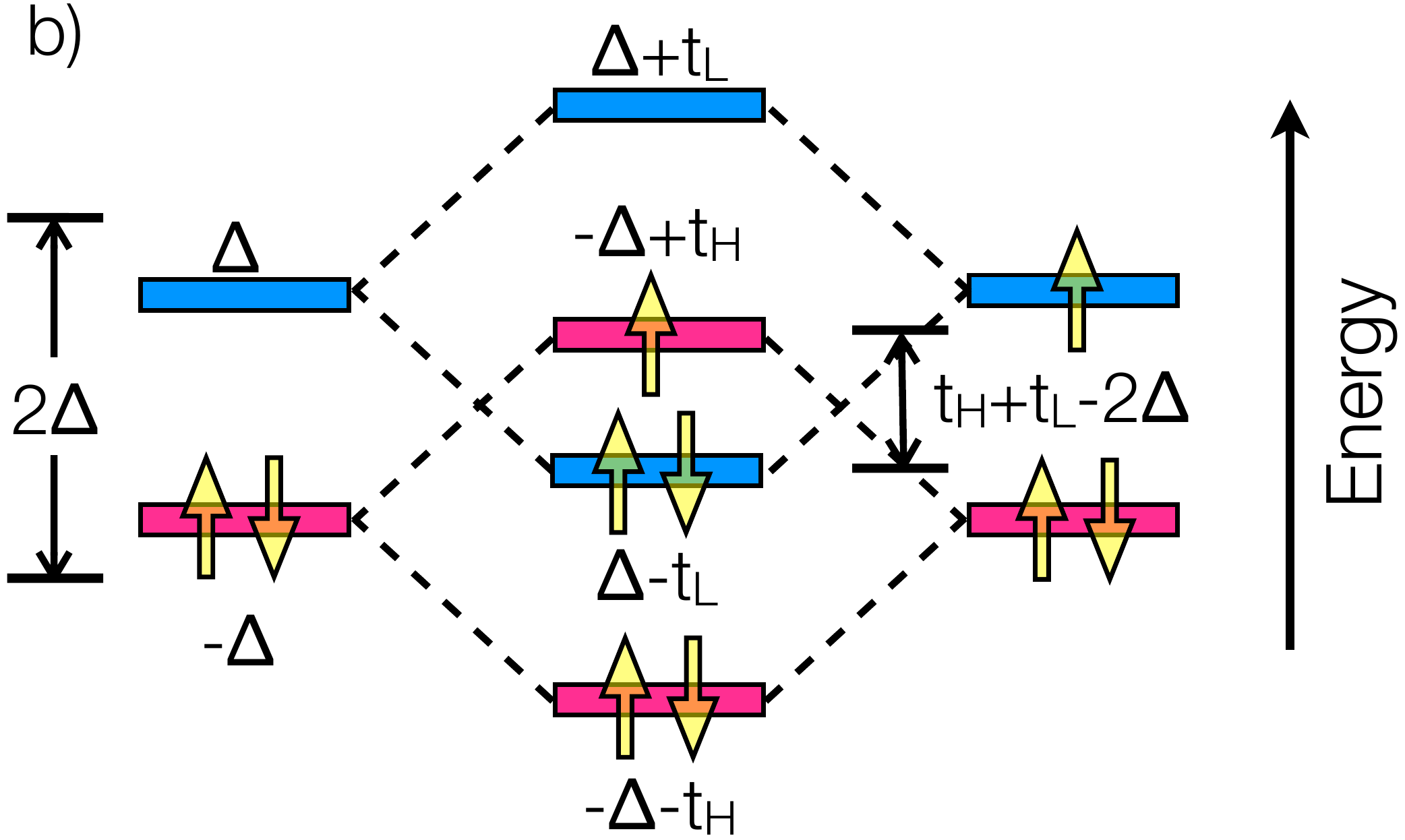}
\caption{Sketch of the solution of the two site--two orbital model for a \dmittwo dimer [Eq. (\ref{eqn:2site2orbital})]. In salts formed with the monovalent cations, which we discuss here, the dimer has, on average, five electrons. (a) For $2\Delta>t_H+t_L$ (weak dimerisation) both the bonding and antibonding combinations of the molecular HOMOs are fully occupied, and the bonding combination of the molecular LUMOs contains one electron. Therefore, in the crystal, the half-filled metallic band will result primarily from the hybridisation of these LUMO antibonding orbitals. This case is believed to be relevant to the salts of Ni(dmit)$_2$ \cite{MiyazakiPRB99}. The shape of the molecular orbitals gives rise to a quasi-one-dimensional band structure in these compounds. (b) For $2\Delta<t_H+t_L$ (strong dimerisation) the antibonding combination of the molecular HOMOs is pushed above the bonding combination of molecular LUMOs. Thus, the half-filled metallic band in the crystal will be formed from predominately from these HOMO antibonding orbitals. This is believed to be the case relevant to the salts of Pd(dmit)$_2$. The structure of the molecular HOMOs gives rise to a quasi-two-dimensional band structure with the topology of the anisotropic triangular lattice, cf. Fig. \ref{fig:crystal-bp}. In order for these single band descriptions to be relevant to the real materials $|2\Delta-t_H-t_L|$ must be larger than the other energy scales relevant to the problem. If this is not the case multiband effects may have important consequences.}
\label{fig:dmit-dimer-non-int}
\end{centering}
\end{figure}

 $M$(dmit)$_2$ is an electron
acceptor molecule   
and    on average the dimer has a net charge of -1 in the crystal. Thus na\"ively one might expect that the extra electron resides in the bonding combination of monomer LUMOs. However, electronic structure calculations find that there is significant hybridisation between the two monomer HOMOs, which complicates this picture. 
A simple, non-interacting model for the dimer is
\begin{eqnarray}
\hat{\cal H}_{[M({dmit})_2]_2} &=& \Delta\sum_{i=1}^2\sum_\sigma \left( \hat L_{i\sigma}^\dagger \hat L_{i\sigma} - \hat H_{i\sigma}^\dagger \hat H_{i\sigma} \right) \nonumber \\&& 
- t_H \sum_\sigma \left(  \hat H_{1\sigma}^\dagger \hat H_{2\sigma} + \hat H_{2\sigma}^\dagger \hat H_{1\sigma}  \right)
 \nonumber \\&& 
- t_L \sum_\sigma \left(  \hat L_{1\sigma}^\dagger \hat L_{2\sigma} + \hat L_{2\sigma}^\dagger \hat L_{1\sigma}  \right),\label{eqn:2site2orbital}
\end{eqnarray}
where $\hat H_{i\sigma}$ creates an electron with spin $\sigma$ in the HOMO of the $i^{th}$ monomer and $\hat L_{i\sigma}$ creates an electron with spin $\sigma$ in the LUMO of the $i^{th}$ monomer. The solution of this model is trivial; and sketched in Fig. \ref{fig:dmit-dimer-non-int}. It is clear that with five electrons, as is approporate for [$M$(dmit)$_2$]$_2^-$, one of the energy levels will be partially occupied. However, which state this is is dependent on the ratio $(t_H+t_L)/\Delta$. There are two regimes: (a) strong dimerisation ($t_H+t_L>2\Delta$): the antibonding combination of monomer HOMOs contains one electron, while the  bonding combination of monomer LUMOs contains two electrons (Fig. \ref{fig:dmit-dimer-non-int}a); and (b) weak dimerisation ($t_H+t_L<2\Delta$): the antibonding combination of monomer HOMOs contains two electrons, while the  bonding combination of monomer LUMOs contains one electron  (Fig. \ref{fig:dmit-dimer-non-int}b).  In either, case a one band description will only be justified if $|t_H+t_L-2\Delta|$ is sufficiently large compared to the other energy scales relevant to the problem. 

Of course, once other orbitals, electron-electron interactions and the weak (almost symmetry forbidden) hybridisation between the HOMO on one monomer and the LUMOs on the other are included the situation becomes significantly more complicated. However, DFT calculations \cite{MiyazakiPRB99} suggest that 
the cartoon sketched above does capture many of the important physical features of both the [Pd(dmit)$_2$]$_2$ and the [Ni(dmit)$_2$]$_2$ dimers. Furthermore, these calculations suggest that [Pd(dmit)$_2$]$_2$ corresponds to case (a) (strong dimerisation)  whereas [Ni(dmit)$_2$]$_2$ corresponds to case (b) (weak dimerisation). Hence the metallic band in salts of [Pd(dmit)$_2$]$_2$ results primarily from the antibonding combination of monomer HOMOs whereas  the metallic band in salts of [Pd(dmit)$_2$]$_2$ results primarily from the bonding combination of monomer LUMOs. This may sound like a trivial detail but it has important consequences for the physics of these salts. In particular, the bands formed (predominately) from the bonding combination of LUMOs in $X$[Ni(dmit)$_2$]$_2$ are quasi-one-dimensional, whereas the bands formed (predominately) from the antibonding combination of HOMOs in $X$[Pd(dmit)$_2$]$_2$ are quasi-two-dimensional. Thus, the profound differences in the observed behaviour of the salts of Ni(dmit)$_2$ and Pd(dmit)$_2$ results from a rather small structural change (the degree of dimerisation). The differences between in the band structures of the Ni and Pd compounds have been nicely illustrated by comparative DFT calculations for Me$_4$N[Pd(dmit)$_2$] and  Me$_4$N[Ni(dmit)$_2$] \cite{MiyazakiPRB99}.

\begin{figure*}
\begin{centering}
\includegraphics[width=14cm,angle=0]{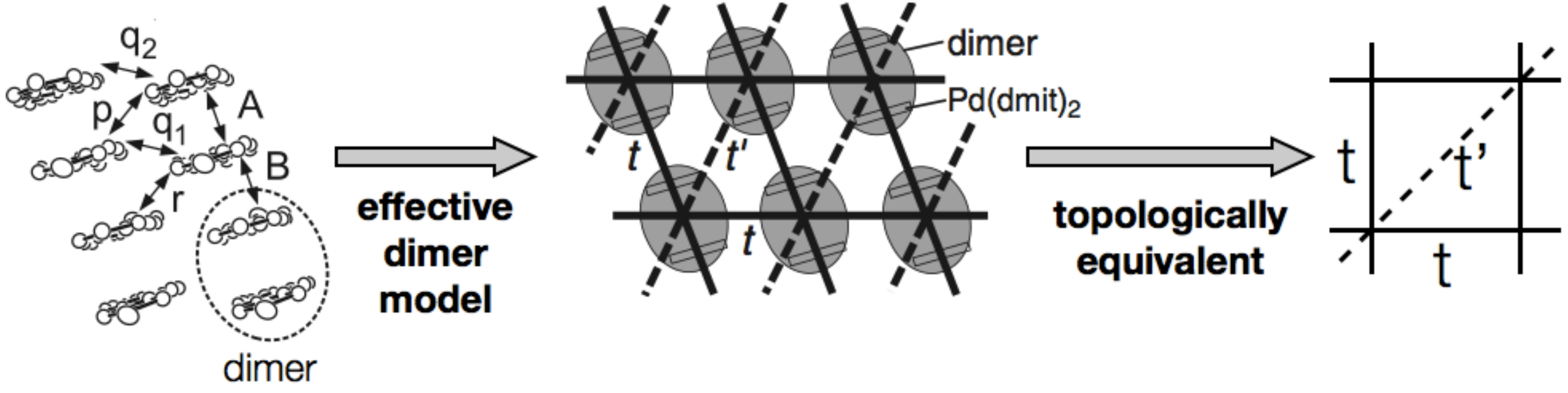}
\caption{The anisotropic triangular lattice in crystals of Et$_n$Me$_{4-n}Pn$[Pd(dmit)$_2$]$_2$. The left panel shows the a cross section of the organic layer, with a dimer ringed. The largest intermolecular hopping integrals are marked following the notation common in the H\"uckel literature. The central panel shows the structure of the effective dimer model. Here the dimers are represented by a single orbital, cf. Fig. \ref{fig:dmit-dimer-non-int}. This can be mapped, without loss of generality, onto the same anisotropic triangular lattice model as we discussed for the \kX salts in Section \ref{ET}. This model is half filled, so a large Hubbard $U$ must be associated with the dimers in order to cause the (Mott) insulating phase observed experimentally.  [Modified from \cite{YamauraJPSJ04} and \cite{ShimizuJPCM07}.]}
\label{fig:tri-lat-dmit}
\end{centering}
\end{figure*}

If one takes the dimers as a basic building block for the electronic structure, then the band structure of the Et$_n$Me$_{4-n}Pn$[Pd(dmit)$_2$]$_2$ salts is described by an anisotropic triangular lattice, cf. Fig. \ref{fig:tri-lat-dmit}. To date the only parameterisations of this anisotropic triangular lattice come from  H\"uckel calculations \cite{CanadellCCR99}. Miyazaki and Ohno \cite{MiyazakiPRB99} reported that their DFT band structures could be described by a fit to
a tight-binding model
for  this lattice, but did not report the values of $t'/t$ obtained from these fits. Given the H\"uckel methods systematic overestimation of $t'/t$ in the \kX salts, (cf. Section \ref{ET-struct}), one should exercise care when dealing with the  H\"uckel parameters for Et$_n$Me$_{4-n}Pn$[Pd(dmit)$_2$]$_2$ salts.

Because of the 2:1 ratio of anions to cations and the monovalency of Et$_n$Me$_{4-n}Pn$ cations the anisotropic triangular lattice model of the Et$_n$Me$_{4-n}Pn$[Pd(dmit)$_2$]$_2$ salts is half filled. Therefore, both  H\"uckel and DFT calculations predict a metallic state. In contrast, these materials are found to be insulating at ambient pressure \cite{KatoCR04}, and many undergo a metal-insulator transition under hydrostatic pressure and/or uniaxial stress. This suggests Mott physics is at play and hence that electron-electron interactions are vitally important. Therefore, the simplest model that may be compatible with the above considerations is the Hubbard model on an anisotropic triangular lattice
(cf. Section \ref{sec:hubbard}). Nothing is reliably known about the importance of longer range electron-electron interactions or electron-phonon interactions.

\subsection{Frustrated antiferromagnetism}

At ambient pressure the $\beta'$-Me$_{4-n}$Et$_nPn$-[Pd(dmit)$_2$]$_2$ salts are Mott insulators \cite{KatoCR04}. A large number of these materials order antiferromagnetically at low temperatures. To date relatively little is known about this antiferromagnetic state, for example, no experiments have investigated the ordering wavevector. Nevertheless, it is known that changing the cation does vary the N\'eel temperature, $T_N$. Indeed Shimizu \etal \cite{ShimizuJPCM07} have argued that there is a correlation between $T_N$ and the ratio $t'/t$ calculated from extended H\"uckel theory. However, given the issues with the values of these parameters obtained from H\"uckel theory, see Sections \ref{ET-struct} and \ref{dmit-struct}, and the small changes in $t'/t$ invoked in this comparison it is not clear how much confidence one should invest in this claimed
correlation at present. Certainly the values calculated from H\"uckel are systematically larger than than those found from DFT in the \kX salts (see Section \ref{ET-struct}). Nevertheless, one might hope that the trend that increasing $t'/t$ supresses $T_N$ may prove to be robust if higher level band structure calculations were performed.

The bulk magnetic susceptibility of many of the antiferromagnetic compounds has been studied at length \cite{TamuraJPCM02}. Fits to high temperature series expansions for the magnetic susceptibility reveal several interesting trends \cite{TamuraJPCM02,ZhengPRB05}. Firstly, for all of the materials for which such fits have been performed, strong frustration is found ($0.85<J'/J<1.15$; one should note however that this method has difficulty in determining whether $J'/J$ is greater or less than one). Secondly, for all of these materials a fit assuming an isotropic triangular lattice (i.e. $J'=J$) gives $J=250-260$ K. 
Hence, it should be noted that in these magnetically ordered materials $k_BT_N\ll J$, consistent with strong geometrical frustration.

\subsection{Spin liquid behaviour in $\beta'$-Me$_{3}$EtSb-[Pd(dmit)$_2$]$_2$ (Sb-1)}\label{sect:dmit-spin-liquid}

The bulk magnetic susceptibility, $\chi$, measured at high temperatures in Sb-1 is remarkably similar to the high temperature magnetic susceptibility in the frustrated antiferromagnets discussed above. $\chi(T)$ has a broad maximum around 50 K \cite{ItouPRB08} and fits well to high temperature series expansions for the isotropic triangular lattice with $J\simeq240$ K. However, no magnetic phase transition has been observed in Sb-1 down to the lowest temperatures studied [1.37 K \cite{ItouPRB08}], i.e., temperatures two orders of magnitude smaller than $J$.

\begin{figure}
\begin{centering}
\includegraphics[width=7cm]{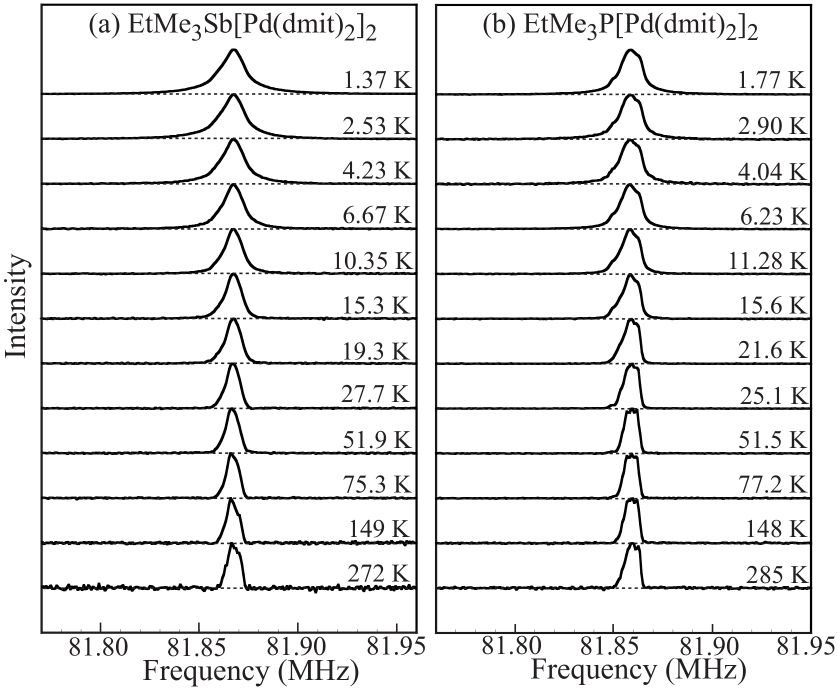}
\caption{Comparison of the $^{13}$C-NMR spectra of (a) Sb-1 which is a spin liquid and (b) P-1 which forms a valence bond crystal at low temperatures. Both sets of spectra look remarkably similar, showing no signs of long range ordering, but an inhomogeneous broadening that increases as the temperature is lowered. A similar broadening is observed in the spin liquid state of \CNn, cf. Fig. \ref{fig:CN-Cl-NMR}. From \cite{ItouPRB08}.
[Copyright (2008) American Physical Society]. }
\label{fig:ItouPRB08}
\end{centering}
\end{figure}

Itou \etal  \cite{ItouPRB08} also observed an inhomogeneous broadening of the NMR spectra at low temperatures, see Fig. \ref{fig:ItouPRB08}. They argued that this broadening is due to static local fields, but that, given the measured value of the hyperfine coupling constant, the broadening is too narrow to be understood in terms of long range magnetic order or spin glass behaviour. This is particularly interesting because it is very similar to what is observed in \CNn, cf. Fig. \ref{fig:CN-Cl-NMR}, raising the possibility of a common origin. At these low temperatures the recovery curve for the  $^{13}$C nuclear magnetic moment is \emph{not} described by a single exponential, suggesting that the nuclei see more than one environment.

It is also interesting to compare the findings of NMR experiments with the measurements of $\chi$. At high temperatures $1/T_1T\propto\chi$ \cite{ItouPRB08}. 
However, at low temperatures $1/T_1$ saturates to a constant; this is very different from what would be expected for a system with a spin gap (where one expects $1/T_1T\rightarrow0$ as $T\rightarrow0$). However, as noted above, in this low temperature regime the recovery of the magnetization is not described by a single exponential, making the estimation of $T_1$ rather difficult and raising questions about the interpretation of the estimated value of $1/T_1T$.


\begin{figure}
\begin{centering}
\includegraphics[width=7cm]{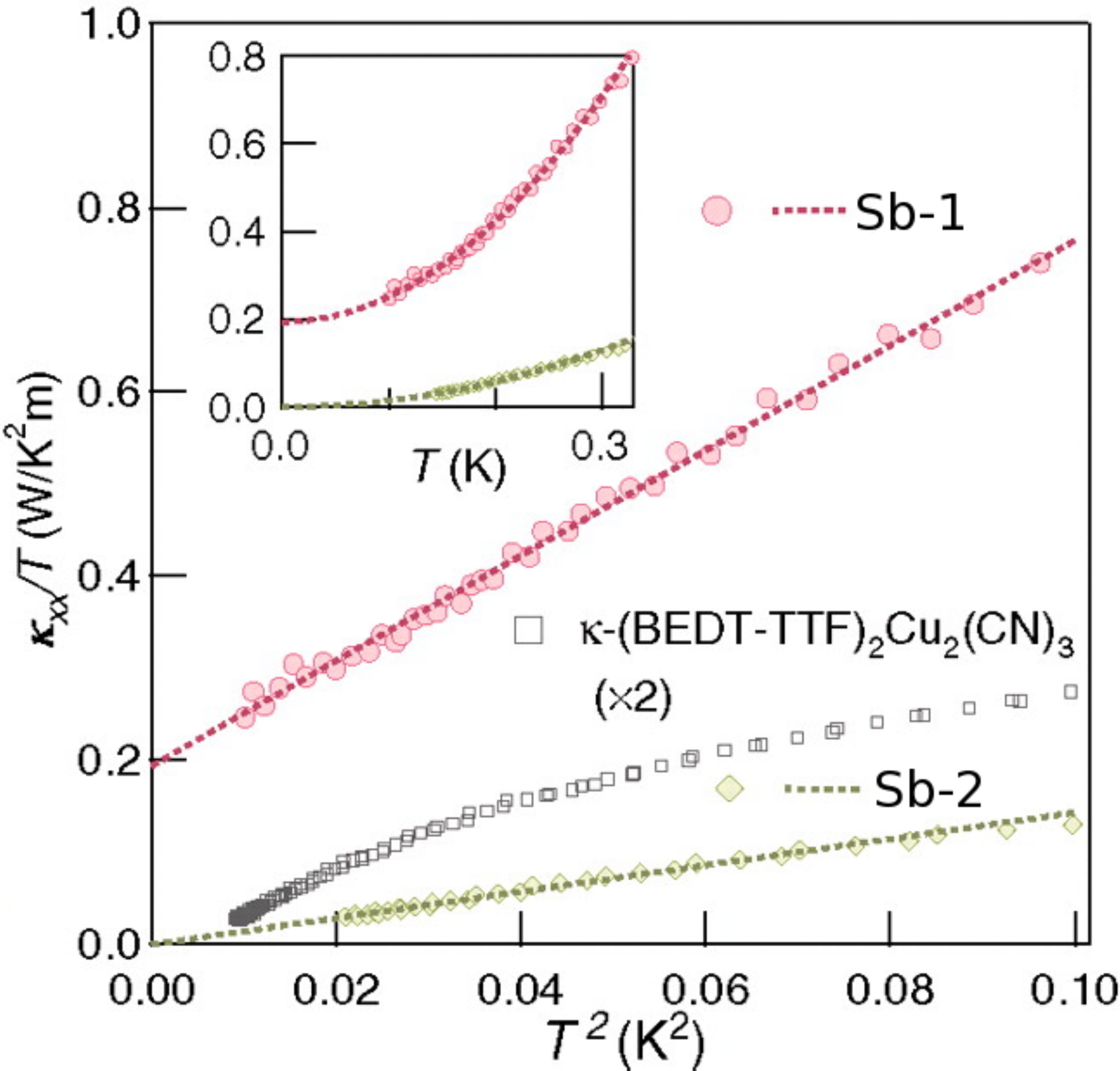}
\caption{
Temperature dependence of the thermal conductivity of several different
 frustrated materials. For \CN and Sb-2 the $\kappa/T\rightarrow0$ as $T\rightarrow0$ indicating that the excitations are gapped. However, for Sb-1 the spectrum appears to be gapless at $\kappa/T$ tends to a finite number as $T\rightarrow0$. Indeed the residual value of $\kappa/T$ is comparable to that observed in the metallic state of other organics \cite{BelinPRL98}. The inset shows the same data on a linear scale From \cite{YamashitaScience10}.}
\label{fig:kappaT}
\end{centering}
\end{figure}

A recent report of measurements of the thermal conductivity of Sb-1 gives several important insights into the nature of the spin liquid state in Sb-1 \cite{YamashitaScience10}.
Figure \ref{fig:kappaT}
shows the temperature dependence of the thermal conductivity
 $\kappa(T)$. The non-zero intercept of $\kappa(T)/T$ as the
temperature approaches zero  is a clear signature of gapless excitations. The magnitude of the intercept is comparable to its value in the metallic phase of other organic charge transfer salts, and an order of magnitude larger than what one gets in the d-wave superconducting state due to nodal quasi-particles \cite{BelinPRL98}.

Another important finding reported in this paper is that a spin gap is observed in the magnetic field dependence of thermal conductivity. At first sight
this is rather puzzling as the 
temperature dependence of the thermal conductivity (Fig. \ref{fig:kappaT}) clearly shows that there are gapless excitations. However, one should note that excitations with any spin state can  be excited thermally, whereas a field does not 
affect 
singlet excitations, cf. Section \ref{sec:spinliquid}. Therefore, these two results combined suggest that there are gapless singlet excitations, but a gap to the lowest lying triplet (or higher spin) excitations. Thus, these results suggest that P-1 is what Normand has called a `Type-II' spin liquid \cite{NormandCP09}, see Sections
\ref{sec:spinliquid} and \ref{sec:heisenberg} for further discussion.

Recently Katsura \etal \cite{KatsuraPRL10} predicted that there would be a
sizeable thermal Hall effect in quantum spin liquids with deconfined spinons.
This motivated measurements of the thermal conductance tensor  in a magnetic field perpendicular to the layers \cite{YamashitaScience10}. Within error the thermal Hall angle was zero. 

\subsection{Is there a valence bond crystal or spin Peierls state in  $\beta'$-Me$_{3}$EtP-[Pd(dmit)$_2$]$_2$ (P-1)?}\label{section:VBC}

\begin{figure}
\begin{centering}
\includegraphics[width=7cm]{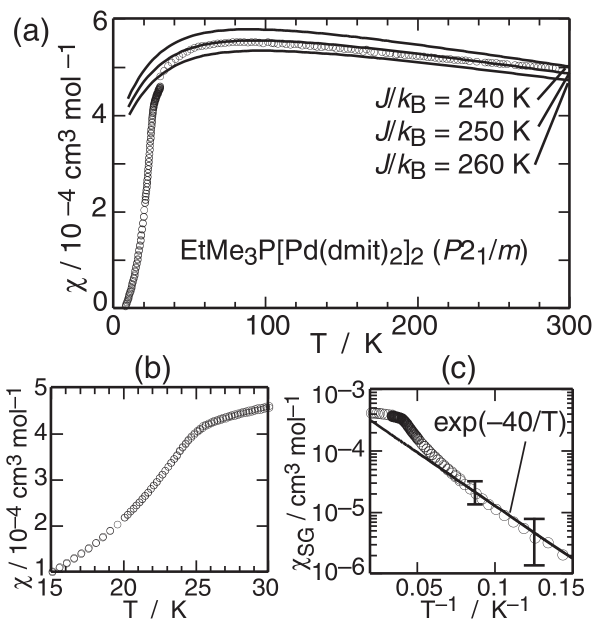}
\caption{Opening of a spin gap  in P-1. (a) At high temperatures the spin susceptibility is similar to the other \dmit salts and can be fit to high temperature series expansions of the Heisenberg model on the isotropic triangular lattice with $J\sim250$ K. But, below $\sim 25$ K a sudden drop is seen (b). At low temperatures the susceptibility appears to be activated consistent with the opening of a spin gap.  From \cite{TamuraJPSJ06}.
}
\label{fig:sus-P-1}
\end{centering}
\end{figure}

P-1 shows an unusual phenomenology at low temperatures.  As with the other \dmit salts discussed above at high temperatures the bulk magnetic susceptibility is well described by high temperature series expansions for the Heisenberg model on the anisotropic triangular lattice, in this case with $J\simeq250$ K, cf. 
Figure \ref{fig:sus-P-1}. One should note here that a Curie term (corresponding to about one $S=1/2$ spin per 300 formula units) and a constant term (the value of which was not reported) have been subtracted from the experimental data before this comparison was made.  However, below 25 K an exponential decrease in $\chi$ is observed \cite{TamuraJPSJ06}, cf. 
Figure \ref{fig:sus-P-1}
. The low temperature susceptibility is consistent with the opening of a spin gap of $\Delta=40\pm10$ K, although Tamura \etal stress that this value is quite sensitive to the details of the values of the Curie and constant terms subtracted from the experimental data. One should also note that this fit was carried out over less than one decade in temperature.

\begin{figure}
\begin{centering}
\includegraphics[width=7cm]{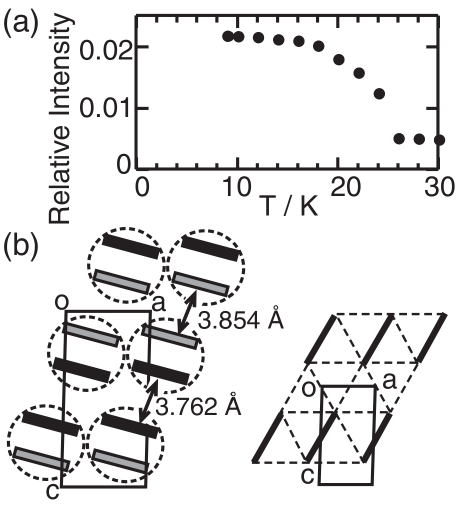}
\caption{Lattice distortion at low temperatures in P-1. At high temperatures all of the dimers in P-1 are uniformly spaced (3.82 \AA~ apart). However, below $\sim25$ K at lattice restructuring occurs and the dimers `pair-up'
 (we will avoid the word dimerise here$^{\ref{foot:dimerisation}}$).
 Below the same temperature significant changes in the magnetic susceptibility are observed, cf. Fig \ref{fig:sus-P-1}, which suggest that a spin gap opens. Both effects are large suggesting that neither is a secondary effect induced by the bilinear coupling between the order parameters. This conclusion would imply that neither the simple valence bond crystal nor the simple spin-Peierls transition is a sufficient description.  Panel (a) shows the growth of the Bragg peak associated with this distortion at low temperatures, while panel (b) shows the distorted crystal lattice.    From \cite{TamuraJPSJ06}.}
\label{fig:VBC-P-1}
\end{centering}
\end{figure}

X-ray crystallography \cite{TamuraJPSJ06} reveals structural changes in P-1 at approximately the same temperature as the spin gap opens. Satellite peaks indicating a doubling of the periodicity in the crystallographic $c$ direction (which lies in the highly conducting plane) grow rapidly below 25 K, see Fig. \ref{fig:VBC-P-1}. In the high temperature phase the distance between neighbouring \dmit molecules in different dimers is, uniformly, 3.82 \AA. In the low temperature, spin gapped, phase there are two crystallographically distinct types of \dmittwo dimer with the distances between molecules in 
the two different dimers
 now being 3.76 \AA~ and 3.85 \AA. Thus the \dmittwo dimers have paired up.\footnote{If we treat the \dmittwo dimers as `sites' one would conventionally say that the sites have dimerised, but this can get a little confusing as we are already using the word `dimer' as a synonym for site. \label{foot:dimerisation}} 

The $^{13}$C NMR spectrum of P-1, Fig. \ref{fig:ItouPRB08} \cite{ItouPRB08}, broadens slightly below 25 K consistent with the increased number of environments for the nuclei. It is interesting to note how similar the NMR spectra for P-1 and Sb-1 are, in marked contrast to the very different behaviours of the low temperature bulk magnetic susceptibility.

There are two natural explanations for this pairing
of dimers
occurring concomitantly with the opening of the spin gap: a valence bond crystal or a spin Peierls distortion. Although these two possibilities are quite different theoretically,  
the experimental consequences of these differences are rather subtle. It is therefore helpful to examine each possibility briefly. The key issue is whether a spin-lattice coupling
is {\it necessary} for formation of the spin gap.
In other words, does the lattice distortion cause the gap 
or does the formation of spin singlets due to frustration
cause the lattice distortion?
Similar issues have been discussed for
Heisenberg models relevant to CaV$_4$O$_9$ \cite{StarykhPRL96}
and Li$_2$VOSiO$_4$ \cite{BeccaPRL02}.

\begin{figure}
\begin{centering}
\includegraphics[width=7cm]{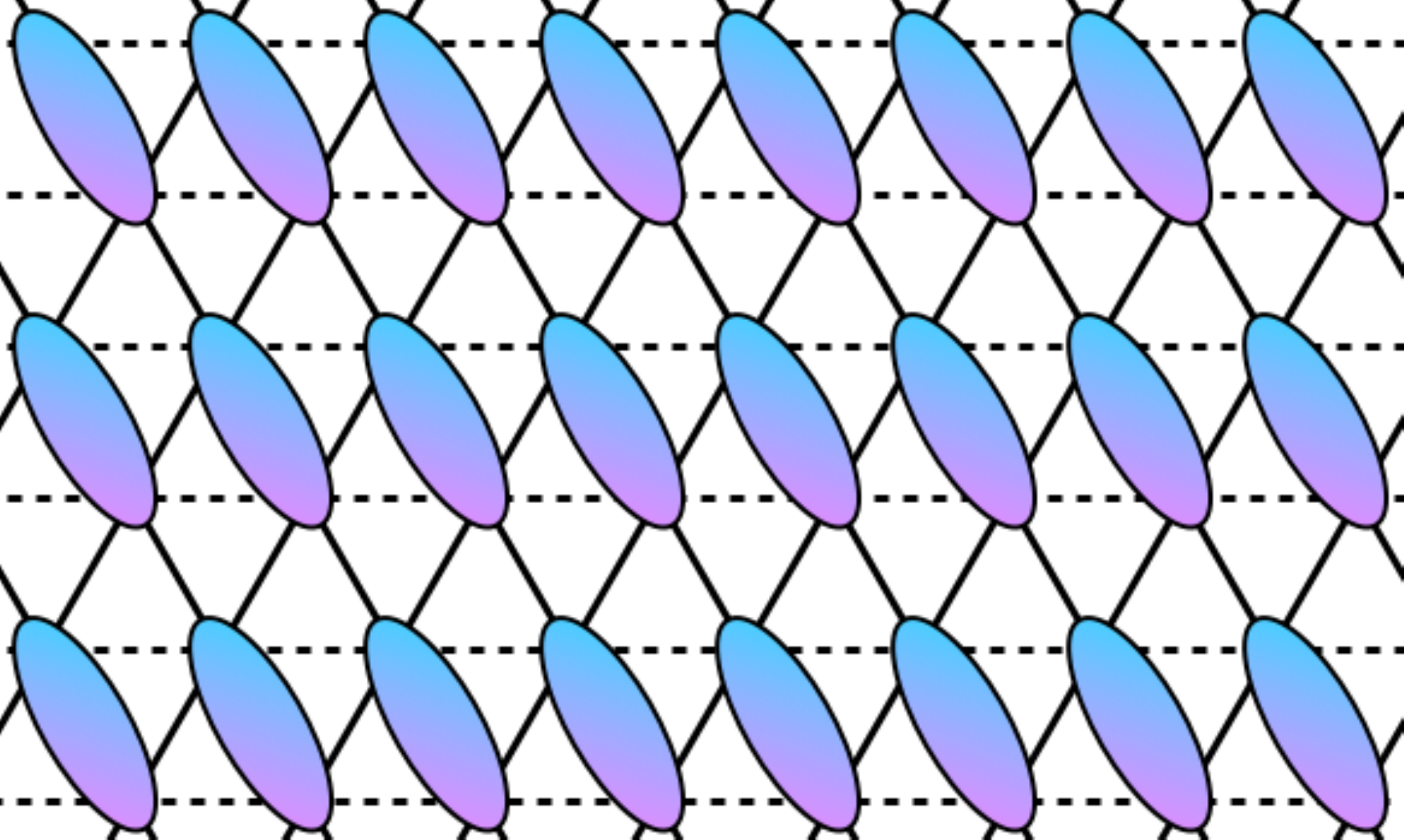}
\caption{Sketch of a valence bond crystal. Coloured elipses indicate singlet correlations between pairs of sights. Note that this state breaks both the rotational and translational symmetries of the underlying lattice, although a particular combination of the two remains unbroken. This should be contrasted with a valence bond liquid, that does not break these symmetries. On possible configuration of such a valence bond liquid is shown in Fig. \ref{fig:dimercover}.}
\label{fig:VBC}
\end{centering}
\end{figure}

The valence bond crystal (VBC) is a purely electronic phenomena. The VBC has been postulated as a possible ground state for various frustrated Heisenberg models (or related spin Hamiltonians) \cite{NormandCP09}
including the model on the anisotropic
triangular lattice that is relevant here (cf. Section \ref{sec:heisenberg}).
 In the VBC state pairs of spins form singlets. These singlets are arranged periodically so as to break the translational symmetry of the underlying lattice (cf. Fig. \ref{fig:VBC}). Note that the lattice degrees of freedom do not play any explicit role in stabilising the VBC state. However, in any real material the spin-phonon coupling will drive a lattice distortion that decreases the distance between the the sites within a singlet.

In contrast to the VBC, the spin Peierls distortion involves the lattice in an essential way. The spin Peierls distortion is usually conceived as a one-dimensional (1D) phenomena. The uniform one dimensional Heisenberg chain has gapless excitations.
However, dimerisation of the lattice opens a gap in the excitation
spectrum and lowers the ground state energy. This decrease in energy is
greater than the cost in elastic energy associated with the 
lattice distortion and so if the spin-phonon
coupling is non-zero there is a spontaneous dimerisation of
the lattice \cite{CrossPRB79}.

An appropriate order parameter for the VBC state is $\phi=\sum_i\sum_{n.n.}\langle \hat{\bf S}_i \cdot \hat{\bf S}_{i+1} -  \hat{\bf S}_i \cdot \hat{\bf S}_{i-1}\rangle$ where ${\bf S}_i$ is the spin operator on site $i$ and the sites $i+1$ and $i-1$ are nearest neighbours of site $i$, but in opposite directions, the sum over n.n. indicates that this sum runs over all such pairs of nearest neighbours. An order parameter for the spin-Peierls transition is $\lambda=\sum_i\sum_{n.n.}\langle \hat{\bf r}_i  \hat{\bf r}_{i+1} -  \hat{\bf r}_i  \hat{\bf r}_{i-1}\rangle$, where ${\bf r}_i$ is the position of site $i$. These order parameters couple at bilinear order in a Landau theory. Thus, the simplest Landau theory will be
\begin{eqnarray}
\Delta F=\alpha\phi^2+\beta\phi^4+a\lambda^2+b\lambda^4+s\phi\lambda,
\end{eqnarray}
where $\Delta F$ is the difference between the energies of the high and low symmetry phases, and $\alpha$, $\beta$, $a$, $b$ and $s$ are the parameters of the theory.
Thus, if one order parameter becomes finite it acts as a `field' for the other implying that the second will take, at least, a small non-zero value. Therefore the observation that both $\lambda$ and the spin gap become non-zero at $\sim25$ K is not overly surprising. However, both the gap and the lattice distortion are large $\Delta/k_BT_c=1.6\pm0.4$,\footnote{Note that this is close to the size of the gap predicted by weak coupling theories: the weak-coupling BCS gap in a superconductor or spin density wave is $\Delta/k_BT=1.76$.} and $\lambda=0.1$ \AA, i.e., 2.4\% of the average interdimer spacing
This is comparable to or larger than the amplitudes of the distortions observed in many organic spin Peierls systems \cite{Foury-LeylekianPRB04,VisserPRB83,DumoulinPRL96,MoretPRL86}.
This may indicate that neither effect is simply parasitic on the other, in which case one would expect the parasitic order parameter to be small, and hence that the low temperature phase is stabilised  cooperatively by both the lattice and electronic degrees of freedom and not  properly characterised as either a spin-Peierls or a VBC state. 

A key question is why this large spin gap is observed in P-1, but not in other \EtMePn salts. Tamura \etal proposed that an essential ingredient is 
the crystal structure of P-1. Most \EtMePn salts form the so-called $\beta'$ structure, which  displays a $C2/c$ space group \cite{KatoCR04}. The unit cells of these materials each contain two organic layers, cf. Fig. \ref{fig:crystal-bp}. These planes are crystallographically equivalent and related by an axial glide along the $c$-axis (interlayer direction). This glide maps the ${\bf a}+{\bf b}$ direction, in which the dimers stack, in one plane onto the  ${\bf a}-{\bf b}$ plane in the next. This results in what is known as a `solid crossing' crystal structure, cf. Fig \ref{fig:crystal-bp}. However, P-1 forms crystals with $P2_1/m$ symmetry \cite{KatoJACS06}. This unit cell also contains two crystallographically equivalent organic layers, but now they are related by $b$-mirror and $b$-screw symmetries (in this structure the $b$-axis is the interlayer direction) rather than a glide plane. Thus, both layers stack in the same (${\bf a}+{\bf c}$) direction. Other than the loss of the glide plane, the crystal structure of P-1 is remarkably close to the $\beta'$ structure.
Thus Tamura \etal argued that the solid crossing in the $\beta'$ phase means that if there were Peierls distortion along the dimer stacking direction in a $\beta'$-\EtMePn salt this would lead to large internal strains within the crystal as the distortion would alternate along the ${\bf a}+{\bf b}$ and  ${\bf a}-{\bf b}$ directions in the neighbouring planes. No detailed calculation has yet investigated whether this could be energetically unfavourable enough to prevent the opening of the spin gap. However, if there are many competing ground states  it is possible that a small increase in the ground state energy of the spin Peierls/VBC phase, such as that caused by the internal strains invoked by Tamura \etaln, could be enough to suppress this phase in favour of some other.

The above proposal requires that the elastic coupling between layers be sufficiently large that it can change the ground state. An alternative hypothesis is that the intralayer anisotropy $J'/J$ varies enough between the different materials that the ground state is different. In Section \ref{sec:heisenberg} (see especially Figure \ref{fig:zhenggap}) it will be seen that a significant spin gap only occurs for a narrow parameter range ($J'/J \simeq 0.7-0.9$) of the Heisenberg model on a triangular lattice.

\subsection{Paramagnetic to non-magnetic transition in \Sbtwo (Sb-2) and Cs\dmittwo (Cs-00)}

\begin{figure}
\begin{centering}
\includegraphics[width=7cm]{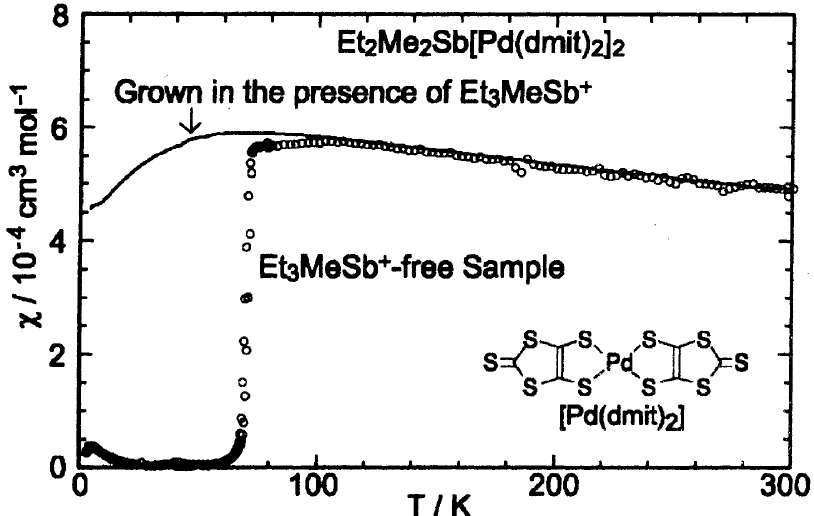}
\caption{The paramagnetic to non-magnetic transition in Sb-2. A rather similar transition is also seen in Cs-00. The microscopic nature of this transition is not clear. However, it is clear that small amounts of non-magnetic impurities in the cation layer completely suppress the non-magnetic phase.   From \cite{TamuraSM05}.}
\label{fig:TamuraSM05}
\end{centering}
\end{figure}

Rather similar phase transitions are seen at $\sim70$ K in Sb-2 and $\sim65$ K in Cs\dmittwo (henceforth Cs-00). 
In Cs-00 the resistance shows a clear metal-insulator transition at this temperature \cite{UnderhillJPCM91}. This is accompanied by a sudden drop in the bulk magnetic susceptibility \cite{UnderhillJPCM91}, which is $\sim3.5$ emu/mol, independent of temperature, for $T>65$ K and zero to within experimental error for $T<65$ K.
In Sb-2  a non-metal to insulator 
 transition corresponding to ``steep rise of resistivity with decreasing temperature'' was reported by Tamura \etal (although they did not show the data) concomitant with the sudden drop in the bulk magnetic susceptibility, shown in Fig. \ref{fig:TamuraSM05}  \cite{TamuraSM05}.

\begin{figure}
\begin{centering}
\includegraphics[width=7cm] {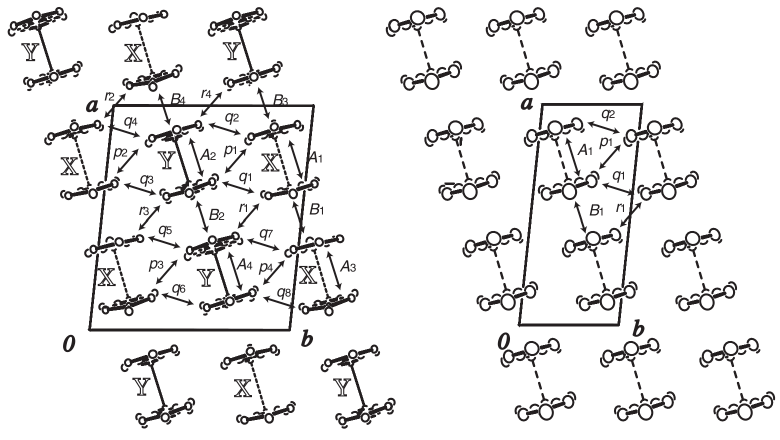}
\caption{Lattice reconstruction in the low temperature non-magnetic phase of Sb-2. At the temperature where the sudden drop in the bulk susceptibility is observed (Fig. \ref{fig:TamuraSM05}) the crystal changes symmetry from $C2/c$ to $P2_1/c$. This results in the low temperature unit cell (left) being double the size of the high temperature unit cell (right). In the high temperature unit cell both dimers are crystallographically equivalent. But in the low temperature unit cell there are two crystallographically inequivalent dimers (labelled X and Y). The bond lengths are distinctly different in the X and Y dimers, suggesting significant charge disproportionation. From \cite{NakaoJPSJ05}.}
\label{fig:NakaoJPSJ05}
\end{centering}
\end{figure}

Nakao and Kato \cite{NakaoJPSJ05} have shown that these phase transitions are both associated with changes in crystal structure. At room temperature both Sb-2 and Cs-00 form crystals with the $C2/c$ crystal structure of the $\beta'$ phase, cf. section \ref{section:VBC} and Fig. \ref{fig:crystal-bp}. At temperatures just above the phase transition critical temperature, $T_c$, additional incommensurate satellite reflections are seen in both materials via x-ray scattering \cite{NakaoJPSJ05}. These become fully developed Bragg peaks below $T_c$, indicative of a change in the crystal structures. In their low temperature phases both P-2 and Cs-00 have crystals with $P2_1/c$ symmetry. The most dramatic change associated with this is a doubling of the unit cell along the (in plane) $b$-axis. This leads to there being two crystallographically distinct dimers, labelled X and Y in Fig. \ref{fig:NakaoJPSJ05}, per unit cell, which are arranged in alternating rows perpendicular to the stacking direction.
The bond lengths within the X dimers are significantly different from those within the Y dimers  \cite{NakaoJPSJ05}.

The optical conductivities of both Sb-2 \cite{TamuraCPL05} and Cs-00 \cite{UnderhillJPCM91} are also very similar and show dramatic changes below $T_c$. Above $T_c$ Cs-00 displays a Drude peak and a much stronger broad Lorentzian peak centred around $\sim 1$ eV. Below $T_c$ the Drude peak is absent, consistent with the metal-insulator transition observed in the dc conductivity. No major qualitative changes are observed in the high energy feature between the spectrum recorded at 80 K and that at 50 K. However, by 20 K this peak has split into two distinct features. This is consistent with there being two distinct dimers in the crystals at these temperatures. Tamura \etal only reported the optical conductivity of Sb-2 for frequencies greater than $5\times10^{3}$ cm$^{-1}$, so it is not possible for us to discuss the Drude peak in this material, although one presumes it will be absent. However, otherwise the high frequency conductivity is remarkably similar to that of Cs-00. At 100 K (and higher temperatures) there is a single broad feature, which can be fit to a single Lorentzian. But, at 50 K and 4 K (the only temperatures below $T_c$ for which the optical conductivity was reported) two narrower Lorentzian peaks were observed.

Both Underhill \etal and Tamura \etal interpret the peaks in the high energy optical conductivity in terms of intradimer transitions (from the HOMO-bonding level to the HOMO-antibonding level and the LUMO-bonding level to the LUMO-antibonding level, cf. Fig. \ref{fig:dmit-dimer-non-int}).  This neglects correlations, which we have seen are important in these materials. However, one can describe the transition as an intra-dimer charge transfer transition, which yields similar results \cite{MerinoPRB00b}.
Underhill \etal argued that the transition involved a charge density wave coupled to a secondary order parameter. Alternatively, Kato's group \cite{TamuraCPL05,NakaoJPSJ05,TamuraCPL04} have argued that the low-energy phase is charge ordered, with complete charge disproportionation between [Pd(dmit)$_2$]$_2^0$ X dimers (cf. Fig. \ref{fig:NakaoJPSJ05}) and [Pd(dmit)$_2$]$_2^{+2}$ Y dimers. One way to test this hypothesis would be to study how the phonon frequencies shift at the transition. Nevertheless, no investigations
of the interesting physics in Cs-00 and Pb-2
using quantum many-body theory have yet been reported.

It is also interesting to note that there is at least two differences between the phase transition in Sb-2 and Cs-00 \cite{NakaoJPSJ05}. Firstly, in Sb-2 the additional Bragg peaks grow sharply at the phase transition and hysteresis is observed, suggesting a first order transition. In contrast, in Cs-00 no hysteresis is observed and the additional Bragg peaks grow continuously suggesting a second order phase transition. Secondly, the paramagnetic to non-magnetic transition in Cs-00 is associated with a metal to insulator transition \cite{UnderhillJPCM91}, whereas Sb-2 is insulating in both phases \cite{TamuraSM05}. 
These differences, and the possible connection between them, have not, yet, been explained.

\subsubsection{Et$_3$MeSb impurities in \Sbtwo (Sb-2)}

The first crystals of Sb-2 to be grown were electrocrystallised out of a solution containing Et$_3$MeSb${^+}$ ions as well as Et$_2$Me$_2$Sb${^+}$ ions \cite{AonumaSM97}. These materials did \emph{not} show the phase transition described above, cf. Fig. \ref{fig:TamuraSM05}. Rather, the bulk magnetic susceptibility, $\chi$, in these samples is remarkably similar to that of the spin liquid Sb-1. In both materials $\chi$ has a broad maximum around 50 K. And the exchange interaction extracted from fits to $\chi(T)$ is $J\simeq240$ K for both Sb-1 \cite{ItouPRB08} and Sb-2 \cite{ZhengPRB05,TamuraJPCM02}. Indeed, no magnetic phase transition was observed in these Sb-2 salts grown in the presence of Et$_3$MeSb down to the lowest temperatures studied [4.3 K \cite{NakamuraJMC01}], which is temperatures two orders of magnitude smaller than $J$.

Nakao and Kato \cite{NakaoJPSJ05} have grown crystals of Sb-2 from electrolytes consisting of different ratios of Et$_2$Me$_2$Sb${^+}$ to Et$_3$MeSb${^+}$. For a 10:1 ratio the $T_c$ of the nonmagnetic phase was supressed by $\sim5$ K relative to samples grown electrolytes free of  Et$_3$MeSb${^+}$. However, a 2:1 ratio completely suppresses the phase transition. 

Note that Et$_2$Me$_2$Sb and Et$_3$MeSb are isoelectronic, so Et$_3$MeSb should be a non-magnetic impurity. Nakao and Kato's results could suggest that the low temperature phase in Sb-2 is extremely sensitive to disorder. However, they
 also found that the inclusion of Et$_3$MeSb leads to a significant expansion of the unit cell, as one would reasonably expect given that an ethyl group is approximately twice as large as a methyl group. Compared to the Et$_3$MeSb free crystal, the unit cell is 0.11\% larger in crystal grown from the 10:1 electrolyte and 0.25\% larger in the crystal grown from the 2:1 electrolyte \cite{NakaoJPSJ05}. They also found that the unit cells of the earlier crystals were 0.35\% larger than those of the  Et$_3$MeSb free crystal suggesting that these first crystals contained significant concentrations of Et$_3$MeSb. However, no direct measurements of the concentrations of Et$_3$MeSb in an Sb-2 crystal have yet been reported. 

It has not yet been possible to grow crystals of Sb-3 as Et$_3$MeSb[\dmitn]$_3$ is preferentially formed \cite{NakaoJPSJ05}. Clearly, it would be very interesting to know how Sb-3 behaved at low temperatures.

Whether impurity physics or the lattice expansion is the dominant effect of Et$_3$MeSb impurities, it is clear that there is a rather subtle competition between the low temperature phase in Sb-2 and spin liquid behaviour. This is rather surprising as the critical temperature in Et$_3$MeSb free Sb-2 is quite large, 70 K.

It would be extremeley interesting to understand this physics. Not only for its intrinsic interest, but also in relation to the spin liquid state in \CN and Sb-1.




\subsection{Mott transition under hydrostatic pressure and uniaxial stress}\label{section:pressurestress}

Most studies of \EtMePn salts under hydrostatic pressure or uniaxial stress have been limited to measurements of the resistivity. Thus, there has been a mapping out of the phase diagram in terms of the insulating, metallic and superconducting states.
In this section we review these phase diagrams and discuss the few pioneering studies to go beyond resistivity measurements.

P-0 and As-0 do not exhibit a metal-insulator transition under even the highest pressures studied ($\sim17$ kbar) \cite{KatoCR04}. This suggests that they are further into the Mott insulating regime than many other salts of \dmitn, which are driven metallic under these pressures. This is rather counterintuitive as these materials have small anions (containing only methyl groups and no ethyl groups), which one would na\"ively associate with a large `chemical pressure'. However, a uniaxial strain along the $b$ axis of $>7$ kbar drives As-0 metallic \cite{KatoPRB02}. In contrast uniaxial strains along the $a$ and $c$ axes do not drive the system metallic \cite{KatoPRB02}, indeed moderate strains ($\sim2$ kbar) the these directions drive As-0 deeper into the insulating state. Kato \etal argued that that applying a stress along the $a$ axis increases the dimerisation and so increases $U$, while strain along the  $b$ axis moves the dimer stacks closer together and thereby increases $W$. This argument make two implicit assumptions: (i) that a single orbital description of the electronic structure
of  these materials is appropriate; (ii) that they are in the limit $(U_m-V_m)\gg2t_A$ and hence that $U\simeq2t_A$, where $U$ is the effective Coulomb repulsion between two electrons on the same \dmittwo dimer, $U_m$ is the Coulomb repulsion between two electrons on the same \dmit molecule, $V_m$ is the Coulomb repulsion between two electrons on different molecules within the \dmittwo dimer, and $t_A$ is the hopping integral between the two molecules in the dimer. It is not clear, at present, that either of these assumptions is valid.

Kato \etal \cite{KatoSSC98} measured the resistivity of P-2 under pressure in order to map out its phase diagram (Fig. \ref{fig:P-2}). They found that the ambient pressure insulating phase can be driven into a metallic state (which superconducts below $\sim4$ K) by pressures  above 9 kbar. However above $\sim12$ kbar an insulating state is observed again. 
Yamaura \etal \cite{YamauraJPSJ04} found that the high pressure insulating phase is driven by a change in the crystal structure of P-2  from the usual $\beta'$ phase with $C2/c$ symmetry and four crystallographically equivalent dimers per unit cell, cf. Fig. \ref{fig:crystal-bp}, to a phase with $P\overline1$, with two crystallographically \emph{inequivalent} dimers per unit cell.  As the two inequivalent dimers are in different
layers this means that the layers are inequivalent. However, no detailed theory of why this causes an insulating state has yet been reported.


It is interesting to note that in the metallic state of P-2  the resistivity \cite{KatoSSC98} does not have the non-monotonic temperature dependence associated with the `bad-metal' regime, so typical of the $\kappa$-ET$_2X$ salts. Nor is there any sign of a bad metal in the metallic phases of Sb-0 [above $\sim9$ kbar \cite{KatoSSC96}]. There is a small peak around 50 K in As-0 under a uniaxial strain $\sim7$ kbar along the $b$ axis \cite{KatoPRB02}, but it appears rather less pronounced than one sees in \kXn.

Both Shimizu \etal \cite{ShimizuPRL07} and Itou \etal \cite{ItouPRB09} have investigated the metal-VBC/spin Peierls insulator transition in P-1, which occurs under $\sim4$ kbar of hydrostatic pressure (Fig. \ref{fig:P-1}). Shimizu \etal found clear evidence that the transition is first order from sharp discontinuities and hysteresis in the resistivity. They also found that a magnetic field destabilises the insulating phase in favour of the metallic phase. Whence, they were able to show, from a Clausius-Clapeyron analysis, that the magnetisation, $M=\chi B$, in the insulating phase is less than that in the metallic phase, consistent with the proposed non-magnetic VBC state. This is consistent with the results of Itou \etal from NMR spectroscopy. They did not observe any signs of magnetic ordering under pressure and concluded that the VBC/spin Peierls state persists over the entirety of the  insulating region of the phase diagram. This means that the VBC/spin Peierls phase directly abuts the superconducting phase, raising interesting questions about the role of spin correlations in mediating this superconductivity. In particular this means that P-1, like the spin liquid compounds Sb-1 and \CNn, provides an interesting contrast to the, more common, case of superconductivity near an antiferromagnetic Mott insulator \cite{ItouPRB09,PowellPRL07}. Nevertheless, both of these insulating states have large singlet fluctuations, and so, may actually have rather similar relationships to their nearby superconducting phases. In the low temperature insulating phase Itou \etal find that the spin lattice relaxation cannot be fit to a single exponential suggesting that the systems is rather inhomogeneous. Similar effects are also observed in the spin-Peierls phase of CuGeO$_3$ \cite{KikuchiJPSJ94,ItohPRB95}. Itou \etal propose that this is because unpaired spins strongly influence the relaxation rate. However, this would appear to be a rather subtle question, and it would be interesting to know if specific microscopic theories can account for this effect.

The resistivity in the metallic state of P-1 \cite{ShimizuPRL07} is remarkably similar to the resistivity in the metallic states of the \kX compounds (cf. section \ref{ET-metal}), suggesting the
same bad metal physics is likely to be at play. However, $1/T_1T$ is constant at low temperatures\footnote{For $P=4.8$ kbar below $\sim20$ K, for $P=6.0$ kbar below $\sim100$ K, for $P=8.0$ kbar below $\sim200$ K. Data extends down to 2 K for all pressures.} \cite{ItouPRB09} suggesting that there is no pseudogap in P-1.

One puzzling result is that in the metallic state of P-1 the in-plane ($a$-axis) resistivity is only described by the usual Fermi liquid form, $\rho_\|(T)=\rho_{\|0}+AT^2$, at the highest pressures studied $\sim8$ kbar \cite{ShimizuPRL07}. At lower pressures and for temperatures in the range of about
2-20 K
the data can be fit to the form $\rho_\|(T)=\rho_{\|0}+AT^\epsilon$ with $2<\epsilon<3$. This may suggest that near the Mott transition the electrons scatter 
off an additional mode as well the direct electron-electron scattering that gives rise to the quadratic temperature dependence of the resistivity in Fermi liquid theory. For example, below their Debye temperature phonons give rise to an electron-phonon scattering rate $1/\tau_{e-ph}\propto T^5$ \cite{Ashcroft76}. If both electron-electron and electron-phonon scattering gave rise to similar scattering rates this could appear as a intermediate power law over a limited temperature range, like those discussed above.
On the other hand, it may be that one is fitting at temperatures above that at which simple Fermi liquid applies. At the lower pressures the resistivity increases rapidly above about 30 K, corresponding to the destruction of quasi-particles, which suggests that the $T^2$ behaviour may not last as high as 20 K even at 8 kbar.

 Shimizu \etal also note that the value of $A$ they observe is 
a factor of about fifty times
smaller than that found in the metallic phase of the \kX salts. They suggest that this is because the electron-electron interactions are weaker. However, this seems unlikely as metallic P-1 is on the border of a Mott transition. A number of material specific factors are important in determining the value of $A$ \cite{JackoNP09}, therefore  these effects are probably responsible for the smaller value of $A$ in P-1. 
Also, caution is in order since accurately measuring the intralayer resistivity in layered materials can be difficult because of uncertainties about the actual current path through the sample.

At low temperature P-1 superconducts. $T_c$ is suppressed from its maximum value, $\sim5$ K, near the Mott transition by the application of greater pressure  \cite{ShimizuPRL07}.
However, very little is known about the superconducting state in any of the Pd(dmit)$_2$ salts. This is at least in part due to the fact that the superconducting state is only observed under pressure, making many experiments difficult. Most of the reports of superconductivity simply consist of resistivity measurements. However, \cite{IshiiJPSJ07} did observe the Meissner effect in P-1, from which they were able to show that the superconductivity is a bulk effect.

%% file: nmr.tex
\section{Nuclear magnetic resonance as a probe of spin fluctuations}



The experimental data for $1/T_1T$ for most of the materials
of interest show two
regimes: $T > T_\nmr$ in which $1/T_1T$ increases as
temperature decreases and $T < T_\nmr$ where $1/T_1T$ is rapidly
suppressed as the temperature is further lowered. The 
spin fluctuation models that 
we discuss below predict that 
$1/T_1T$ 
 is a monotonic decreasing function
of temperature and so  cannot describe            
the data below $T_\nmr$. Hence our discussion is confined
to the $T > T_\nmr$ regime, where
the data for a wide range of materials can be fit  to the form \cite{YusufPRB07,PowellPRB09}
\begin{equation}
\frac{1}{T_1T} =     
\left(\frac{1}{T_1}\right)_\infty
\frac{1}{(T+T_x)}.
\label{cw_t1t}
\end{equation}
This temperature dependence is obtained for three
different spin fluctuation models described below.
This then raises the question of which model gives the physically
appropriate picture.

\subsubsection{Long-range antiferromagnetic spin fluctuation model}

A phenomenological antiferromagnetic spin fluctuation model
was introduced by Moriya in his self-consistent
renormalization (SCR) theory \cite{MoriyaRPP03} and was used by Millis,
Monien and Pines (MMP) \cite{millis} to describe NMR in the 
metallic state of the cuprates. 
Together with Yusuf we recently applied this model to describe NMR relaxation
in the metallic phase of several superconducting organic charge transfer
salts from the family, $\kappa$-(BEDT-TTF)$_2X$ \cite{YusufPRB07,PowellPRB09}.

The dynamic susceptibility is assumed to have the form
\begin{equation}
\chi({\bf q},\omega) = \frac{\chi_Q(T)}{1+\xi(T)^2|{\bf q}-{\bf
Q}|^2-i\omega/\omega_\sf(T)} 
\label{dynamic_af}
\end{equation}
where $\chi_Q(T)$ is the static spin susceptibility
at a non-zero wavevector ${\bf q}={\bf Q}$, $\omega_\sf(T)$ is the
characteristic spin fluctuation energy which represents damping in the
system near ${\bf q}={\bf Q}$, and $\xi(T)$ is the
temperature dependent correlation length. 
When there are long-range antiferromagnetic fluctuations,
(i.e., $\xi(T) \gg a$) 
the spin relaxation rate (\ref{t1t}) is given by
\begin{equation}
\frac{1}{T_1T} = 
\frac{2\pi k_B |A|^2 }{\gamma_e^2\hbar^4 q_c^2}
\frac{\chi_Q(T)}{\omega_\sf(T)\xi(T)^2}
\label{t1t_af1}
\end{equation}
where $q_c \sim  \pi/a$ is the cut-off wavevector
when one integrates over the Brillouin zone.
This expression can be simplified further by making the scaling assumptions
$\chi_Q =\alpha
(\xi/a)^{2-\eta}$ and $\omega_\sf = \alpha' (\xi/a)^{-z}$ where
$\alpha$ and $\alpha'$ are temperature independent constants and $a$ is
the lattice spacing. Following MMP \cite{millis}, we assume a
relaxational dynamics of the spin fluctuations, which are described by
a dynamic critical exponent $z=2$, and the mean-field value
of the anomalous critical exponent
$\eta=0$. Within
these approximations, the relaxation rate, can be written as
\begin{equation}
\frac{1}{T_1T} = 
\frac{ k_B |A|^2 }{\gamma_e^2\hbar^3}
\frac{\chi_Q(T)}{T_0}
\label{t1t_af2}
\end{equation}
where the temperature scale, $T_0$ defined by Moriya 
and Ueda \cite{MoriyaRPP03} is
\begin{equation}
T_0 \equiv\frac{\omega_\sf (q_c\xi(T))^2}{2 \pi} = 
\frac{\alpha'}{2 \pi} (q_c a)^2.
\label{t0def}
\end{equation}
In passing we note that this temperature scale
is of particular physical significance because
Moriya and Ueda \cite{MoriyaRPP03} find that  
for a wide range of unconventional superconductors their
transition temperature increases monotonically with $T_0$.

 We further assume that the temperature dependence of the correlation
length $\xi(T)$ is \cite{MoriyaRPP03,millis}
\begin{equation}
\frac{\xi(T)}{\xi(T_x)} = \sqrt{2T_x \over (T+T_x)}.
\label{xi-temp}
\end{equation}
 For this form, 
 $T_x$ represents a natural temperature scale and
$\xi(T)$ is only weakly temperature dependent for $T\ll T_x$.
The static susceptibility associated with the antiferromagnetic
fluctuations then has the temperature dependence 
\begin{equation}
\chi_Q(T) = \chi_Q(T_x) 
\frac{2T_x }{T+T_x}.
\label{chiq-temp}
\end{equation}
Then the relaxation rate has the temperature dependence (\ref{cw_t1t}) with
\begin{equation}
\left(\frac{1}{T_1}\right)_\infty
=
\frac{2 k_B |A|^2 }{\gamma_e^2\hbar^3}
\chi_Q(T_x) 
\frac{T_x}{T_0}.
\label{limit_t1}
\end{equation}

\subsubsection{Quantum critical spin fluctuation model}

Sachdev has interpreted the observed temperature dependence
of  $1/T_1$ in the cuprates La$_{2-x}$Sr$_x$CuO$_4$ in terms of 
quantum criticality
(see Figure 4 of \cite{SachdevScience00}).
He notes that theoretical calculations 
for non-linear sigma models associated with the
Heisenberg model on the square lattice that
in the quantum critical regime $1/T_1$ becomes
independent of temperature. 
Making this identification requires that there is a quantum
phase transition as a function of the doping near $x=0.075$.
A connection can be made to the MMP model if
we assume that $T_x$ defines the temperature scale above which
the crossover to quantum critical behavior occurs (cf. Figure \ref{fig:qcp}).

\subsubsection{Local spin fluctuation model}

The local spin fluctuation model presents a physically different
picture  for the high temperature relaxation rate
because the spin fluctuations are local,
in contrast, to the long-range fluctuations in the two models above.
It can be shown that, in the high-temperature limit,
 the uniform magnetic susceptibility of a spin-1/2 Heisenberg antiferromagnetic system is given by a Curie-Weiss expression.
For the triangular lattice this form holds down to much lower
temperatures than for the square lattice
because the frustration increases the domain of validity
of the single site approximation associated with the
Curie-Weiss theory \cite{ZhengPRB05,MerinoPRB06} (cf. Figure \ref{fig:susc}).
In the same high temperature
limit the NMR relaxation rate is given by \cite{MoriyaPTP56,SinghPRB90,gulley}
\begin{equation}
\left(\frac{1}{T_1}\right)_\infty
={ \sqrt{\pi} |A|^2 \over z \hbar J}
\label{t1-highT}
\end{equation}
where $A$ is the hyperfine interaction and $z$ is the co-ordination number of the lattice.
The derivation of this result involves a short time expansion of the
electronic spin correlation function, which is assumed
to decay in a Gaussian manner.
For Sb-1 it was shown that the magnitude of $1/T_1$ is
consistent with the  above expression
with independent estimates of $A$ (from the scale
of the susceptibility and the Knight shift) and $J$ 
(from the temperature dependence and magnitude of
the susceptibility) \cite{ItouPRB08}.


%% file: lattice.tex
\section{Quantum many-body lattice Hamiltonians}
 
\subsection{Heisenberg model for the Mott insulating phase}
\label{sec:heisenberg}

In Section \ref{ET-struct} we argued that from a quantum chemical perspective that
the simplest possible effective Hamiltonian for the organic
charge transfer salts is a Hubbard model on the anisotropic triangular
lattice at half filling. This means that in the Mott insulating
phase the spin degrees of freedom can be described by
a Heisenberg model on an anisotropic triangular lattice \cite{McKenzieCCMP98}.
This lattice can also be viewed as a square lattice with interactions along
one diagonal.
The Hamiltonian is
\begin{equation}
\hat{\cal H}=
J \sum_{\langle ij\rangle} \hat{\bf S}_i\cdot\hat{\bf S}_j
+
J' \sum_{\langle\langle ik\rangle\rangle} \hat{\bf S}_i\cdot\hat{\bf S}_k
\label{HamJJ'}
\end{equation}
where $J$ describes the exchange interaction in the vertical and horizontal 
directions and $J'$ is the interaction along the diagonal (compare
equation (\ref{h4site})).
This model interpolates between the square lattice ($J'=0$), the isotropic
triangular lattice ($J'=J$), and decoupled chains ($J=0$).
We will also consider  the above Hamiltonian
with an additional  ring-exchange interaction $J_\square$ 
for every square plaquette and given by (\ref{ringexch}).

Extensive studies of the above Hamiltonian have been made of
the case $J_\square=0$ and for 
$J'=J, J_\square \neq 0$,
which we review below.
We are unaware of any studies of the full two-dimensional model with 
$J' \neq J, J_\square \neq 0$.
Below we do briefly discuss some very recent studies of related two- and four-rung ladder models for these parameters.

Below we discuss studies giving a ground state with no magnetic order when
$J'/J \simeq 0.7-0.9, J_\square =0$, and
$J'=J, J_\square >0.05 J$.
In both cases there is an energy gap to the lowest lying triplet state.
In the former case the state
is a valence bond crystal with dimer order on horizontal (or vertical)
bonds.
In the second case there is believed to be no dimer order.
The observed ground state of P-1 is consistent with the
valence bond crystal state.
It is not possible to clearly identify the ground state of \kcn3
with either of these two non-magnetic  states. We discuss this 
in Section \ref{sec:aniso}, below.

\subsubsection{RVB states}

A recent field-theoretic perspective on RVB states and their
excitations has been given \cite{Sachdev09}. 
Becca {\it et al.} have given a nice review of
variational RVB
wave functions for frustrated Heisenberg spin models \cite{Becca09}.
They show that these wave functions become close to the true
ground state wave function as the frustration increases.
Figure \ref{fig:becca} shows this for the 
case of the frustrated Heisenberg
model (the $J_1-J_2$ model) on the square lattice. 
The upper panel shows the energy difference
between a projected BCS wave function and the exact ground state
for a lattice of $6 \times 6$ sites.
The lower panel shows the magnitude of the overlap of these two states.
A number of different numerical techniques find that there is
Neel order for $0 \leq J_2/J_1 \leq 0.5$
 and that there is no long-range magnetic order for 
 $0.5 \leq J_2/J_1 \leq 0.7$ \cite{Becca09}.

\begin{figure}\begin{centering}
\includegraphics[width=7.5cm]{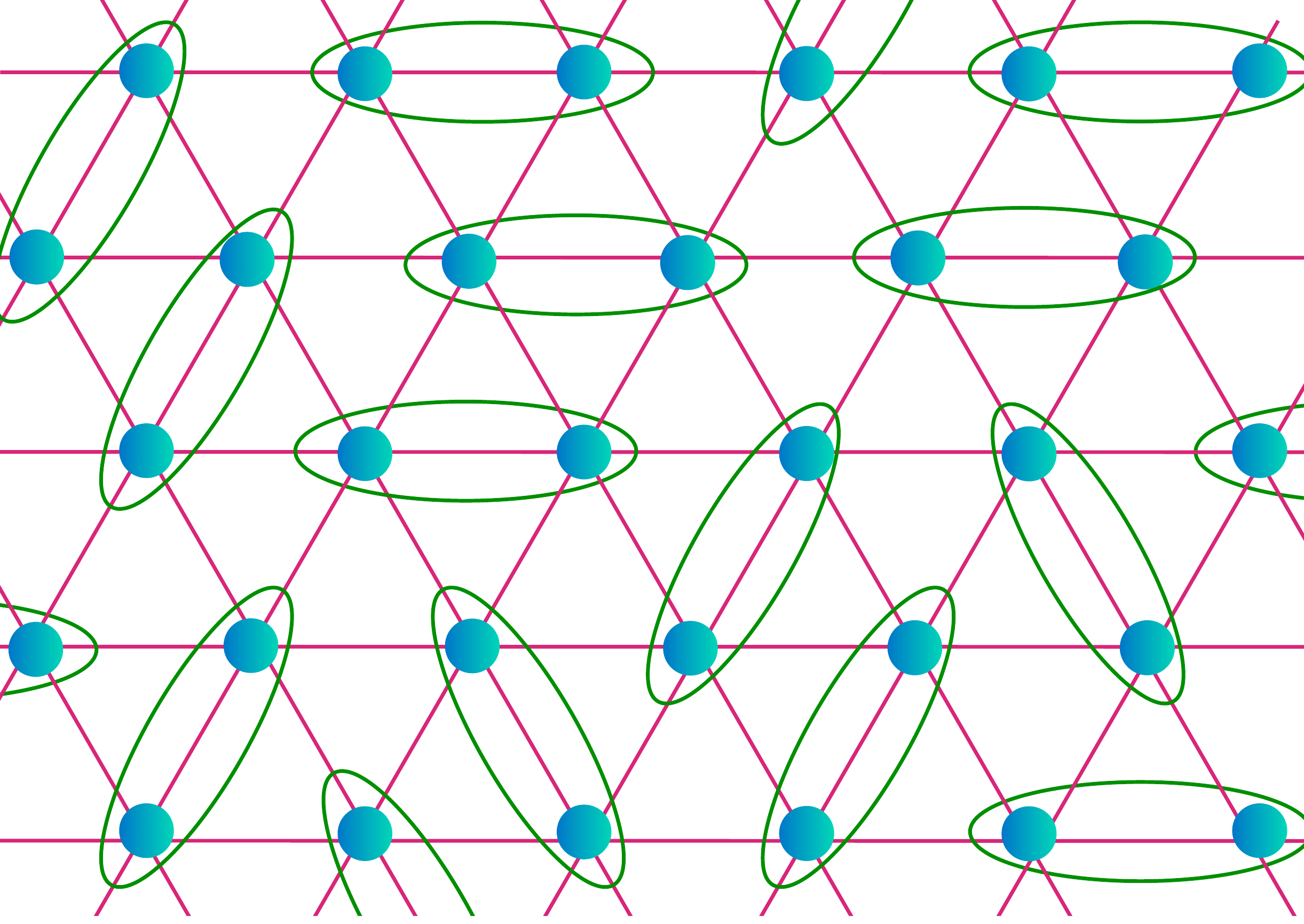}
\caption{
One possible dimer covering of the triangular lattice.
Each oval represents a singlet pairing of the spins
on the two sites
enclosed within the oval. Short range RVB states consist
of superpositions of such states.
(Figure provided by S. Sachdev).
}
\label{fig:dimercover}
\end{centering}
\end{figure}

\begin{figure}\begin{centering}
\includegraphics[width=\columnwidth]{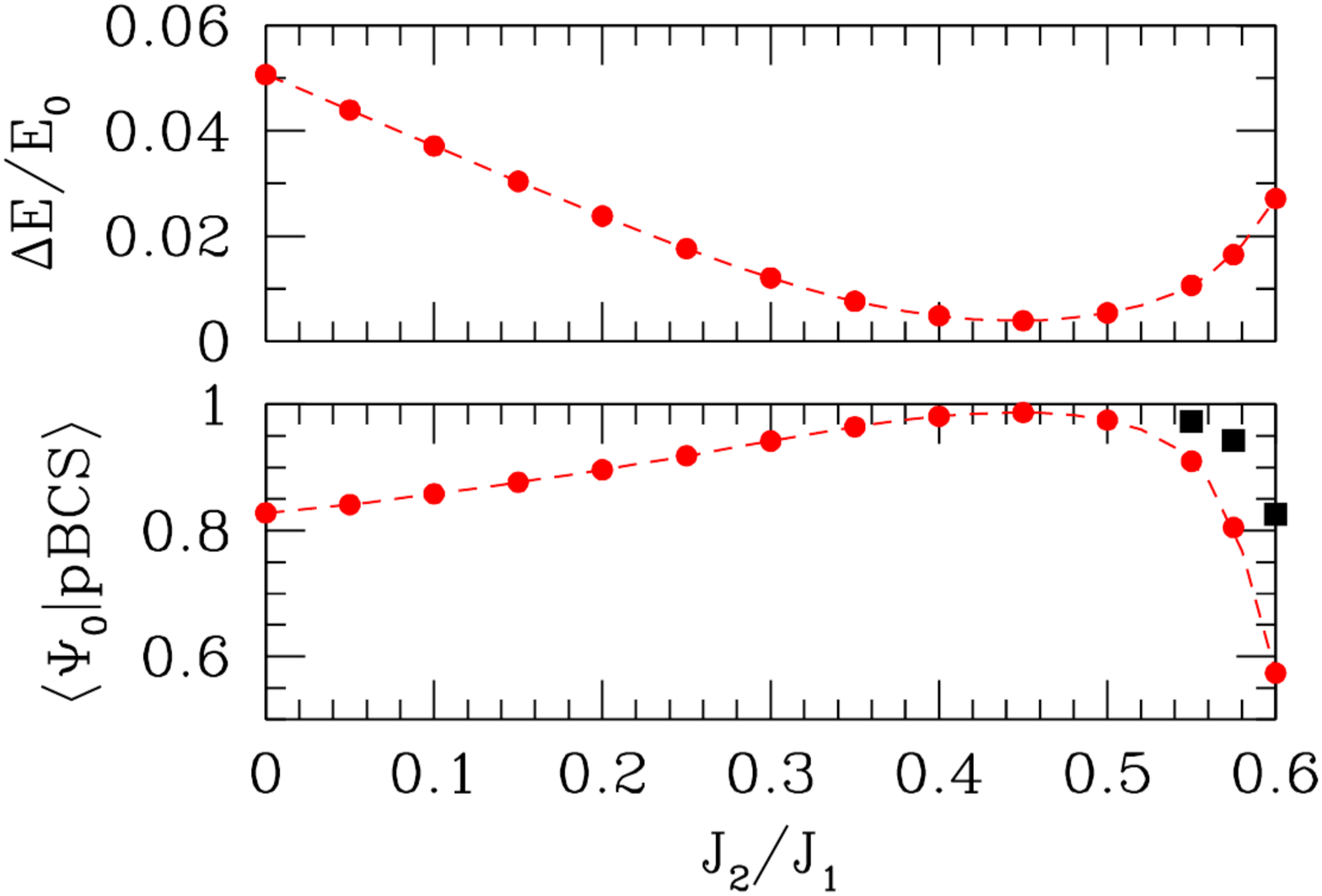}
\caption{
Frustration stabilises RVB and spin liquid states \cite{Becca09}. 
The upper panel shows the energy difference
between a projected BCS wave function and the exact ground state
for a lattice of $6 \times 6$ sites for the $J_1-J_2$ model on a square lattice.
($J_1$ and $J_2$ are the nearest and next-nearest neighbour interactions,
respectively).
The lower panel shows the magnitude of the overlap of these two states.
}
\label{fig:becca}
\end{centering}
\end{figure}

There are two main classes of RVB wave functions \cite{Becca09}.
States in the first class are sometimes called short-range RVB states.
They are similar    to the RVB states introduced by 
Pauling into quantum chemistry \cite{AndersonPT08,Shiak08}.
The simplest possible
state  consists of an equal superposition of all possible dimer coverings
of the lattice, $\{c\}$, where each dimer corresponds to a local spin singlet,
\begin{equation}
|\Psi_{SRVB}\rangle = \sum_{c} |\Psi_c\rangle.
\label{eqn:rvb}
\end{equation}
Figure \ref{fig:dimercover} shows one possible dimer covering of the triangular lattice.
Generalisations of the wavefunction (\ref{eqn:rvb})
 have unequal coefficients for the
different dimer coverings and also longer range
singlet pairings.
It is non-trivial but possible to show that
certain parametrisations of this
class become equivalent to the second class below \cite{Becca09}. 

The second class of RVB wave functions
consist of Gutzwiller projected BCS states similar to that
 first introduced by Anderson \cite{AndersonScience87}
\begin{equation}
|\Psi_{pBCS} \rangle = 
\Pi_i (1 - \alpha n_{i\uparrow} n_{i\downarrow})
|BCS\rangle  
\label{eqn:rvb2}
\end{equation}
where 
$|BCS\rangle $ is a BCS state with a variational pairing function
and $\alpha$ is a Gutzwiller variational parameter ($0 \leq \alpha \leq 1$)
which determines the number of doubly occupied sites.
$\alpha=1$ when no doubly occupied states are allowed.

\subsubsection{Isotropic triangular lattice}

The clear consensus from a wide range
of studies is that the true ground state for $J'=J$, $J_\square=0$ is 
not a spin liquid but a Neel antiferromagnet
with 120 degree order.
Table III in \cite{ZhengPRB06} gives a summary
of the results and relevant references
from studies using a wide range of numerical methods and approximation schemes.
Thus, the conjecture of Anderson and Fazekas
\cite{AndersonMRS73,FazekasPM74}
that this model has a spin liquid ground state turns out to be
incorrect. Nevertheless,
variational short-range RVB states have been found to be 
close to the exact ground state for small lattices \cite{SindzingrePRB94}.
Specifically, they give comparable short-range spin correlations.
For example, for a 12-site lattice the nearest, next-nearest neighbour,
and next-next-nearest neighbour spin correlations for
the exact ground state are 
$\langle \vec{S}_r \cdot \vec{S}_0 \rangle =$
 -0.2034, 0.1930, and -0.0511, respectively.
For comparison an equal superposition RVB state gives values
of -0.2032, 0.2065, and -0.075.
Furthermore, we will see below that small perturbations of the Hamiltonian, such as spatial anisotropy or ring exchange terms can lead to a ground state
with no magnetic order.

\subsubsection{Role of spatial anisotropy ($J' \neq J$)}
\label{sec:aniso}

\begin{figure}\begin{centering}
\includegraphics[width= \columnwidth]{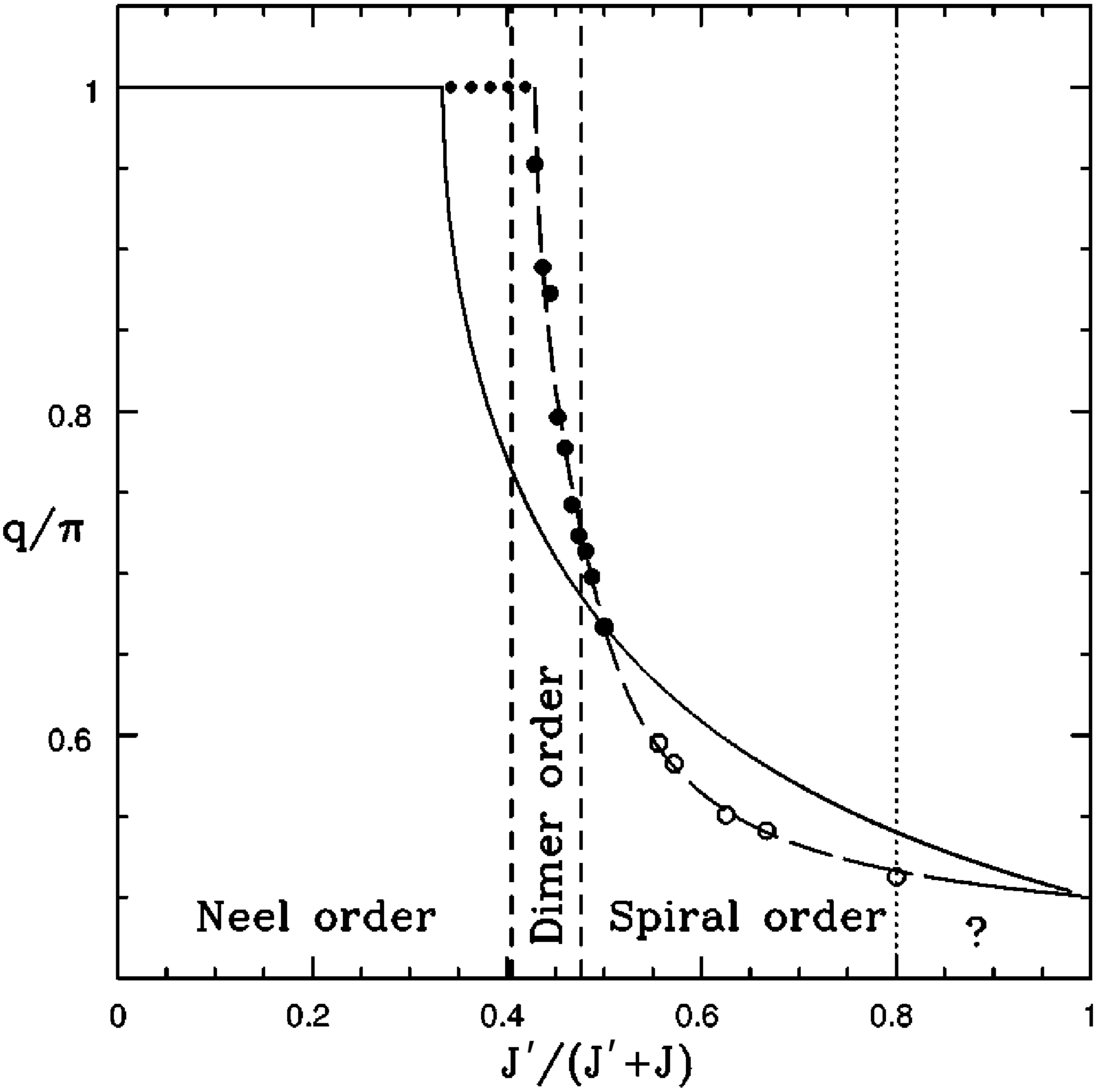}
\caption{
Phase diagram and incommensurability
of spin correlations for
 the Heisenberg model on the anisotropic triangular lattice \cite{ZhengPRB99}. 
The solid  curve shows the classical ordering wavevector $(q,q)$ 
 as a function of the diagonal interaction $J'/(J'+J)$.
The circles show the value of $q$ which defines the location
of the minimum energy gap to a triplet excited state.
These are  calculated from  series expansions relative 
to different reference states \cite{ZhengPRB99}.
Incommensurate (spiral) antiferromagnetic order occurs for $J' >0.95J$.
The antiferromagnetic order is unstable for the range $0.6 <  J'/J < 0.95$.
The series cannot determine the ground state for weakly coupled chains
with $J'/J > 4$. Modified from \cite{ZhengPRB99}.
[Copyright (1999) by the American Physical Society.]
}
\label{fig:zhengphased}
\end{centering}
\end{figure}

It turns out that spatial anisotropy can destroy the magnetic
order present when $J'=J$.
The model Hamiltonian (\ref{HamJJ'})  has been studied by a wide range of techniques:
linear spin-wave theory \cite{MerinoJPCM99,TrumperPRB99},
series expansions \cite{ZhengPRB99,FjaerestadPRB07}, 
large-N expansion of an sp(N) Schwinger boson theory \cite{ChungJPCM01},
mean-field RVB theory \cite{HayashiJPSJ07},
variational Monte Carlo of Gutzwiller projected
BCS states \cite{YunokiPRB06,HeidarianPRB09}, 
pseudo-fermion functional renormalistion group \cite{Reuther10}
and the density matrix renormalisation group \cite{WengPRB06}.
The weakly coupled chain regime $J' \gg J$ has
been studied by perturbing about the exact
ground state single chains \cite{KohnoNatPhys07}.
Most studies agree that for $0 \leq J'/J \lesssim 0.5$ the Neel state
with ordering wavevector $(\pi,\pi)$ is stable and that a spiral
ordered state is stable for $J' \sim J$. However, whether the ground
state is a spin liquid in the regimes $J' \gg J$ and $J'/J \sim 0.6-0.9$ is controversial.
The phase diagram of the model deduced from
series expansions \cite{ZhengPRB99} is shown 
in Figure \ref{fig:zhengphased}.
Comparing the solid and dashed curves in
 Figure \ref{fig:zhengphased} shows that quantum
fluctuations tend to make the excitation spectrum more
commensurate than the order found in the classical Hamiltonian.
In particular, deviations of the wavevector
$\vec{Q}=(q,q)$ from the commensurate values
$q=\pi, 2\pi/3, \pi/2$, are reduced.
The reduction of deviations
from commensurability by quantum fluctuations is also found in renormalisation group
analysis of the corresponding non-linear sigma models \cite{ApelZPB92}.
There, it is found that quantum fluctuations
drive the system towards a fixed point with  $O(4)$ symmetry,
and at which the spin wave anisotropy is reduced.
Hence, quantum fluctuations will reduce the incommensurability.
This is an example of ``order through disorder'' as the ``disorder"
due to the quantum fluctuations stabilises the ``order" associated
with commensurate spin correlations \cite{ChandraJPCM90,ChubukovPRB92}.

\begin{figure}\begin{centering}
\includegraphics[width=\columnwidth]{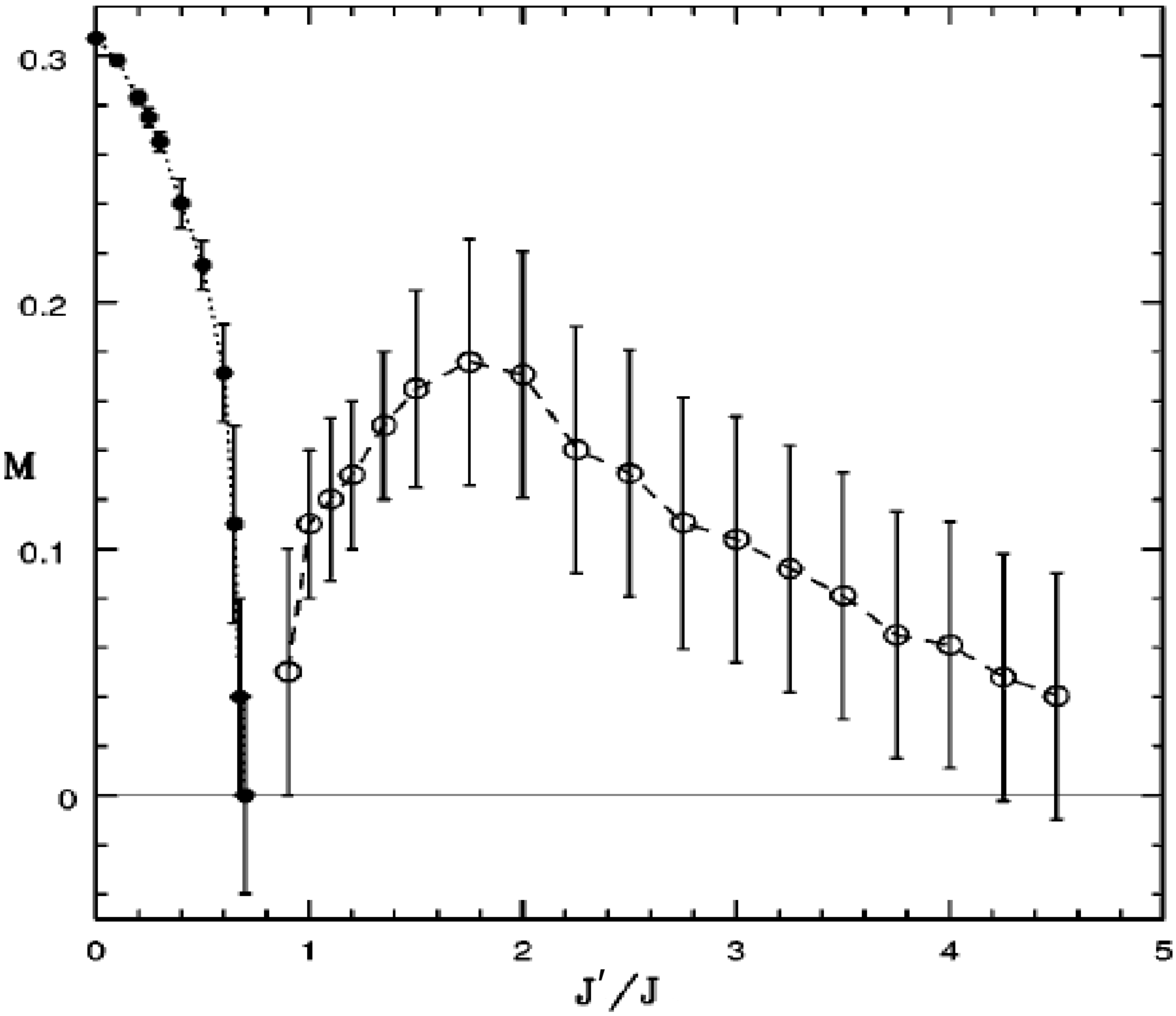}
\caption{
Magnitude of the magnetic order parameter (magnetic moment)
 as a function of the diagonal interaction $J'/J$.
[In our notation $J_1/J_2=J'/J$.]
This is calculated from a series expansion \cite{ZhengPRB99}. 
For $J' < 0.6J$ the frustration associated with $J'$ reduces the magnitude of the Neel ordering.
Incommensurate (spiral) antiferromagnetic order occurs for $J' >0.95J$.
The antiferromagnetic order is unstable for the range $0.6 <  J'/J < 0.95$.
This behaviour is also qualitatively 
reproduced by linear spin-wave theory \cite{MerinoJPCM99,TrumperPRB99}.
This shows just how sensitive the ground state is to the
spatial anisotropy.
This is the parameter regime relevant
to  many of the compounds discussed in this review.    
[Copyright (1999) by the American Physical Society.]
}
\label{fig:zhengmoment}
\end{centering}
\end{figure}

\begin{figure}\begin{centering}\includegraphics[width= \columnwidth]{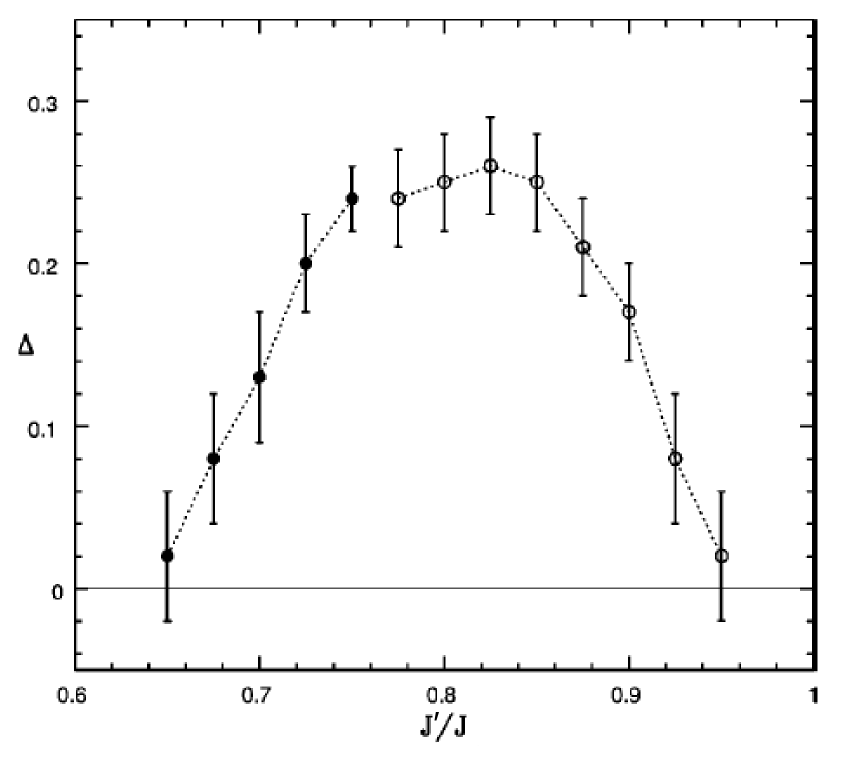}
\caption{
Dependence of the triplet energy gap on the diagonal coupling in the valence bond solid phase (with dimers in either the horizontal or vertical direction relative to the "square" lattice, i.e., dimers are associated with the $J$ interaction), calculated in a dimer series expansion \cite{ZhengPRB99}. 
 For $0.65 < J'/J < 0.75$ the gap occurs at
the wavevector $(0,\pi)$, whereas for larger $J'/J$ it is at an incommensurate wavevector. Together with Figure \ref{fig:zhengmoment} 
this suggests that the model
undergoes quantum phase transitions at $J'/J \simeq$ 0.65 and 0.95. Modified from  \cite{ZhengPRB99}.
[Copyright (1999) by the American Physical Society.]
}
\label{fig:zhenggap}
\end{centering}
\end{figure}

Figure \ref{fig:zhengmoment} shows the magnitude of the magnetic moment
associated with the magnetic order,
calculated  from series expansions.
For the range $0.65 < J'/J < 0.95$ the most stable state has
dimer order, a valence bond crystal with bonds along either
the horizontal or vertical direction (i.e., associated with the $J$ interaction). There is an energy gap to the lowest triplet excited state, and its magnitude
as a function of $J'/J$ is shown in Figure \ref{fig:zhenggap}.

The ground state of the dmit material P-1 is reminiscent of this
valence bond crystal state. The pattern of bond ordering 
is consistent with that deduced from x-ray scattering (see Figure \ref{fig:VBC-P-1})
and the measured energy gap [$40 \pm 10 K \simeq (0.15 \pm 0.05)J$
(see Figure \ref{fig:sus-P-1})]
is comparable to that shown in Figure \ref{fig:zhenggap}.
This identification would require that the hopping integrals
in the corresponding Hubbard model have $t'/t \simeq 0.8-0.95$,
which is significantly smaller than the value of
1.1 \cite{ShimizuJPCM07} estimated from H\"uckel calculations.
However, DFT calculations for the $\kappa$-(ET)$_2$X family \cite{KandpalPRL09,NakamuraJPSJ09} 
find that the H\"uckel method tends to over-estimate this ratio.
This underscores the need for DFT calculations on the dmit materials.

It is not clear that the possible spin liquid state in \kcn3 can
be identified with this valence bond crystal (VBC) state.
The NMR, specific heat, and thermal conductivity show no evidence
of a spin gap at temperatures less than 1  K which is two orders of magnitude
smaller than $J$. The only possibility consistent with the existence
of the VBC phase would be that this material lies
close to one of the quantum critical points associated with the transition
between the VBC and the magnetically ordered states. 

\subsubsection{Ring exchange}
\label{sec:ring}

As the band-width controlled Mott metal-insulator transition
is approached from the metallic side the charge fluctuations
increase and the average double occupancy increases (see, for example,
Figure 4 in \cite{OhashiPRL08}).
Hence, for a Mott insulating state close to the metallic phase
in the Hamiltonian one needs to include
the ring exchange terms which arise from the charge fluctuations.
A strong coupling expansion to fourth order in $t/U$ for the Hubbard model gives
$J_\square/J=5(t/U)^2$ \cite{YoshiokaPRB90,DelannoyPRB05,Yang10}.


Even well into the Mott insulating phase the ring exchange terms
can have both qualitative and quantitative effects. For example,  in La$_2$CuO$_4$ ring exchange interactions modify the dispersion relation of triplet spin excitations near the zone boundary \cite{ColdeaPRL01}.

The Heisenberg  model on the isotropic triangular lattice
(i.e. $J'=J$) with ring exchange has been studied using exact diagonalisation \cite{LimingPRB00},
a Gutzwiller projected Fermi sea of spinons \cite{motrunich},
and variational Monte Carlo for projected BCS states \cite{GroverPRB10}. 
Exact diagonalisation calculations on lattices of up to
36 sites suggest these ring exchange terms
 can lead to a spin liquid ground state for $J_\square > 0.05J$ \cite{LimingPRB00,slreviews}.
Furthermore, in this state there are a large number of singlet
excitations below the lowest energy triplet excited state \cite{LimingPRB00}.
The presence of a spin gap for triplet excitations
(estimated to be about $0.07J$ for $J_\square \simeq 0.1J$)
implies that the NMR relaxation rates would approach
 zero exponentially with decreasing temperature.

It has also been proposed that the ground state of \CN
is a spin liquid with a spinon Fermi surface \cite{motrunich,leelee}, with
spinons that are coupled to gauge field fluctations.
This is discussed further in Section \ref{sec:signature} below.

Variational Monte Carlo calculations were recently
performed for Gutzwiller-projected BCS states where different
possible types of pairing were considered \cite{GroverPRB10}.
 A mean-field theory on these states for the 
Heisenberg model without ring exchange
gives a BCS state with broken time-reversal symmetry (known as the chiral spin liquid). The form of the fermion pairing function is $d_{x^2-y^2} + i d_{xy}$ which belongs to the $E$ representation of the $C_{6v}$ point group symmetry of the lattice.
We now summarise some of the main results of the variational Monte Carlo
 study \cite{GroverPRB10}.
(i) In contrast to mean-field theory, it is found that
for  ring exchange strengths in a small range 
the pairing function is purely $d_{x^2-y^2}$.
(ii)
The authors suggest that under increasing 
pressure (increasing $t/U$) the Mott insulator will be destroyed leading to a superconducting state with the same
 $d_{x^2-y^2}$ pairing. This is qualitatively different to what one gets with a mean-field RVB theory of the model without the ring exchange \cite{PowellPRL07}.
(iii) This ground state
 cannot explain why the observed low temperature specific heat of 
\kcn3
 is weakly dependent on magnetic field.
(iv)
The ``Amperean pairing'' theory proposed earlier \cite{leelee2}
 does not have this problem, but has difficulty describing the superconducting state which develops under pressure.

Several important issues are not addressed in this paper.
First, in the range $J_\square/J \simeq 0.05-0.1   $
exact diagonalisation calculations on small lattices  give
 a different spin liquid ground state, one with an energy gap to triplet excitations, and many singlet excitations inside the gap \cite{Yang10}.
Second, the NMR relaxation rate $1/T_1$ for 
\kcn3 is observed to have a power law temperature dependence consistent
 with gapless excitations.
 Third, the estimate of the spinon scattering rate 
due to impurities ($\sim 1.5$ K) can be
 compared to estimates of the scattering rate associated with
charge transport in the metallic phase of 
similar organic charge transfer salts. Table I in \cite{PowellPRB04}
 gives estimates of this scattering rate
 which are an order of magnitude smaller
than the proposed spinon scattering rate \cite{GroverPRB10}.
However, given that the quasi-particles being scattered are different we should
not necessarily expect them to have the same scattering rate, but a complete theory should describe this difference.

For a Hubbard model the requirement
$J_\square/J > 0.05 $ needed for a spin liquid ground state
corresponds to $U <10|t|$. A crucial
question is then how large is the critical value  $U_c/|t|$ at
which the metal-insulator transition occurs?
In order for a Heisenberg model to be relevant the system
must still be  in the insulating phase for $U <10|t|$.
Below, we see that most estimates of $U_c/|t|$ lie in the range
5-8, depending on the numerical method used.
Spatial anisotropy (i.e., $t'<t$) reduces the critical value.
A comparison \cite{MerinoPRL08}
 of the measured optical conductivity for the alloy
$\kappa$-(BEDT-TTF)$_2$Cu[N(CN)$_{2}$]Br$_{x}$Cl$_{1-x}$
with $x=0.73$ (which lies just on the metallic side of the Mott transition)
with that calculated from dynamical mean-field field theory (DMFT)
found good agreement for $U/|t| \simeq 10$ (see, also, the discussion in section \ref{sect:opt-cond}).
Hence, this is roughly consistent with the ring exchange term being
sufficiently large to     produce a spin liquid state.

Motrunich found    a low
variational energy for a  projected spinon Fermi sea state \cite{motrunich}.
He combined this with a slave particle-gauge theory analysis, to argue that the
 spin liquid has spin correlations that are singular along surfaces in momentum space, i.e., ``Bose surfaces.''
A density matrix renormalisation group (DMRG) study of a frustrated
 ladder model (corresponding to two coupled chains in the anisotropic
triangular lattice model)
 with ring exchange has produced a rich phase diagram \cite{ShengPRB09}.
 In particular, when $J \sim J'$ a ring
exchange interaction $J_4 \sim 0.2J$ can lead to a ``spin-Bose metal'' phase.
This is a spin liquid state with gapless excitations at specific wave vectors.
This underscores the need for a study of the full Heisenberg model
with both $J' \neq J$ and $J_\square \neq 0$.
A recent study was made of the model on a four rung ladder
with $0 \leq J'/J \leq 1$ and $0 \leq J_\perp \leq J$
using DMRG and variational Monte Carlo of a projected Fermi sea \cite{Block10}. 
The phase diagram contained rung, VBC, and spin-Bose metal phases.
The latter has three gapless modes and power law spin correlations
at incommensurate wavevectors.
Spatial anisotropy increased the stability of
the VBC state.

A recent definitive study \cite{Yang10}
of the Hubbard model on the isotropic triangular lattice
 used  a high powered perturbative continuous unitary transformation to derive an effective spin Hamiltonian in the Mott insulating phase, up to 
twelth order in $t/U$.
They find that as $U/t$ decreases, at $U/|t| \simeq 10$,
there is a first-order phase transition from the 120 degree Neel ordered phase to a spin liquid phase (no net magnetic moment) and large numbers of singlet excitations below the lowest lying triplet excitation. 
This spin liquid state is identified
 with the "spin Bose metal" proposed by Motrunich \cite{motrunich}.
The first-order transition from the magnetically ordered state to the spin liquid is also associated with a small jump in the double site occupancy.
It is also found that the transition to the metallic state does not occur until $U/t$ decreases to about $6-8$. Hence, there is a significant range of $U/t$ for which the Mott insulator is a spin liquid.

\subsubsection{Dzyaloshinski-Moriya interaction}

In crystals which lack inversion symmetry relativistic effects
lead to an additional interaction between spins which breaks spin-rotational 
invariance and is known as the Dzyaloshinski-Moriya interaction.
The DM interaction has been characterised in the insulating phase
of \Cl 
and has a magnitude of about $D \simeq 5$ K \cite{SmithPRL04}.
Even though it is small compared to the nearest neighbour
exchange the DM interaction can have a significant effect on frustrated systems.
For example, for the kagome lattice it can induce a quantum phase transition
from a spin liquid state to an ordered state for $D>0.1J$ \cite{CepasPRB08}.
For the anisotropic triangular lattice,
even when $D  \sim J/20$ the DM interaction induces energy changes in the spectrum of energies as large as $J/3$, including new energy gaps \cite{FjaerestadPRB07}.
A detailed analysis of the effect of the DM interaction in the weakly
coupled chain limit has also been given \cite{StarykhPRB10}.


\subsubsection{The effect of disorder}

Gregor and Motrunich \cite{GregorPRB09} studied the
 effects of nonmagnetic impurities in the Heisenberg
model on the triangular lattice with the goal
of understanding the large broadening of $^{13}$C NMR lines in 
$\kappa$-(BEDT-TTF)$_2$Cu$_2$(CN)$_3$.            
They used a high-temperature series expansion to 
 calculate the local susceptibility near a nonmagnetic impurity,
for temperatures  down to $J/3$.
At low temperatures they assumed a gapless spin liquid described by
a Gutzwiller projected spinon Fermi sea.
In both temperature regimes, they found that the value of
the local susceptibility decays to the uniform value
within a few lattice spacings. Hence   a low density of impurities cannot
explain the observed line broadening.
This analysis needs to be combined with independent estimates
of the strength of disorder in these materials \cite{ScrivenJCP09}.

\subsection{Hubbard model on the anisotropic triangular lattice}
\label{sec:hubbard}

The Hamiltonian (\ref{eq:aniso}) depends on three parameters: $t$, $t'$, and $U$.  Estimates of values for these parameters
from quantum chemistry and electronic structure
calculations were discussed in Section \ref{sec:dimer}.
The key open questions concerning the model are whether
it has superconducting and spin liquid ground states
for physically reasonable parameter values.

\subsubsection{Phase diagram}

We have already discussed the phase diagram at non-zero temperature in section \ref{sec:undergrad}. 
The zero temperature phase diagram of the Hubbard model on the anisotropic triangular lattice has also been studied by a wide range of techniques including:
exact diagonalization \cite{KoretsuneJPSJ07,ClayPRL08}, slave bosons/RVB mean-field theory \cite{zhang,PowellPRL05,PowellPRL07},
large-N expansion of a sp(N) theory \cite{ChungJPCM01},
weak-coupling renormalisation group \cite{TsaiCJP01},
variational quantum Monte Carlo on Gutzwiller projected BCS states \cite{trivedi,WatanabePRB08,TocchioPRB09}
cluster and cellular dynamical mean-field 
theory \cite{ParcolletPRL04,KyungPRL06,KyungPRB07,OhashiPRL08,LiebschPRB09},
slave rotor representation \cite{leelee,leelee2},
path-integral renormalisation group \cite{MoritaJPSJ02},
cluster variational approach \cite{SahebsaraPRL08},
and dual fermions \cite{LeePRB08}.

There is little consensus on the phase diagram in the physically
important region near the Mott transition, and particularly where there
are several competing magnetic phases (i.e., $0.7t <t'<t$).
This lack of  consensus arises for two reasons: one mundane and the other profound. The first is that not all approaches allow for all possible 
states. The second is that
there are very small differences in energy
between the competing phases.
Different computational methods and approximation schemes
 will get different values for these small differences
in energy and so produce different phase diagrams.

The fact that in the organic charge transfer salts there
 is a first order phase transition between superconducting
and Mott insulating states shows that these two very different states can 
have identical energy.
This is actually why these materials are so ``tuneable''
 (i.e. one can induce transitions between different phases with ``small'' changes in the pressure, temperature or magnetic field, and by chemical substitution).

\begin{figure}\begin{centering}
\includegraphics[width=9cm]{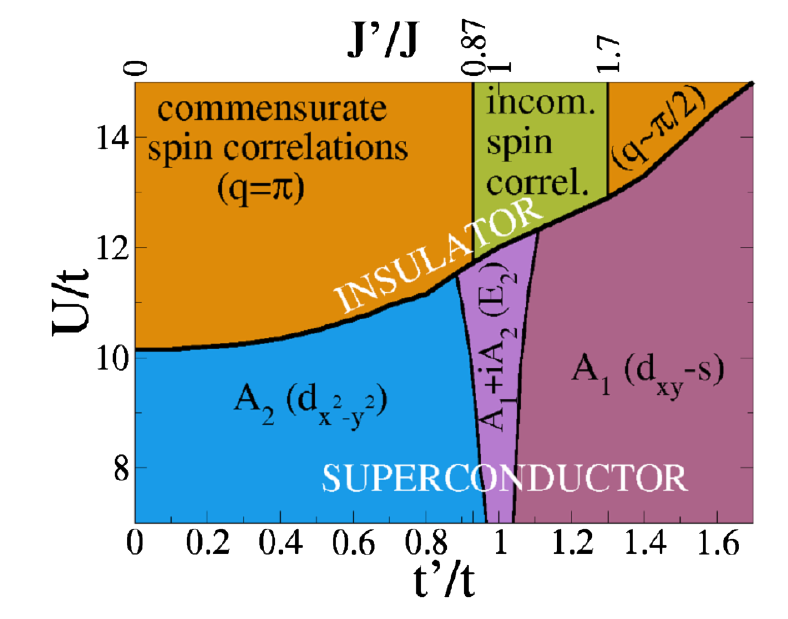}
\caption{
Phase diagram of the Hubbard model on the anisotropic triangular lattice 
at half filling and zero temperature calculated from an RVB variational wave function \cite{PowellPRL07}.
Similar results are found for other strong coupling approaches. 
The $t'/t=0$, $t'/t=1$ and $t'/t\rightarrow\infty$ correspond to the square lattice, the isotropic triangular lattice and
decoupled chains, respectively.
For large $U/(t+t')$ the ground state is a Mott insulator.
The type of magnetic ordering varies with $J'/J=(t'/t)^2$.
Near the metal-insulator phase boundary superconductivity occurs and
the symmetry of the superconducting order parameter is correlated with the
dominant spin correlations in the neighbouring Mott insulating phase. 
[Copyright (2007) by the American Physical Society.]
}
\label{fig:rvbphasediag}
\end{centering}
\end{figure}

\subsubsection{Ladder models}

Ladder models provide a means to investigate in a controlled
manner (e.g., via DMRG, bosonisation, and weak-coupling renormalisation
group) 
physics which it is hoped may be related to what occurs in 
the two-dimensional limit.

One can characterise different ground states on ladders by $n$ and $m$, the
number of gapless charge and spin modes, respectively. This leads to the 
notation $CnSm$ and the following identifications. 
$C2S2$ is the ladder analogue of Fermi liquid metal,
$C1S0$ is a superconductor,
$C1S2$ is a spin Bose metal,
$C0S0$ is a spin gapped Mott insulator,
and
$C0S2$ is a spin liquid Mott insulator.
At half filling a two-leg ladder without frustration
has a $C0S0$ ground state, which upon doping changes
to $C1S0$ consistent with Anderson's RVB ideas \cite{BalentsPRB96}. 

A weak-coupling renormalisation group analysis has been
performed on the zig-zag ladder with longer-range Coulomb repulsion \cite{LaiPRB10}.
The longer-range interaction stabilises the $C2S2$ phase and
leads to a subtle competition between all the different phases listed above.
Indeed it     is interesting to compare Figure 8 in \cite{LaiPRB10} with our
Figure \ref{fig:rvbphasediag}.

%% file: gauge.tex
\section{Emergence of gauge fields and fractionalised quasi-particles}
\label{sec:gauge}

For a given phase the key question (or assumption) is:
what are the quantum numbers of the quasi-particles describing
the lowest lying excited states? The answer determines the nature and symmetries of an effective field theory for the low-energy physics.
Field theories for magnons (bosonic triplets), spinons (spin-1/2 bosonic 
or fermionic excitations), and visons (bosonic singlets) have all been considered in various different theories of the organic charge transfer salts.

In a ``round table discussion''
 about the theory of the cuprate superconductors \cite{ZaanenNP06}
Patrick Lee stated that the genuinely new idea that has been developed
is ``the notion of emergence of gauge fields and fractionalized particles 
as low-energy phenomena in systems that did not contain them in the starting model.''
He suggested that this idea is of comparable importance in
condensed matter theory to that
of Goldstone bosons associated with spontaneously broken symmetry.

Gauge fields emerge when the electron or spin operators are represented
 in  terms of Schwinger bosons \cite{Auerbach}, slave fermions,
 slave bosons \cite{LeeRMP06}, or slave rotors \cite{FlorensPRB04}.
These alternative representations introduce an over-complete
description of the problem which requires a constraint so that
the canonical commutation (or anti-commutation) relations are preserved.

A nice discussion (for the specific case of Schwinger
bosons) is contained in a review \cite{SachdevNP08}, which
we now follow closely. It shows clearly how the effect of gauge field
fluctuations leads to qualitative differences in
the quantum disordering of commensurate and incommensurate
magnetic states. In particular, deconfinement of bosonic spinons is possible
in the latter but not the former.

We illustrate this now by showing how, if the spin-1/2 operators are represented by Schwinger bosons \cite{Auerbach},
there is a redundancy because the phase of each boson field can be shifted by an arbitrary amount without affecting the spin degree of freedom.
The spin 1/2 field ${\bf N}$ can be written in terms of a
$S=1/2$ complex spinor boson field $z_\alpha$,
where $\alpha = \uparrow,\downarrow$ by
\begin{equation}
{\bf N} = z_\alpha^\ast \boldsymbol{\sigma}_{\alpha \beta} z_\beta
\label{Phiz}
\end{equation}
where $\boldsymbol{\sigma}$ are the $2\times2$ Pauli matrices.
The spin commutation relations  are
preserved provided that\footnote{See page 70 of \cite{Auerbach}.} 
\begin{equation}
\sum_\alpha z_\alpha^\ast z_\alpha = 1.
\label{constraint}
\end{equation}
As an aside, we note that there is no problem having spin-1/2 bosons.
The spin statistics theorem in relativistic quantum field theory (which 
requires bosons to have integer spin) does not apply here because
in this field theory there is no Lorentz invariance.
Also, in what follows we are always considering ${\bf N}$ to
be a slowly varying field which defines the  
spin at site $j$ relative to the commensurate wave vector 
for Neel ordering ${\bf Q}=(\pi,\pi)$,
\begin{equation}
\langle {\bf S}_j \rangle = 
 {\bf N} \cos ({\bf Q} \cdot {\bf r}_j). 
 \label{SNn}
\end{equation}
where ${\bf r}_j$ is the position of site $j$.

But, note that the representation (\ref{Phiz}) of ${\bf N}$ in
terms of  $z_\alpha$ has
some redundancy. In particular, a  change in the phase of
both fields $z_\alpha$ by the same
space and time dependent field $\theta(x,\tau)$
\begin{equation}
z_\alpha \rightarrow
z_\alpha 
\exp(i \theta)
\label{gauge}
\end{equation}
leaves  ${\bf N}(x,\tau)$ unchanged.
 All physical properties must then be invariant under 
the transformation (\ref{gauge}), and so any effective Lagrangian
for the field $z_\alpha$ has a U(1) gauge invariance, 
similar to that      in quantum electrodynamics. 
This leads naturally to
 the introduction of an `emergent' U(1) gauge field 
$A_\mu$,  where the index $\mu$ describes the 2+1 space-time components.
Under the gauge transformation (\ref{gauge}), 
$A_\mu \to A_\mu - i\partial_\mu \theta$.
It should be stressed that this gauge field is
\emph{not} related to the physical electromagnetic field but rather
is an alternative way of describing the interactions
between the spinor fields due to the antiferromagnetic fluctuations.
Describing the system in terms of the field $\vec{N}$ or 
the two fields $z_\alpha$ and $A_\mu$ is a matter of choice.

The low-energy and long-wavelength action of the quantum field theory for 
$z_\alpha$ and $A_\mu$ is determined by 
the constraints of spin rotational symmetry and gauge invariance to be
\begin{eqnarray}
\mathcal{S}_z &=&  \int d^2 r d \tau \biggl[
|(\partial_\mu -
i A_{\mu}) z_\alpha |^2 + s_z |z_\alpha |^2  + u (|z_\alpha |^2)^2 \notag\\&&\hspace{2cm}+ \frac{1}{2e_0^2}
(\epsilon_{\mu\nu\lambda}
\partial_\nu A_\lambda )^2 \biggl]
 \label{Sz}
\end{eqnarray}
where there is an implicit summation over all indices, 
$ \epsilon_{\mu\nu\lambda}$ is the anti-symmetric tensor and
$e_0$ is the coupling constant, which determines the strength
of the coupling between the $z_\alpha$ field and the gauge field.

If the coefficient $s_z <0$ then the mean-field theory of the action gives 
a ground state with
 $\langle z_\alpha \rangle \neq 0$.
Substituting this into (\ref{Phiz}) and (\ref{SNn}) we see that 
this corresponds to a state
with commensurate antiferromagnetic order. This also
leads to a gap in the spectrum 
of the   $A_\mu$ gauge field, and reduces its fluctuations.
The case $s_z >0$ gives $\langle z_\alpha \rangle = 0$.
A rather sophisticated analysis is required to show that the 
gauge field fluctuations are associated with Berry's phases which
lead to VBC order \cite{SachdevNP08}.

A key point is that U(1) gauge fields in 2+1 dimensions are always confining \cite{KogutRMP}.
 This is because of instantons which describe the
quantum tunneling of the gauge field between alternative classical 
ground states.
 The physical consequence of this for commensurate antiferromagnets is
that the spinons are always bound together so that the elementary
excitations are spin-1 bosons.

\subsection{Spinons deconfine when incommensurate
phases are quantum disordered}

A ground state with incommensurate magnetic order
can be described by 
two orthogonal vectors, ${\bf N}_1 $ and $ {\bf N}_2$
so that the magnetic moment  at site $j$ (with position 
${\bf r}_j$) is \cite{ChubukovNPB94}
\begin{equation}
\langle {\bf S}_j \rangle = 
 {\bf N}_1 \cos ({\bf Q} \cdot {\bf r}_j) + 
{\bf N}_2 \sin ({\bf Q} \cdot {\bf r}_j) \label{SN}
\end{equation}
where ${\bf Q}$ is the incommensurate ordering wavevector.

The analog of the spinor representation in Eq.~(\ref{Phiz}), is to 
introduce another spinor $w_\alpha$, which parameterizes ${\bf N}_{1,2}$
 by \cite{ChubukovNPB94}
\begin{equation}
{\bf N}_1 + i {\bf N}_2 = \varepsilon_{\alpha\gamma} w_\gamma \boldsymbol{\sigma}_{\alpha\beta} w_\beta,
\label{Nw}
\end{equation}
where $\varepsilon_{\alpha\beta}$ is the antisymmetric tensor.
The physical spin is then invariant under the $Z_2$ gauge transformation 
\begin{equation}
w_\alpha \rightarrow \eta w_\alpha \label{z2w}
\end{equation}
where $\eta (r,\tau) = \pm 1$.
This $Z_2$ gauge invariance is key to stabilising a spin liquid ground
state because it reduces the magnitude of the U(1) gauge field
fluctuations which confine the spinons in antiferromagnets
with commensurate interactions. 
In contrast to U(1) gauge theories a $Z_2$ gauge theory
 can have a deconfined phase in 2+1 dimensions  \cite{KogutRMP}.
We now introduce  a Higgs scalar field, the condensation of which, 
$\langle \Lambda \rangle \neq 0$ 
can break the U(1) symmetry, in
a similar manner to that in which the BCS superconducting 
state   breaks the U(1) gauge invariance associated with electromagnetism. 
In particular, to break U(1) down to $Z_2$, requires a Higgs scalar,  
that carries U(1) charge 2, i.e.  $\Lambda \rightarrow e^{2i\theta} \Lambda$,
 under the transformation (\ref{gauge}) \cite{FradkinPRD79}.

The physical interpretation of the field $\Lambda$
becomes clearer
 by writing down the effective action for $\Lambda$.
This is constrained only
by symmetry and gauge invariance, including its couplings to $z_\alpha$. 
One adds to the action (\ref{Sz}) the action for the Higgs field,
\begin{eqnarray}
\mathcal{S}_\Lambda &=&  \int d^2 r d \tau \Big[
|(\partial_\mu - 2 i A_\mu) \Lambda_a |^2 + \tilde{s} |\Lambda_a |^2 + \tilde{u} |\Lambda_a|^4 \notag\\&&\hspace{2cm}-i 
\Lambda_a \varepsilon_{\alpha\beta} z_\alpha^\ast \partial_a z_\beta^\ast + \mbox{c.c.} \Big].
 \label{SL}
\end{eqnarray}
Multiple fields $\Lambda_a$, with spatial indicies $a$, 
are necessary to account
for the space group symmetry of the underlying lattice.
The crucial term is the last one coupling $\Lambda_a$ and $z_\alpha$.
%

A mean-field treatment of $\mathcal{S}_z + \mathcal{S}_\Lambda$,
gives two possible condensates, and hence four possible phases
(i.e. neither, either, or both fields condensed), depending
on the sign of the two parameters $s_z$ and $\tilde{s}$
(compare Figure \ref{fig:sachdevphasediag}).

\begin{enumerate}[i.]

\item $s_z<0$, $\tilde{s}>0$: This state has $\langle z_\alpha \rangle \neq 0$ and $\langle \Lambda \rangle =0$.
The modes of the $\Lambda$ field are gapped and so not relevant.
This is a  N\'eel state.

\item $s_z>0$, $\tilde{s}>0$: This state has $\langle z_\alpha \rangle = 0$ and $\langle \Lambda \rangle  = 0$.
Again the $\Lambda$ modes are gapped and so not relevant.
This is the VBC state.

\item  $s_z<0$, $\tilde{s}<0$: This state has $\langle z_\alpha \rangle \neq 0$ and $\langle \Lambda \rangle  \neq 0$. Because of the $z_\alpha$ condensate, this state breaks spin rotation invariance, and we determine the spin configuration by finding the lowest energy $z_\alpha$ mode in the background of a non-zero $\langle \Lambda \rangle$ in 
Eq.~(\ref{SL}), which is
\begin{equation}
z_\alpha = \left( w_\alpha e^{i \langle \Lambda \rangle \cdot r} + \varepsilon_{\alpha\beta} w_\beta^\ast e^{-i \langle
\Lambda \rangle \cdot r} \right)/\sqrt{2}, \label{zw}
\end{equation}
with $w_\alpha$ a constant spinor.
Inserting  the above expression
(\ref{zw}) into Eq.~(\ref{Phiz}) gives a local
moment that is space-dependent
so that $\langle {\bf S}_i \rangle$ 
is given by Eq.~(\ref{SN}) with 
${\bf N}_1$ and ${\bf N}_2$
 given by Eq.~(\ref{Nw}) and 
the wavevector ${\bf Q} = (\pi,\pi) + 2\langle {\bf \Lambda} \rangle$.
Hence, the field $\Lambda$ measures the deviation of the spin fluctuations
from commensurability.

\item $s_z>0$, $\tilde{s}<0$: This state has $\langle z_\alpha \rangle = 0$ and $\langle \Lambda \rangle  \neq 0$.
This a $Z_2$ spin liquid.
Spin rotation invariance is preserved, and there is no VBC order 
because monopoles are suppressed
by the $\Lambda$ condensate \cite{SachdevNP08}.
\end{enumerate}


We also note that the last term in (\ref{SL}) which couples
the incommensurability to the gradient of a field has some similarity
to that which occurs in other field theories of incommensurate systems \cite{KleeNPB96}.

\subsection{sp(N) theory}

A specific realisation of the above considerations
 was given \cite{ChungJPCM01} in a study of the 
Heisenberg model on the anisotropic triangular lattice
[compare equation (\ref{HamJJ'})], in the large
$N$ limit of an $sp(N)$ approach \cite{ReadPRL91,SachdevIJMPB91,SachdevPRB92}.
In the $N \to \infty$ limit, mean-field theory is exact
and the bosonic spinons are deconfined.
The calculated mean-field phase diagram as a function of $J'/J$ and the
magnitude of the quantum fluctuations (which can be tuned by varying the
ratio of $N$ to the total spin $S$), exhibits four distinct phases as
in Figure \ref{fig:sachdevphasediag}.

Fluctuations at finite $N$, however, allow for 
U(1) gauge field fluctuations, which modify the mean-field results. 
In particular, the Berry's phase associated
with the instantons in the gauge field confine the spinons 
in the commensurate phase with short-range magnetic order.
However, they do not for the incommensurate phase with short-range order,
because their is a non-zero spinon pairing field in the diagonal ($J'$) direction.
This field carries a gauge charge of $\pm 2$ making it equivalent to a Higgs field, which prevents confinement in 2+1 dimensions \cite{FradkinPRD79}.

The dimerization pattern seen
 in near the decoupled chain limit ($J' \gg J$)
is similar to that found \cite{WhitePRB96} for a ladder         with zigzag coupling. Furthermore, spinon excitations are confined into pairs by the U(1) gauge force.
This phase is believed to be an analogue
of the  RVB state
 found on the isotropic triangular lattice quantum dimer model \cite{MoessnerPRB01}.
 The phase has ‘topological’ order; i.e., 
if the lattice is placed on a torus, the ground state becomes four-fold degenerate in the thermodynamic limit.

\subsection{Experimental signatures of deconfined spinons}
\label{sec:signature}

Thermal properties will reflect the presence of 
deconfined spinons and fluctuating gauge fields.
These have been calculated for a spinon Fermi surface
coupled to a $U(1)$ gauge field \cite{NavePRB07}.
The low-temperature  specific heat is dominated by a term
$\sim T^{2/3}$ due to gauge field fluctuations \cite{motrunich}.
The thermal conductivity is dominated by the  contribution
due to spinons, which give a term $\sim T^{1/3}$ \cite{NavePRB07}.

The low temperature specific heat data for \kcn3
\cite{YamashitaNP08}
 can be fit to  either the form expected for gauge fluctuations
or for a spinon Fermi surface without the gauge fluctuations \cite{RamirezNP08}.
Hence, it is not possible from the 
experimental data to definitely conclude that there are
gapless fermionic spinon excitations.

One question is: is the spin susceptibility simply related to the spinon susceptibility? This would imply that gauge fluctuations do not modify the spin susceptibility. If so one might expect that the NMR relaxation may exhibit a Korringa-like temperature dependence, i.e., $1/T_1 \sim T$, a temperature independent $1/T_2$ and Knight shift,                and a Korringa ratio of unity if there are deconfined spinons.


It has recently been argued that a definitive signature
of deconfined spinons in a Mott insulator 
would be a sizeable thermal Hall effect \cite{KatsuraPRL10}.
In an external magnetic field for the Hubbard model on a triangular lattice
there is an orbital interaction
 between the field and the spin chirality \cite{SenPRB95,motrunich}.
This leads to a thermal Hall effect \cite{KatsuraPRL10}
which is estimated to be larger than that due to conventional mechanisms
by a factor of order $(J \tau/\hbar)^2$, where $\tau$ is the spinon
scattering lifetime.
The latter is estimated to be about $10^{-12}$ s from the
 magnitude of the low temperature
thermal conductivity. 
However, as noted in Section \ref{sec:dmit} 
this thermal Hall effect is not seen in the
candidate spin liquid material Sb-1 \cite{YamashitaScience10}.


\subsection{Non-linear sigma models for magnons}
\label{sec:nlsigma}

The schematic phase diagram shown in Figure \ref{fig:qcp}
provides a means to understand the different qualitative behaviours that
can occur in non-linear sigma models, resulting from the presence of a
quantum phase transition between ordered and disordered (i.e., spin
liquid) phases.

Antiferromagnets which classically exhibit non-collinear magnet order,
such as the Heisenberg model on the triangular lattice,
may be described by a non-linear sigma model with $SU(N) \times O(2)$ symmetry \cite{ChubukovNPB94}.
A large $N$ expansion treatment has been given of such a
non-linear sigma model,
including fluctuations to order $1/N$. The physical spin-1/2 model has $N=2$.
The temperature dependence of the correlation length $\xi(T)$
in the renormalised classical regime \cite{ChubukovNPB94,azaria},
 is given by
\begin{equation}
\xi(T) = 0.021 \left(c \over \rho_s \right)
\left({4 \pi \rho_s \over T} \right)^{1/2}
\exp \left({4 \pi \rho_s \over T} \right)
\label{eqn:xi}
\end{equation}
where
$c$ is a spin wave velocity and $\rho_s$
 is a zero-temperature spin stiffness.
The static structure factor at the ordering wavevector is \cite{ChubukovNPB94}
\begin{equation}
S(Q) \simeq 0.85
\left({T \over 4 \pi \rho_s } \right)^4
\xi(T)^2.
\label{eqn:sq}
\end{equation}
The above equations show that $\rho_s$
 sets the temperature scale for the development of 
antiferromagnetic spin correlations.
For the isotropic triangular lattice non-linear
spin wave theory (to order $1/S^2$) \cite{ChubukovJPCM94} gives
$c=Ja$ and $\rho_s=0.06J$.
The above expressions are quite similar to those for
the $O(3)$ non-linear sigma
model that is relevant to the Heisenberg model 
on the square lattice \cite{chn89,Auerbach},
but with the factor $2\pi \rho_s$ replaced by $4\pi \rho_s$.

A monotonic increase in the NMR relaxation rates
with decreasing temperature occurs for a  non-linear sigma model
in the renormalised classical regime \cite{ChubukovNPB94,azaria},
which occurs as a magnetically ordered state is approached
at zero temperature.
It is found that $1/T_1 \propto T^{7/2} \xi(T)$
and $1/T_2 \propto T^{3} \xi(T)$,
when the correlation length $\xi(T) \gg a$.

In the quantum critical regime, close to a quantum critical
point, the temperature dependence of the NMR rates is \cite{ChubukovNPB94}
\begin{equation}
 1/T_1 \sim T^\eta, \ \ \ \ \ 1/T_2 \sim T^{(\eta-1)},
\label{qcrit}
\end{equation}
 where $\eta$ is the anomalous
critical exponent associated with the spin-spin correlation
function. Generally, for $O(N)$ non-linear sigma models
(e.g., that are appropriate for collinear antiferromagnets),
this exponent is much less than one \cite{chn89}.
For example, for $N=3$, $\eta=0.04$.

\subsection{Field theories with deconfined spinons}
\label{sec:qcdecon}

Field theories with deconfined spinons can
have  $\eta > 1$ \cite{ChubukovNPB94,alicea,IsakovPRB05}.
 Then $1/T_1 T$ {\it decreases} with decreasing
temperature, {\it opposite} to what occurs when the spinons are confined,
because then $\eta \ll 1$.
Hence, this is a significant experimental signature.
To leading order in $1/N$,
the $SU(N) \times O(2)$ model \cite{ChubukovNPB94}, has $\eta = 1 + 32/(3 \pi^2 N)$.
For $N=2$ this gives $\eta \simeq 1.5$,
comparable to the value deduced from NMR experiments
on \CN (Figure \ref{fig:shimNMR}).

\subsection{Field theories with bosonic spinons and visons}

\begin{figure}
\begin{centering}
\includegraphics[width=8cm]{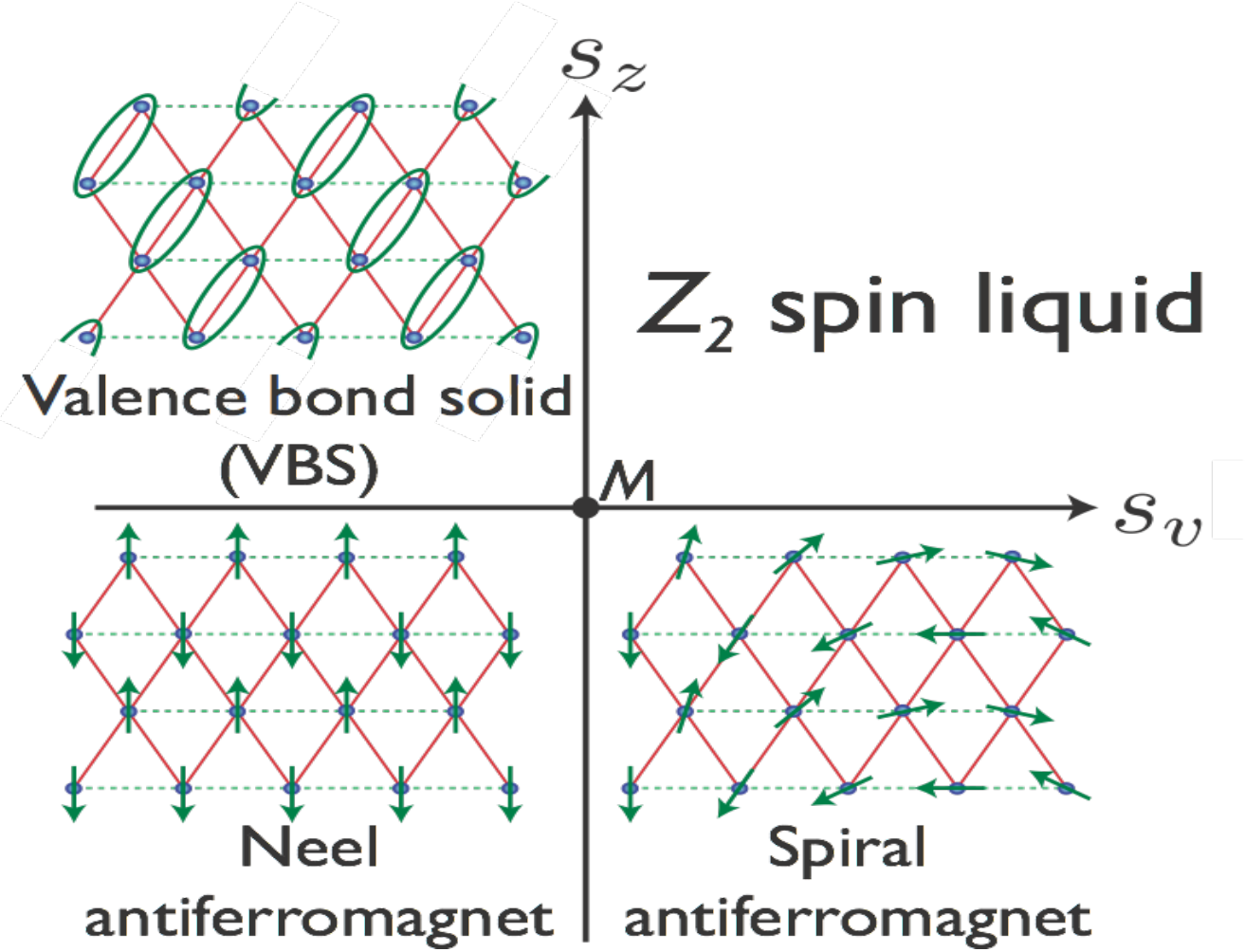}
\caption{
Phase diagram of a specific field theory with spinons (spin-1/2 bosons)
and visons (spinless bosons) \cite{XuPRB09}.
The vertical and horizontal axes describe the ``mass" of the visons and spinons, respectively. A similar mean-field phase diagram
is obtained for the field theory described by the action 
$\mathcal{S}_z + \mathcal{S}_\Lambda$ given by (\ref{Sz}) and
(\ref{SL}) with $s_v=\tilde{s}$.
  M denotes a multi-critical point.  One sees competition between phases similar to those found from series expansion studies of the Heisenberg model on the anisotropic triangular lattice \cite{ZhengPRB99}
 (cf. Figure \ref{fig:zhengphased}).
[Copyright (2009) by the American Physical Society].
}
\label{fig:sachdevphasediag}
\end{centering}
\end{figure}

Qi, Xu, and Sachdev \cite{QiPRL09,XuPRB09}  propose that the ground state 
of $\kappa$-(BEDT-TTF)$_2$Cu$_2$(CN)$_3$             
is a $Z_2$ spin liquid close to a quantum critical point with quasiparticles that are spin-1/2 bosons (spinons) and spinless bosons (visons).
The visons correspond to low-energy singlet excitations and can be viewed as vortices in the $Z_2$ gauge field
 associated with a liquid of resonating valence bonds.
They showed that at low temperatures
spinons dominate the NMR relaxation rate
and that visons dominate the thermal conductivity.
The visons form a dilute Boltzmann gas with a bandwidth of about 8 K, which the authors claim corresponds to the peak observed in the heat capacity and thermal conductivity. Note that this bandwidth is only about 3 per cent of the exchange interaction $J$, which sets the energy scale for the spinons.
Figure \ref{fig:sachdevphasediag} shows the phase diagram of one
of the field theories considered \cite{XuPRB09}.
The ``doubled Chern-Simons theory'' used implements
the mutual semionic statistics of the visons and spinons.

\subsection{Field theories with fermionic spinons and gauge fields}

An alternative approach \cite{leelee}
 starts with the slave rotor representation \cite{FlorensPRB04} and 
derives an effective Lagrangian describing fermionic spinons and `X bosons'
coupled to U(1) gauge fields.
In  the Mott insulating phase  the X bosons are gapped
and the ground state has a spinon Fermi surface coupled to a 
U(1) gauge field.
Although it is known that a compact U(1) pure gauge field is confining,
it is controversial as to whether such a field coupled to a matter
field can be deconfining.
At low temperatures there is the possibility of an Amperean instability
\cite{leelee2} which leads to pairing of the spinons.

%% file: quasiparticles.tex
\subsection{Effective field theories for quasi-particles in the metallic phase}
\label{sec:quasiparticle}

There have been many
papers written about possible field theories for the
metallic phase of doped Mott insulators \cite{LeeRMP06}.
A common ingredient of many of these theories is that the
quasi-particles interact strongly via fluctuations in
the gauge field associated with representions of the electron
operators in terms of the slave bosons.
However, in contrast little work has been done for the metallic
state associated with the bandwidth controlled Mott metal-insulator transition.

Qi and Sachdev \cite{QiPRB08}
consider field theories on the triangular lattice
which describe transitions from an insulating $Z_2$ spin liquid
state (with bosonic spinon excitations) to 
 metallic states with Fermi surfaces. 
They argue that near this insulator-metal transition
an excitonic condensate can form. This condensate involves pairing of 
 charge neutral pairs of charge $+e$ and charge $−e$ fermions. This condensate breaks the lattice space group symmetry.
They propose this state as an explanation of   an anomaly in thermodynamic
properties seen near temperatures
of about 6 K in $\kappa$-(BEDT-TTF)$_2$Cu$_2$(CN)$_3$.             
They also discuss the superconductivity associated
with the pairing of fermions of the same charge.

An alternative approach \cite{leelee,leelee2} 
 starts with the slave rotor representation \cite{FlorensPRB04} and 
derives an effective Lagrangian describing fermionic spinons and X bosons
coupled to U(1) gauge fields.
Compared to traditional (Kotliar-Ruckenstein) slave bosons the X boson
is relativistic. In the Mott insulating phase it is gapped 
and the holon and doublon are bound.

The corresponding field theory has been used to describe a 
continuous transition from a Fermi liquid to paramagnetic
Mott insulator with a spinon Fermi surface \cite{SenthilPRB08}.
 At the critical point
the quasiparticle weight $Z$ vanishes and the effective mass $m^*$ diverges.
 Nevertheless,
there is still a sharply defined Fermi surface.
Also, the product $Z m^*$ tends to zero as the transition is approached, whereas for
dynamical mean-field theory it tends to a non-zero constant.
As the temperature increases on the metallic side
 there is a crossover from the Fermi liquid
to a marginal Fermi liquid and then to a quantum critical non-Fermi liquid.
A universal jump in the intralayer resistivity of order  $h/e^2$ is predicted.
It is suggested that this theory is particularly relevant to \CNn.
However, as discussed in Section \ref{sec:ET} the transition appears
to be first order experimentally.
A logarithmic correction to the Fermi liquid
quadratic temperature dependence of
the resistivity is found, whereas a power closer to 2.5 is
observed experimentally (cf. section \ref{section:pressurestress}).

\section{Relation to other frustrated systems}

In the search for general organising principles we briefly review 
other classes for frustrated materials and models.
Some of the systems discussed below have been more 
extensively reviewed elsewhere \cite{NormandCP09,BalentsNat10}.

\subsection{$\beta$-(BDA-TTP)$_2X$}

A combined experimental and theoretical study was made of these 
organic charge transfer salts with the two anions $X$=SbF$_6$ and AsF$_6$
\cite{ItoPRB08}.
An extended H\"uckel calculation was used to argue that the
relevant effective Hamiltonian was a Hubbard model on an anisotropic
triangular lattice with three unequal hopping integrals, $t_0, t_1, t_2$.
(If two of these three hopping integrals are equal one obtains the
$t-t'$ model discussed extensively in this review).
These were calculated for the different crystal structures obtained as 
a function of uniaxial stress. All were found to vary within the range,
0.03-0.05 eV. Hence, these materials involve significant frustration.
The superconducting transition temperature $T_c$ measured  
as a function of uniaxial stress was compared to that calculated from
from a fluctuation exchange approximation 
\cite{ItoPRB08}.
The parameterisation of the band structure needs to be compared to the actual Fermi surface determined from angle-dependent magnetoresistance \cite{ChoiPRB03}.

\subsection{$\lambda$-(BETS)$_2X$}             

This family of materials has attracted considerable interest due to the
discovery of magnetic-field induced superconductivity
in the $X$=FeCl$_4$ material \cite{UjiNature01}.
At ambient pressure and zero magnetic field
it has a Mott insulating ground state, whereas the $X$=GaCl$_4$ material
is a superconductor.
Applying a magnetic field parallel to the layers  creates a metal, and for sufficiently high magnetic fields, superconductivity. This can be explained in terms of the exchange interaction between the localised magnetic
 Fe$^{3+}$ ions in the anion layer and the 
itinerant electrons in the layers of BETS molecules. When this exchange interaction is cancelled by the applied field 
 the electron spins effectively see zero magnetic field \cite{BalicasPRL01,CepasPRB02}.
One can also tune between Mott insulating, metallic, and superconducting states by varying the temperature or the relative concentration of magnetic 
 Fe$^{3+}$ ions and non-magnetic Ga$^{3+}$ ions (which effectively tunes
the magnitude of the exchange interaction). (For a review see \cite{UjiJPSJ06}).

The simplest possible lattice model Hamiltonian to describe this family of
materials is a Hubbard-Kondo model
with a Hubbard model on an anisotropic triangular lattice at half
filling with an exchange interaction between the electrons and localised
spin-5/2 spins  \cite{CepasPRB02}.
However, there are questions about the role of dielectric
fluctuations and charge ordering in these materials \cite{ToyotaCRC07}.
The fact that one can tune between ground states with perturbations
involving energy scales of the order of 1 meV (e.g. exchange
interactions, fields of order 10 Tesla
and temperatures of order 10 Kelvin) underscores how the interplay of
frustration and strong correlations leads to competition between
different ground states with very similar energies.
Given this tuneability more systematic studies of the role of frustration and 
possible spin liquid states in this family is worthy of further study.

\subsection{Sodium cobaltates}             

The material Na$_x$CoO$_2$ has attracted considerable
interest because of its large thermopower and rich
phase diagram which contains metallic, superconducting, insulating,
charge ordered, and various magnetic phases \cite{OngScience04}.
The $x=0$ member of the family should be described by a
single band Hubbard model on the isotropic triangular lattice, at
half filling \cite{MerinoPRB06}.
Due to the large geometric frustration of magnetic ordering
and the absence of Fermi surface nesting the ground state is
metallic below a critical value of
$U/t \simeq 8$ (Section \ref{sec:hubbard}).
NMR measurements on CoO$_2$ found that the Knight shift is
weakly temperature dependent and the spin relaxation rate
$1/T_1$ could be fitted to a Curie-Weiss form \cite{VaulxPRL07}.
There is also significant particle-hole symmetry and properties
of the model depend significantly on the sign of $t$.

It turns out that to describe the $x \neq 0$ materials, particularly
those with $x$ a rational number (e.g., $x=1/3,1/2,2/3$), one needs to take into account the spatial ordering of the Na$^+$ 
ions and the associated periodic potential experienced by electrons in the
cobalt layers \cite{MerinoPRB09,MerinoPRB10,PowellPRB10}.

\subsection{Cs$_2$CuCl$_4$}

The best evidence for deconfined spinon excitations
in an actual quasi-two-dimensional material is for this one.
Both Cs$_2$CuCl$_4$ and Cs$_2$CuBr$_4$
 can be described by a Heisenberg model
on the anisotropic triangular lattice.
From a range of experiments it is estimated that the value of
$J'/J$ is about 3 and 2 for 
Cs$_2$CuCl$_4$ and Cs$_2$CuBr$_4$, respectively \cite{ZhengPRB05,FjaerestadPRB07}.
A very detailed analysis of the effect of small residual interactions
such as a
Dzyaloshinskii-Moriya interaction and an external magnetic field
on the ground state has been performed \cite{StarykhPRB10}.

\subsection{Monolayers of solid $^3$He}             

A single monolayer of helium atoms can be adsorbed on graphite
plated with HD molecules.
At the appropriate areal density the atoms form
a solid with a hexagonal lattice. The $^3$He atoms have nuclear 
spin-1/2 and the spin degrees of freedom can be described by 
a Heisenberg model on the triangular lattice with multiple ring exchange.
No spin gap was observed down to temperatures as low at 10 $\mu$K \cite{MasutomiPRL04}.

At the density at which the monolayer solidifies into a 
$\sqrt{7} \times \sqrt{7}$ commensurate solid, 
a Mott-Hubbard transition between a  Fermi liquid and a magnetically disordered solid  is observed. This is signified by a diverging
linear co-efficient of the specific heat and
 and a diverging magnetization \cite{CaseyPRL03}.
This transition has also been investigated in bilayers; it
is found that  the interband coupling associated with the two layers
vanishes as the insulating phase is approached \cite{NeumannScience07}.
The experimental results are well described by a cluster DMFT
treatment of a bilayer Hubbard model on the triangular lattice \cite{Beach}.

\subsection{Pyrochlores}             

The pyrochlore lattice consists of a three-dimensional network of
corner sharing tetrahedra.
In a number of transition metal oxides the metal ions are located
on a pyrochlore lattice.
The ground state of the antiferromagnetic Heisenberg model on a 
pyrochlore lattice is a gapped spin liquid \cite{CanalsPRB00}.
The ground state consists of weakly coupled RVB (resonating valence bond) states on each tetrahedra.
However, Dzyaloshinskii-Moriya interactions have
a significant effect, leading to the formation of long-range magnetic
order \cite{ElhajalPRB05}.
The conditions necessary for 
deconfined spinons has been explored in
Klein type models \cite{NussinovPRB07}.
The repeat unit in this lattice consists of a tetrahedron
of four spins (giving an integer total spin) and
so the Lieb-Schultz-Mattis-Hastings theorem \cite{HastingsPRB04}
which can preclude gapped spin liquid ground states does not apply. 

The material KOs$_2$O$_6$ has a pyrochlore structure and is
found to be superconducting with a transition temperature of
about 10 K \cite{YonezawaJPCM04}.
Originally it was thought that the superconductivity
might be intimately connected to RVB physics \cite{AokiJPCM04}.
However, it now seems that the superconductivity
is s-wave and can be explained in
terms of strong electron-phonon interactions which
arise because of anharmonic phonons associated with 
``rattling" vibrational modes of the K ions which 
are located inside relatively large spatial 
regions within the cage of Os and O ions \cite{HattoriPRB10}.

\subsection{Kagome materials}             

The material herbertsmithite ZnCu$_3$(OH)$_6$Cl$_2$
has generated considerable interest as a realisation of the 
spin-1/2 Heisenberg model on the Kagome lattice.
However, it turns out that analysis of the experimental results
is significantly complicated by the presence of a small number
of impurities and by the Dzyaloshinskii-Moriya interaction \cite{GregorPRB08}.

Na$_4$Ir$_3$O$_8$ is a material in which the Ir ions have
spin-1/2 and  are located on a three-dimensional ``hyperkagome'' lattice of corner-sharing triangles. It has been proposed that the ground state of the Heisenberg model on this lattice may be
 a quantum spin liquid with spinon Fermi surface \cite{LawlerPRL08}.

The antiferromagnetic spin-1/2 Heisenberg 
model on the Kagome lattice has at 
times been thought to be a prime candidate for
a quantum spin model with a spin liquid ground state.
This is partly because the classical model has an infinite number of degenerate ground states. However, a few years ago a series expansion study \cite{SinghPRB07} found that the ground state was actually a valence bond crystal with a unit cell of 36 spins. This result was confirmed by a completely different numerical method based on entanglement renormalisation \cite{EvenblyPRL10}.
However,  very recent numerical results using the
density matrix renormalisation group (DMRG) \cite{Yan10} found
a spin liquid ground state, with a gap to both singlet and triplet excitations.

\subsection{Spin-1 materials}             

The spin-1 Heisenberg model on the anisotropic triangular 
lattice has been studied in the weakly coupled chain
limit ($J' \gg J$) using zero-temperature series
expansions about magnetically ordered spiral states \cite{PardiniPRB08}.
There is a critical interchain coupling $J/J'\sim 0.3-0.6 $ required to overcome
the Haldane spin gap (which occurs in the decoupled
chain limit, $J=0$). This critical coupling is an order of
magnitude larger than that required for the case of unfrustrated
coupling between chains (i.e. an anisotropic square lattice).
Hence, it may be that a Haldane phase can exist in a two-dimensional 
system. This raises an interesting question about whether this model
has topological order.

The family of materials LiV$X_2$  ($X=\textrm{O, S, Se}$) can be viewed as
 spin-1 systems on a triangular lattice.
The $X=\textrm{O}$ material has an insulating valence bond solid (VBS) ground state.
Upon cooling the $X=\textrm{S}$ compound undergoes a first-order phase
transition from a paramagnetic metal (possibly with a pseudogap)
 to a VBS insulator at 305 ~K \cite{KatayamaPRL09}.
The $X=\textrm{Se}$ material is a paramagnetic metal down to 2 K.

The material NiGa$_2$S$_4$ can be described by spin-1 antiferromagnet
on a triangular lattice. There is no sign of magnetic
order \cite{NakatsujiScience05} and it has been
proposed that the ground state is a spin nematic phase
which is stabilised by bilinear--biquadratic interactions \cite{TsunetsuguJPCM07}.

\subsection{Cuprates}             

One might not expect frustration to be important in these materials,
particularly because the parent material clearly undergoes antiferromagnetic
Neel ordering. 
However, a correlation has been found between
 the magnitude of next-nearest neighbour hopping on the square lattice, which frustrates the system, and the superconducting transition temperature, $T_c$
 \cite{PavariniPRL01}.

\subsection{$J_1-J_2$ model           }             

This a Heisenberg model on a square lattice where
$J_1$ and $J_2$ are the nearest- and next-nearest- neighbour
interactions, respectively. Thus, $J_2$ acts along {\it both} diagonals of each placquette
and is a frustrating interaction.
The model has been very widely studied with diverse techniques,
motivated by the hope that it would be model case of
a two-dimensional model where frustrations produce a spin liquid
ground state.
For small and for large $J_2/J_1$ the model has Neel order
with wave vector $(\pi,\pi)$
and $(0,\pi)$, respectively.
For intermediate $0.5 < J_2/J_1 < 0.7$ various studies have found a ground statewith no magnetic order; some are VBC states \cite{Becca09}.
Figure \ref{fig:becca}
shows how the true ground state is close to an RVB state without magnetic order.

The corresponding Hubbard model exhibits a 
subtle competition between $d_{x^2-y^2}$ superconductivity,
a Mott insulating phase, different magnetic orders, and a spin liquid state.
See for example
\cite{NevidomskyyPRB08}.

\subsection{Shastry-Sutherland lattice}             

\begin{figure}
\begin{centering}
\includegraphics[width=9cm]{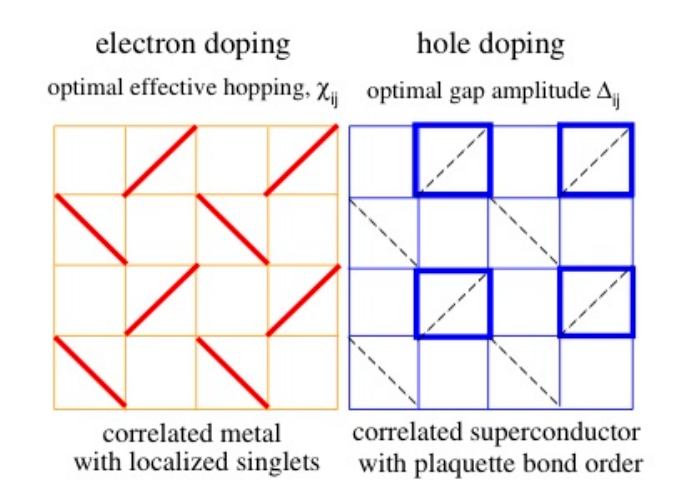}
\caption{
Particle-hole asymmetry of doped Mott insulators on the Shastry-Sutherland lattice \cite{LiuPRL07}.
The ground state of the Mott insulator is a valence bond crystal
with spin singlets along the diagonals shown as heavy lines 
on the left part of the figure. Electron doping produces a
metal with  these same spin singlet correlations preserved.
In contrast, hole doping produces a d-wave superconductor
with co-existing plaquette bond order.
[Copyright (2007) by the American Physical Society.]
}
\label{fig:shastry}
\end{centering}
\end{figure}

It has been argued that  SrCu$_2$(BO$_3$)$_2$
is a Mott insulator on this lattice  \cite{ShastryPTPS02}.
 The corresponding Heisenberg model has  an exchange interaction $J$ along all vertical and horizontal
bonds and a diagonal interaction $J'$ along every other plaquette 
\cite{ShastryPTPS02}.
It can be shown that for $J'/J > 1.44 \pm 0.02$ that the exact ground
state is a product of singlets along the
same diagonals that the $J'$ interaction occurs \cite{KogaPRL00}.
A variational Monte Carlo study \cite{LiuPRL07}
was made of  the corresponding $t-J$ model (including
three site hopping terms) away from half-filling
using a projected BCS wave function
with $t'=\pm 1.25 t$ and $J=0.3t$. 
The results are summarised in Figure \ref{fig:shastry} and
in the four points below.

\begin{enumerate}[i.]
\item There is significant particle-hole asymmetry. The
authors point out that 
one sign of $t'$ corresponds to electron doping and the other to hole doping while $t$ does not change. This is because, for $t'=0$ the lattice is bipartite. The sign of the hopping integrals only matters when the electrons (or holes) can traverse closed loops consisting of an odd number of lattice sites. On the Shastry-Sutherland lattice (and, indeed the triangular lattice) all closed loops consisting of an odd number of hops contain an odd number of hops with amplitude $t'$ (and hence an even number of hops with amplitude $t$; cf. Fig. \ref{fig:shastry}). Hence, only the sign of $t'$ matters for determining particle-hole asymetry. 

\item Hole doping produces d-wave superconductivity.
But, this is {\it not} the result of delocalisation of the pre-existing 
singlets in the Mott insulator since the
latter were along the diagonals.

\item Electron doping does not produce superconductivity, but only
a correlated metal with singlet pairing along the diagonal,
as in the parent Mott insulator.

\item The hole-doped superconducting state co-exists with plaquette bond
order where all the nearest neighbour spins have antiferromagnetic correlations. Thus the spin correlations are qualitatively different from
those in the parent Mott insulator.
\end{enumerate}

This shows that the competition between superconductivity and antiferromagnetism and resonating valence bonds that occurs when doping a frustrated Mott insulator is more subtle (and confusing) than suggested by Anderson's 
original conjecture \cite{AndersonScience87}.
On the other hand, one might argue that the parent Mott insulator is very different from the cuprates and organics because there are {\it no}
resonating valence bonds in the parent insulators for those classes of material.

\subsection{Surface of 1T-TaSe$_2$}             

This can be described by the Hubbard model on the isotropic
triangular lattice \cite{PerfettiPRB05}.
As the temperature decreases the bandwidth also decreases leading to a
metal-insulator transition. The observed ARPES spectrum was found
to be comparable to that calculated from the Hubbard model using
dynamical mean-field theory \cite{PerfettiPRB05}.

\subsection{Honeycomb lattice}             

A recent study presented the results of Quantum Monte Carlo simulations on the Hubbard model at half-filling on the honeycomb lattice \cite{MengNature10}.
This  is the relevant lattice for graphene and Pb and Sn on Ge(111).
As $U/t$ increases there is a phase transition from a semi-metal (which has gapless excitations at corners of the Brillouin zone, Dirac fermions) to a Mott insulating phase, for $U\simeq 3.5t$. More importantly,
 the authors also find that there is a spin liquid phase with a spin gap before entering a phase with antiferromagnetic order. The latter 
is what one expects from a strong coupling expansion (i.e. $ U \gg t$) which is described by an unfrustrated Heisenberg model \cite{PaivaPRB05}.
The spin liquid state has dimer-dimer correlations similar to that in a single hexagon which can be described by the RVB states of benzene.

Although the honeycomb lattice is bi-partite and so is 
not frustrated the authors suggest that near the Mott transition
 effective frustrating interactions occur.
For example, the ratio of the next-nearest neighbour exchange 
interaction  to the nearest-neighbour interaction
is $(t/U)^2$ \cite{DelannoyPRB05}.

In passing we note that 
the spin gap is very small, $\Delta_s \simeq t/40 \simeq J/40$.
The single-particle charge gap is also quite small in the spin liquid state
 being about $t/10 \simeq U/40$.
This illustrates the emergence of new low-energy scales due to the
presence of large quantum fluctuations.

%% file: quarter.tex
\section{Alternative models of organic charge transfer salts}
 
We have presented above the evidence that organic charge transfer salts are an experimental realisation of the half-filled anisotropic triangular lattice. We have argued that all of the important phenomena observed can be explained in terms of frustration and strong electronic correlations.
This, of course, requires  some objective judgement. For example, in which experimental results one views as important and which one views are mere details. Therefore, it is both natural and healthy that others working in these fields have introduced a number of alternative hypotheses. In this section we briefly discuss some of these ideas.
 
\subsection{Quarter filled models}

In order to construct effective low-energy half-filled models of $\kappa$-(\ETn)$_2X$ or Et$_n$Me$_{4-n}Pn$[Pd(dmit)$_2$]$_2$ one has to integrate out all of the internal degrees of freedom within the (\ETn)$_2X$ or [Pd(dmit)$_2$]$_2$ dimer. Several authors have considered models where one of these internal degrees of freedom is retained, i.e., models where a lattice site is a single \ET or Pd(dmit)$_2$ molecule and the lattice is quarter filled with holes. We note that such models must still integrate out all of the internal degrees of freedom within the molecule. Thus, it is not clear \emph{a priori} that even these models will contain all the physics relevant to the materials. However, all of the phenomena that are correctly described by half-filled models should be contained in the corresponding quarter-filled model. Therefore, one would not wish to argue that there is no description of these materials in quarter filled models. But, it may be that such a description is unnecessarily complicated. On the other hand, some of the papers discussed below argue that the correct description of the relevant physics is not captured by half-filled models, and that quarter-filled models are essential for the correct description of the low-energy physics.

Hotta has recently presented a model that interpolates between a range
of different polymorphs of (\ETn)$_2X$ in terms of the degree of
dimerisation and the splitting of the two bands nearest to the Fermi
energy \cite{HottaJPSJ2003}. This model is, in principle,
quarter-filled, but becomes half-filled in appropriate limits. Hotta
studied  this Hamiltonian in the mean-field approximation. Her
calculations found antiferromagnetic, charge ordered and metallic
states, but lacked superconductivity and exotic insulating states such
as spin-liquids and valence bond crystals. This may, of course, be due
to the inadequacies of the Hartree-Fock approximation.

Very recently Li \etal \cite{LiJPCM10} have proposed that a number of the exotic phases (spin-liquids, valence bond crystals, etc.) observed in the charge transfer salts can be understood in terms of single phase, which they call the `paired electron crystal'. The paired electron crystal phase has both charge order and spin order, and is reminiscent of the spin-Peierls phase observed in 1D chains and ladders.
This proposal is based on the results of exact diagonalisation calculations for a quarter-filled model on the anisotropic triangular lattice, which is not dissimilar from Hotta's model. This model has a large number of free parameters, including the hopping integrals, on-site and neighbouring site Coulomb repulsion, intra- and inter-site electron-phonon couplings and the spring constants of the relevant phononic modes. Li \etal only reported numerical results for a limited parameter set but state that similar results were obtained for a ``broad range'' of parameters. 
Li \etal have given a qualitative description of how a number of experimental results in $\kappa$-(\ETn)$_2X$, Et$_n$Me$_{4-n}Pn$[Pd(dmit)$_2$]$_2$ and other organic charge transfer salts might be explained in terms of the paired electron crystals. It will be interesting to see whether this idea can be developed into a fully quantitative theory of the experiments in the coming years.

\subsection{The role of phonons}

The role of phonons and the interplay between electron-phonon coupling and electronic correlations have received less attention. Other than the work of Li \etal \cite{LiJPCM10}, discussed above, and studies that conclude that the phonons play only a relatively minor role \cite{MerinoPRB00-phonon,HassanPRL05} most of the discussion of phonons has focused on the superconducting state \cite{VarelogiannisPRL02,MazumdarPRB08}. 
A proposal to use Raman scattering to rule out pairing via electron-phonon coupling
in the cuprates \cite{ChubukovPRB06} may also be relevant to the organics.

\subsection{Weak-coupling, spin fluctuations, and the Fermi surface}\label{sect:weak}

We have taken a strong coupling (i.e. large $U$) 
perspective where the key physics
is that associated with the RVB spin singlet
fluctuations in the Mott insulating phase.
From this perspective, geometric frustration destabilises magnetic
order and enhances RVB correlations.
The opposite weak-coupling (i.e. small $U$) perspective starts from
a Fermi liquid metallic state which becomes unstable due to
enhanced spin fluctuations associated with imperfect nesting
of the Fermi surface.
Theoretical work on the organic charge transfer
salts, which has taken such a weak-coupling
point of view, has been reviewed previously \cite{MoriyaRPP03}.

A weak-coupling spin fluctuation treatment
(e.g., the fluctuation-exchange approximation (FLEX)) of the
relevant Hubbard model can produce
some aspects of the phenomenology observed in the organic charge transfer salts.
These include a transition from a Fermi liquid metal, to
d-wave superconductivity, to an antiferromagnetic
Mott insulator \cite{SchmalianPRL98,KinoJPSJ98,KondoJPSJ98}.
But it is not clear  that the weak coupling approach can produce the following:

\begin{enumerate}
\item  Large effective masses associated with proximity to the Mott insulating state;

\item The first-order phase transition from a superconductor to a non-magnetically
ordered Mott insulator;

\item The first-order phase transition between an antiferromagnetic insulator
with a large magnetic moment (as  opposed to a small moment spin-density wave)
to a d-wave superconductor;

\item A $d + id$ superconductor near $t'=t$;

\item A Mott insulating valence bond crystal insulator;

\item A Mott insulating spin liquid.
\end{enumerate}
In contrast, the strong coupling approach gives a natural description of these phenomena, cf.  Figure \ref{fig:rvbphasediag}.

However, a widely held view is that the RVB and spin-fluctuation theories are just the strong and weak coupling limits of a more general theory that has yet to be articulated. This argument certainly has some merits, for example the weak coupling theory seems to give a reasonable account of the  cuprates in the overdoped regime, where correlations are        weaker than in more lightly doped cuprates.

This issue of a weak versus strong-coupling 
perspective is intimately connected with
the question of a ``glue" for superconductivity \cite{AndersonS07}.
 The issue can be nicely summarised as follows \cite{MaierPRB08}:
\begin{quote}
The question of whether one should speak of a “pairing glue” in the Hubbard and $t-J$ models is basically a question about the dynamics of the pairing interaction. If the dynamics of the pairing interaction arises from virtual states, whose energies correspond to the Mott gap, and give rise to the exchange coupling $J$, the interaction is instantaneous on the relative time scales of interest. In this case, while one might speak of an “instantaneous glue”, this interaction differs from the traditional picture of a retarded pairing interaction. However, if the energies correspond to the spectrum seen in the dynamic spin susceptibility, then the interaction is retarded and one speaks of a spin-fluctuation glue which mediates the d-wave pairing.
\end{quote}
Norman has reviewed the difficulty of distinguishing between these points of 
view in the cuprates, particularly with regard to the observation of Fermi surface
like properties in the underdoped state \cite{NormanP10}.

%% file: conclusions.tex
\section{Conclusions}

We have reviewed the significant progress that has been made in  understanding  frustrated materials in general and of organic charge transfer salts in particular. We are now in a position to partially answer some of the questions
posed in the introduction:

\begin{enumerate}
\item {\it Is there a clear relationship between superconductivity in 
organic charge transfer salts
 and in other strongly correlated electron systems?}

Yes. Superconductivity occurs in proximity to a Mott insulating phase.
There is substantial evidence that the superconducting state is unconventional 
in that there are nodes in the energy gap. The superfluid stiffness becomes vanishingly small at high pressures in the organics and at low dopings in the cuprates.

\item{\it Are there materials for which  the ground state of the Mott insulating phase is a spin liquid? }

Yes. The strongest candidate materials are \CN and Sb-2.
Neither of these materials show any evidence of magnetic ordering down
to temperatures four orders of magnitude smaller than the antiferromagnetic coupling between neighbouring spins. 

\item{\it  What is the relationship between spin liquids and superconductivity?
In particular, does the same fermionic pairing occur in both?}

With increasing pressure there is a first order-phase transition from
the spin liquid state to a superconducting state.
There is no definitive evidence yet that the same fermionic pairing
occurs in both states.
A possible hint that this is the case is the similarity between the temperature
dependence and magnitude of the thermal conductivity
in the spin-liquid phase of \CN and the superconducting state of \NCSn.

\item{\it What are the quantum numbers (charge, spin, statistics) of the quasiparticles in each phase?}

These appear to be quite conventional in the Neel ordered Mott insulating states, the superconducting states, and the metallic state away from the Mott transition. This question remains open in the spin liquid phases of \CN and Sb-1.

\item{\it Are there deconfined spinons in the Mott insulating 
spin liquid phase? }

The strongest evidence comes from the temperature dependence
of the NMR relaxation rate and the thermal conductivity at low temperatures. This seems to suggest that there are deconfined spinons in Sb-1, but that \CN is fully gapped.
However, the statistics of these spinons is an open question.

\item{\it Can spin-charge separation occur in the metallic phase?
}

There is no evidence of spin-charge separation in the metallic state yet.

\item{\it  In the metallic phase close to the Mott insulating phase 
is there an anisotropic pseudogap, as in the cuprates?}

NMR measurements
 suggest there is a pseudogap in the less frustrated materials. The 
anisotropy of this pseudogap in momentum space
 has not yet been mapped out experimentally.
How the formation of the pseudogap may
be  related to the crossover with decreasing temperature from a bad metal 
to a Fermi liquid metal is not clear.

\item{\it What is  the simplest low-energy
effective quantum many-body Hamiltonian on a lattice that can
describe all possible ground states of these materials?}

There is no evidence yet that one needs to go beyond the Hubbard
model on the anisotropic triangular lattice at half filling.

\item{\it  Is a RVB variational wave function 
 an appropriate theoretical description of
the competition between the Mott insulating and the superconducting phase?}

The Gossamer-RVB hypothesis is qualitatively consistent with experimental data reported so far.

\item{\it Is there any significant difference between destroying the Mott insulator by hole doping and by reducing correlations?}

Perhaps.
This question is only beginning to receive attention.
It does seem that in the organics that the effective mass
of the quasi-particles $m^*$ increases significantly as the Mott
insulator is approached whereas in the cuprates there
is little variation in $m^*$ with doping.

\item{\it  For systems close to the isotropic triangular lattice, does the superconducting state have broken time-reversal symmetry?}

There are no experimental studies of this question yet.
Resolving the question theoretically will require high level
computational studies beyond what is currently possible.
To put this in perspective, there is still no consensus
as to whether the doped Hubbard model on the square lattice
has a superconducting ground state \cite{Scalapino06}.

\item{\it  How can we quantify the extent of frustration? Are there  differences between classical and quantum frustration? If so what are the differences?}

A number of different measures of frustration have been proposed. A clear example of quantum frustration is kinetic energy frustration in, say, the tight binding model, which has no classical analogue. For spin models the differences between quantum and classical frustration are less clear cut and may be a purely taxonomic question.

\item{\it What is the relative importance of frustration and 
static disorder due to impurities?}

This question has not yet received significant attention. The destruction of the non-magnetic state in Sb-2 by, non-magnetic, Et$_3$MeSb$^+$ impurities provides a particularly dramatic case to study.

\item{\it Is the ``chemical pressure" hypothesis valid?}

For the weak frustrated \ET salts a number of experimental features collapse onto a single curve, to within experimental error when plotted against the superconducting critical temperature for a range of materials \cite{PowellPRB09}. This is a success for the chemical pressure hypothesis. The more strongly frustrated \CN behaves differently.

Recent DFT calculations are also consistent with the hypothesis.
A definitive microscopic explanation of the chemical pressure hypothesis  will require further characterisation of the pressure 
and anion dependence of the Hubbard model Hamiltonian parameters 
$t$, $t'$, and $U$.
A powerful approach to this problem would be to combine 
state-of-the-art band structure calculations with experimental
characterisation of the Fermi surfaces using AMRO.

\item{\it Is there quantum critical behaviour associated with  quantum phase transitions    in these materials?}

This is not clear. The most compelling evidence may be the temperature dependence of the NMR relaxation rate in \CNn (Figure \ref{fig:shimNMR}).

\item{\it  Do these materials illustrate specific ``organising principles" that are
useful for understanding other frustrated materials?}

\begin{enumerate}

\item Frustration suppresses long range fluctuations, which improves the the accuracy of mean field theories, such as DMFT, in the normal state.

\item In frustrated systems small changes in parameters can lead to dramatic changes in physical properties of the system. For example, a wide range of insulating phases are seen in the Et$_n$Me$_{4-n}Pn$\dmittwo salts, despite their similar chemistry.

\end{enumerate}

\end{enumerate}

\subsection{Some open questions}

There remain many questions still to be answered. Here we outline some of the most important issues still to be resolved:

\begin{enumerate}

\item Does the excitation 
spectrum change as one moves between phases? And, if so, how?
There is significant evidence in the cuprates that     
the excitation spectrum has essentially 
the same form ``d-wave" form in the pseudogap and
superconducting phases.
This is seen in ARPES \cite{ShiEPL09}, STM \cite{JLeeScience09},
 and thermal conductivity \cite{DoironPRL06}.

\item Quite different physical pictures of the
spin liquid state has been proposed for \CNn.
In particular, Sachdev 
and collaborators argue that the spinons are bosonic, whereas Lee and collaborators argue that the spinons are fermions and there is a well-defined 
Fermi surface.
We need a ``smoking gun" experiment to distinguish these
two proposals.

\item The observation of a valence bond crystal in
EtMe$_3$P[Pd(dmit)$_2$]$_2$
 (P-1 in our notation) is exciting.
On the one hand, this may be a realisation of a long sought after state
of matter.
The fact that this state can be transformed into a superconducting 
state with pressure is even more interesting.
On the other hand, there remains  an open question as to 
whether coupling to the lattice is necessary for stabilisation of
this state. Therefore, understanding the role of the lattice in stabilising the VBC phase is a clear pirority.

\item Thermal conductivity measurements provide a sensitive probe of
the quasi-particle excitation spectrum.
Measurements in materials such as \Br which are close
to the Mott transition should be a priority. 


\item Observation of deviations from the Weidemann-Franz law
(which gives a universal value for the ratio of the thermal
and charge conductivities in a Fermi liquid metal)
is a potential signature of spin-charge separation. However,
both theoretically and experimentally finding such deviations has proven to
 have 
a convoluted and confusing history \cite{SmithPRB05,SmithPRL08}. A careful study of the Weidemann-Franz law in the organic charge transfer salts could, however, provide significant new insights into the question of spin-charge separation in these materials.

\item What is the origin of the very different temperature dependences of the Nernst effects in \Br and \NCSn? What are the roles of superconducting fluctuations, electronic nematic order and proximity to the Mott transition?

\item Is the superfluid stiffness at high chemical pressures as small as $\mu$SR experiments suggest? Does hydrostatic pressure have the same effect? 

\item What is the underlying physical cause of this small superfluid stiffness?
Is it the same as for the overdoped cuprates, where the decreasing stiffness 
with increasing doping has been proposed to be due to pair breaking from
impurities \cite{TallonPRB06}?

\item What is the symmetry of the superconducting state in the superconducting
states derived from the spin liquid or a valence bond crystal? There are strong correlations between ferromagnetic fluctuations and p-wave superconductivity and nascent Ne\'el order and d-wave superconductivity. Presumably the spin fluctuations are rather different in the spin liquid and VBC phases. Therefore, it is possible that they would lead to different superconducting orders.

\item Is time-reversal symmetry broken in the superconducting state of any of these frustrated materials? The superconducting phases which
occur upon applying pressure to the spin liquids \CN and
EtMe$_3$Sb[Pd(dmit)$_2$]$_2$
and the 
valence bond crystal
EtMe$_3$P[Pd(dmit)$_2$]$_2$
would seem to be particularly promising systems to exhibit  superconductivity that breaks time reversal symmetry.

\item If large enough single crystals could be grown inelastic
neutron scattering could provide direct measurement of
the spin excitation spectrum and the signatures of 
deconfined spinons such as a high energy continuum.
Also, observation of an analogue of the neutron resonance mode
seen in the cuprate and pnictide superconductors \cite{ChristiansonNature08}
 could be important.

\item We have seen that the precise value of the
parameter $t'/t$ has       a dramatic effect
on the ground state of the system.
Hence, it is desirable to have  DFT
calculations   for the \dmit family of charge transfer salts. Experimental measurements that test the accuracy of such calculations, such as AMRO in the normal state, would also be of significant value.

\item The deviation of $t'/t$ from unity is a measure
of how far the electronic structure deviates
from the isotropic triangular lattice.
Hence, it is worth asking
whether there is  some structural parameter (e.g. deviation of the 
shape of the first
Brillouin zone from a hexagon) which can be correlated with
this ratio. There have been previous attempts to provide a unified
view of structural trends
 (cf. \cite{MoriBCSJ98,MoriCR04,YamochiJACS93,ShaoCM09})
but more work is needed        to relate these trends
in a definitive manner to electronic properties.

\item In the presence of a constant magnetic field $B$ perpendicular to the layers
a fluctuating $U(1)$ gauge field will modify the effect of 
$B$ on transport properties. A significant amount of analysis of the related
problem for the fractional quantum Hall liquid near
filling factor $\nu=1/2$ \cite{EversPRB99,WolfleASSP00}  has been performed.
Recently, corrections to the Lifshitz-Kosevich form for the temperature
dependence of the magnitude of quantum oscillations were calculated \cite{Fritz}.
More general results for a non-Fermi liquid
associated with quantum criticality 
were then derived using the holographic correspondence \cite{HartnollPRB10}.
A similar analysis of the 
effect of gauge fluctuations on AMRO may provide measurable signatures
of a fluctuating $U(1)$ gauge field in these materials.

\item It is desirable to obtain a better understanding of the
thermal expansion anomalies associated with the superconducting,
pseudogap, and spin liquid transitions \cite{MannaPRL10}.
These anomalies
 may reveal the spatial symmetry breaking associated with the transitions.
With this goal, a  Ginzburg-Landau theory
for the acoustic anomalies associated with the superconducting transition
has been developed \cite{DionPRB09}.

\end{enumerate}

Finally, we stress that in seeking to explain the rich physics still to be understood in frustrated materials in general and organic charge transfer salts in particular an important task for the community is to generate multiple hypotheses that may explain the data \cite{PlattScience64}. It is then important to design and execute experiments that clearly distinguish between these hypotheses.